\documentclass[twocolumn,10pt]{IEEEtran}
\usepackage{indentfirst}
\usepackage{amsmath}
\usepackage{amsthm}
\usepackage[final]{graphicx}
\usepackage{dsfont}
\usepackage{amsfonts}
\usepackage{amssymb}
\usepackage{bm}
\usepackage{subfigure}
\usepackage{booktabs}
\usepackage{graphicx}
\usepackage{caption}
\usepackage{epstopdf}
\usepackage{CJK}
\usepackage{cite}
\usepackage{geometry}
\usepackage{pifont}
\usepackage{hyperref}
\usepackage{color}

\newcommand{\ba}{\begin{array}}
\newcommand{\ea}{\end{array}}
\newcommand{\bena}{\begin{eqnarray}}
\newcommand{\eena}{\end{eqnarray}}

\newcommand{\ec}{\end{center}}

\newcommand{\ei}{\end{itemize}}
\newcommand{\benu}{\begin{enumerate}}
\newcommand{\eenu}{\end{enumerate}}
\newcommand{\bdes}{\begin{description}}
\newcommand{\edes}{\end{description}}

\newcommand{\et}{\end{tabular}}

\newcommand{\circlambda}{\mbox{$\Lambda$
             \kern-.85em\raise1.5ex
             \hbox{$\scriptstyle{\circ}$}}\,}

\def\cC{\mbox{$\mathcal C$}}
\def\cC{\mbox{$\mathcal C$}}

\def\cN{\mbox{$\mathcal N$}}
\def\cN{\mbox{$\mathcal N$}}

\newcommand{\bbm}{\begin{bmatrix}}
\newcommand{\ebm}{\end{bmatrix}}
\newcommand{\bit}{\begin{itemize}}
\newcommand{\eit}{\end{itemize}}
\newcommand{\ben}{\begin{enumerate}}
\newcommand{\een}{\end{enumerate}}
\newcommand{\bdesc}{\begin{description}}
\newcommand{\edesc}{\end{description}}
\newcommand{\bea}{\begin{array}}
\newcommand{\eea}{\end{array}}
\newcommand{\beqa}{\begin{eqnarray}}
\newcommand{\eeqa}{\end{eqnarray}}
\newcommand{\be}{\begin{equation}}
\newcommand{\ee}{\end{equation}}

\newcommand{\Comment}[1]{}

\newcommand{\bor}{{\mbox{\boldmath $r$}}}

\newcommand{\bv}{{\mbox{\boldmath $v$}}}
\newcommand{\bx}{{\mbox{\boldmath $x$}}}

\newcommand{\bz}{{\mbox{\boldmath $z$}}}

\newcommand{\bD}{{\mbox{\boldmath $D$}}}

\newcommand{\bI}{{\mbox{\boldmath $I$}}}

\newcommand{\bM}{{\mbox{\boldmath $M$}}}

\newcommand{\bQ}{{\mbox{\boldmath $Q$}}}
\newcommand{\bR}{{\mbox{\boldmath $R$}}}
\newcommand{\bS}{{\mbox{\boldmath $S$}}}

\newcommand{\bU}{{\mbox{\boldmath $U$}}}
\newcommand{\bV}{{\mbox{\boldmath $V$}}}
\newcommand{\bX}{{\mbox{\boldmath $X$}}}

\newcommand{\bZ}{{\mbox{\boldmath $Z$}}}

\newcommand{\btheta}{{\mbox{\boldmath $\theta$}}}

\newcommand{\bLambda}{{\mbox{\boldmath $\Lambda$}}}

\newcommand{\dmin}{\begin{displaystyle}\min\end{displaystyle}}

\IEEEoverridecommandlockouts

\begin{document}
\title{Classification Schemes for the Radar Reference Window: Design and Comparisons}

\vspace{0.1cm}

\author{Chaoran Yin, Linjie Yan, Chengpeng Hao, \IEEEmembership{Senior Member, IEEE},
Silvia Liberata Ullo, \IEEEmembership{Senior Member, IEEE}, Gaetano Giunta, \IEEEmembership{Senior Member, IEEE},
Alfonso Farina, \IEEEmembership{Life Fellow, IEEE}, and Danilo Orlando, \IEEEmembership{Senior Member, IEEE}
\thanks{This work was supported
by the National Natural Science Foundation of China under Grant 61971412 and
Grant 62201564. ({\em Corresponding author: Chengpeng Hao.})}
\thanks{Chaoran Yin, Linjie Yan, and Chengpeng Hao are with the Institute of Acoustics, Chinese Academy of Sciences,
Beijing, 100190, China. Chaoran Yin is also with the University of Chinese Academy of Sciences, Beijing, 100049, China. E-mail: {\tt yinchaoran18@mails.ucas.ac.cn; yanlinjie16@163.com; haochengp@mail.ioa.ac.cn}.}
\thanks{Silvia Liberata Ullo is with Universit\`a degli Studi del Sannio, 82100 Benevento, Italy. E-mail: {\tt ullo@unisannio.it}.}
\thanks{Gaetano Giunta is with  the Department of Industrial, Electronic, and Mechanical Engineering, Universit\`a degli Studi
Roma TRE, 00146 Rome, Italy. E-mail: {\tt gaetano.giunta@uniroma3.it}.}
\thanks{A. Farina is with Selex ES (retired), a consultant, Roma, Italy. E-mail: {\tt alfonso.farina@outlook.it}.}
\thanks{Danilo Orlando is with Universit\`a degli Studi ``Niccol\`o Cusano'', 00166 Roma, Italy. E-mail: {\tt danilo.orlando@unicusano.it}.}
}

\maketitle

\begin{abstract}
In this paper, we address the problem of classifying data within the radar reference window in terms
of statistical properties. Specifically, we partition these data into statistically homogeneous subsets
by identifying possible clutter power variations with respect to the cells under test (accounting for possible range-spread targets)
and/or clutter edges. To this end, we consider different situations of practical interest and
formulate the classification problem as multiple hypothesis tests comprising several models
for the operating scenario. Then, we solve the hypothesis testing problems by resorting to suitable
approximations of the model order selection rules due to the intractable mathematics associated
with the maximum likelihood estimation of some parameters. Remarkably, the classification results provided by the proposed architectures represent an advanced clutter map since, besides the estimation of the clutter parameters,
they contain a clustering of the range bins in terms of homogeneous subsets. In fact,
such information can drive the conventional detectors towards more reliable estimates
of the clutter covariance matrix according to the position of the cells under test.
The performance analysis confirms that the conceived architectures represent a viable means
to recognize the scenario wherein the radar is operating at least for the considered simulation parameters.
\end{abstract}


\IEEEpeerreviewmaketitle

\section{Introduction}
Target detection is a task that is ubiquitous in radar systems
\cite{richards2013principles,HaoSpringer,Farina-Handbook} and,
due to its vital importance, it has attracted and continues to attract
the attention of the radar community. More importantly, the advances in technology
and the consequent spread of radars in several aspects of real life have led to more challenging
operating scenarios that might no longer meet the design requirements of the classical
detection architectures. In fact, the most common assumption is that the statistical characterization
of the interference (i.e., clutter and noise) in secondary data,
collected in the vicinity of the Cell Under Test\footnote{ Recall that secondary data are exploited to estimate
the interference parameters in the CUT and, then, such estimates are used to ``equalize'' the interference.
} (CUT),
is the same
as that in the CUT (homogeneous environment)
\cite[and references therein]{Kelly1986GLRT,AMF1992,kelly-tr,Bandiera2009,Conte-RangeSpread,KL2021II,
Hao2016Joint,Cai1992,9372970,Yan2017Symmetric,DD,771034,7126163,596885,1605248,301849,SD1,9165125,DeMaioSymmetric,LiuWeiJian2021,
Hao2016Polar,Hao2015Persy,Spillover2011Danilo,Orlando-RAO,MaioOrlando-Persy}.
The first contributions in the context of space-time adaptive processing where
the homogeneous environment is exploited at the design stage, are \cite{Kelly1986GLRT,AMF1992,kelly-tr}.
Actually, they paved the way for a plethora of other works based upon the same interference model but
with additional design assumptions accounting for different situations of practical value.
For instance,\footnote{The interested reader is cautioned that the list of references
provided here is not exhaustive and other contributions, not reported here for brevity, exist.}
when the target signature experiences a slightly different direction from the nominal one due
to imperfect modeling of the nominal steering vector, calibration/beampointing errors, or target position
within the mainlobe, the subspace paradigm represents an effective means to face these situations
as shown in \cite{Bandiera2009,DD,SD1,Orlando2022UnifiedSubspace,301849}.
Another useful signal model in radar arises from the fact that, depending on the system range resolution,
a target can be resolved into multiple scatterers that are distributed over the range leading
to the so-called range-spread or extended target model \cite{Conte-RangeSpread,DD,SD1,771034,596885,1605248}.
Actually, this model comes also in handy to account for the possible spillover of target energy
between adjacent range bins \cite{Spillover2011Danilo} or when the system oversamples the
backscattered echoes \cite{7126163}.
Finally, it is worth mentioning that specific structures for the clutter covariance matrix
can improve the estimation quality of the unknown parameters. As a matter of fact, a
radar system equipped with a symmetrically spaced linear
array or using symmetrically spaced pulse trains could collect
data statistically symmetric in forward and reverse directions.
This results into an interference covariance matrix which
shares a so-called ``doubly'' symmetric form, i.e., Hermitian
about its principal diagonal and persymmetric about its cross
diagonal. The number of unknowns associated with a persymmetric covariance matrix
is lower than that for the general Hermitian case leading to enhanced estimation and, hence, detection performance
as shown in \cite{Cai1992,Hao2015Persy,MaioOrlando-Persy,Hao2016Joint}.
Another useful covariance structure is related to ground clutter whose power spectral density is symmetric
with respect to zero-Doppler frequency. As a result, the associated covariance matrix is real symmetric
\cite{Hao2016Joint,Yan2017Symmetric,DeMaioSymmetric}.

However, in many situations of practical interest, the homogeneous environment does not represent
a good scenario approximation due to the presence of
clutter regions with different properties leading to architectures
with poor detection performance and/or that cannot
keep under control the number of false alarms \cite{guerci2010cognitive,guerci2014space}.
Thus, it appears clear that new insights are necessary
and more articulated models have to be designed starting from the well-known
partially-homogeneous environment \cite{Kraut-PHE,Hao2014Pers,Hao2015oversample,YLJ2020parametric,
Foglia2017Symmetry,SD1,Hao-PerRao2012,Liuw-DoubleSpaceII,Hao2016Symmetry,GaoPACE,Maio2014},
where interference in the secondary data shares the same structure of the covariance
matrix as in the CUT but a different power level,
up to the most general case where secondary data are
heterogeneous in terms of statistical properties.
A widely used clutter model for heterogeneous environments assumes that
each range bin shares the same structure for the clutter covariance matrix but different
scaling factors that are representative of different clutter types (and reflectivity).
The interested reader is referred to \cite[and references therein]{DaniloHete,recursive2002Conte,rs13091628,COLUCCIA2022108401}
for examples related to this model.

From the above discussion it is quite evident that most of the existing adaptive detection architectures
are devised assuming a specific operating environment and clutter model.
Therefore, clutter classification becomes a relevant issue to select the detector and/or training data
that are perfectly matched to the current operating scenario. Such a choice would be a key factor for achieving
higher performance. In this context, pioneering solutions based upon the neural networks
can be found in \cite{Haykin1991Neural,Bouvier1995ICASSP}, where the classifier
distinguishes several types of environment returns including birds, weather, ground, and sea clutter.
The same problem is recently addressed in \cite{rs13224588} by means of an algorithm
based upon the support vector machine \cite{Murphy2012}.
Clutter classification can be also accomplished in terms of covariance structure, which is related
to specific clutter properties and types, as shown in \cite{VC2017CMC,Vincenzo2019clasplusdete},
where the so-called Model Order Selection (MOS) rules \cite{MOS1,MOS2} are applied to deal with
multiple (possibly nested) hypotheses. In \cite{Vincenzo2019Hetegeneous} the above approach is extended to the case of
heterogeneous environments. The common assumption behind these important contributions is
that data under test are characterized by the same covariance structure as also in \cite{Liujun2019Training}
where a binary hypothesis test is solved to discriminate between the homogeneous and the heterogeneous (with respect
to the clutter power) environment. In \cite{XU2021108127}, the aforementioned assumption is no longer used at the design stage
and an architecture grounded on the Generalized Likelihood Ratio Test (GLRT) is devised to declare
the presence of a possible clutter edge within a window of consecutive range bins. The extension of \cite{XU2021108127}
to the case of specific covariance structures can be found in \cite{WangTianqi2022Edge}.
In \cite{Pia2021Learning}, range bin clustering algorithms based upon clutter covariance matrix are conceived
to partition the entire region of interest into homogeneous subsets of (not necessarily contiguous) range cells.
Finally, it is worth mentioning the knowledge-aided paradigm as another
effective means to guide the system towards reliable clutter parameter estimates
\cite{1593334,4014432,KB-melvin}. It consists
in accounting for all the available {\em a priori} informations about the region of interest to
exclude inhomogeneities from the computation of the sample covariance matrix \cite{guerci2010cognitive}.

All the above examples show that the {\em a priori} knowledge about the radar operating environment is
a precious information that can be suitably exploited to come up with reliable estimates
of the clutter parameters. Besides, highly reliable estimates allow for an enhanced adaptivity
with respect to the clutter in the CUT improving the detection capabilities of the system.
From this perspective, in the present paper, we focus on the traditional reference window
used to collect secondary data and formed by the CUT, lagging plus leading windows, and
guard cells \cite{richards2013principles}. Then, all these data are used to
conceive innovative classification schemes capable of identifying different situations
of practical interest.
Since two hypotheses only are not enough to account for all such situations, it naturally follows that
this problem cannot be modeled in terms of
a conventional binary hypothesis test as in \cite{Liujun2019Training,XU2021108127,WangTianqi2022Edge}
but a multiple hypothesis test is required.
As a matter of fact, each hypothesis corresponds to one of the following scenarios:
primary and secondary data are statistically homogeneous;
clutter in primary data have a different power level with respect to that in secondary data
(partially-homogeneous environment); finally,
one or two clutter edges are present in the secondary data unlike \cite{XU2021108127,WangTianqi2022Edge}, where
only one clutter edge is assumed.

Three remarks are now in order. First, the partially-homogeneous model developed in this paper
is different from the conventional one since the overall covariance matrix
originates from the sum of thermal noise and clutter components. Second, training data partitioning
is performed with the constraint that the homogeneous regions are formed by contiguous range bins
unlike \cite{Pia2021Learning} where misclassification errors within a given region might occur.
Third, from an operating point of view, a radar system can perform a preliminary scan by moving the reference
window over the region of interest to build up a kind of clutter map. This map can be used for secondary data selection
once the detection function is active.

Thus, as stated before, the classification problem at hand is
formulated in terms of a multiple hypothesis test and solved by resorting to decision schemes
relying on the MOS rules.
Such a design choice is dictated by the fact that in the presence of nested hypotheses
the likelihood function monotonically increases with the number of parameters. Otherwise
stated, the Maximum Likelihood Approach (MLA) is inclined to select the model with the
highest number of unknown parameter \cite{MOS1,1420342}.
In order to contrast this natural inclination of the MLA in the presence of nested hypotheses,
the MOS rules exploit suitable penalty terms
that promote low-dimensional models \cite{MOS1,1420342}. In fact, the general structure of a MOS decision statistic
is given by the
negative of the compressed log-likelihood function under a specific model order
plus a penalty term that depends on the number of unknown parameters under that model.
Here, ``compressed'' means that the unknown parameters are
replaced by their respective maximum likelihood estimates (MLEs).
The selection of the model order
is accomplished by minimizing the above structure\footnote{In \cite{MOS1}, it is shown
that such a structure arises from the minimization of suitable approximations of Kullback-Leibler divergence
between the true data density and a family of candidates. It is clear that other approaches can be
pursued as in \cite{1420342}.}
with respect to the number of parameters (model order).
The herein considered MOS rule are the Akaike Information Criterion (AIC),
the Generalized Information Criterion (GIC), and the Bayesian Information Criterion (BIC).
However, in this work, we derive suitable approximations
of the MOS rules due to the difficult mathematics related to the maximum likelihood estimation of some parameters.
In addition, we consider two models for the clutter covariance matrices which differ in
the power variation over the range bins (a point better explained in Section \ref{section2}).
Finally, the classification capabilities of the proposed architectures are assessed using synthetic data.
The illustrative examples reveal that they can return the correct operating hypothesis with
high probability at least for the considered parameter values.

The remainder of this paper is organized as follows. The next section provides a formal
statement of the problem at hand along with the preliminary definitions that will be used
in the ensuing derivations. Section \ref{section3} contains the design of the classification architectures,
whereas Section \ref{section4} shows the classification performances through numerical examples.
Concluding remarks and hints for future research lines are discussed in Section \ref{section5}.
Finally, some mathematical derivations are confined to the appendix.

\subsection{Notation}
In the sequel, vectors (matrices) are denoted by boldface lower (upper) case letter.
Superscripts $(\cdot)^T$ and $(\cdot)^\dag$ denote transpose and complex conjugate transpose, respectively.
We denote by $\bS(i,j)$ the $(i,j)$th element in the matrix $\bS$.
$\mathds{R}^{m\times n}$ and $\mathds{C}^{m\times n}$ are real and complex matrix spaces of dimension $m\times n$.
Given a vector $\bx\in\mathds{C}^{N\times 1}$, we denote by $diag(\bx)\in\mathds{C}^{N\times N}$ a diagonal matrix
whose nonzero entries are the elements of $\bx$. $\bI$ stands for an identity matrix with suitable dimension.
The notation $\sim$ means ``be distributed as''. $\cC\cN_N(\bm{\mu},\bX)$
denotes the $N$-dimensional circular complex Gaussian distribution with mean $\bm{\mu}$ and covariance matrix $\bX$.
$\succ$ denotes positive-definite and $\prec$ denotes negative-definite.
$\det(\cdot)$ represents the determinant of a matrix and $|\cdot|$ is the modulus of a scalar. $\mbox{Tr}(\cdot)$ denotes the trace of a square matrix.
$\jmath=\sqrt{-1}$ and $U(0,1)$ the uniformly distributed random variable on the interval from 0 to 1.

\section{Problem Formulation and Preliminary Definitions}
\label{section2}
Consider a radar system equipped with $N\geq 2$ space and/or time channels illuminating the surveillance area.
The measurements of the backscattered echoes from the range cells are properly processed and sampled to form
$N$-dimensional complex vectors. Now, notice that when a point-like target is present in a given range bin, the range cells in the neighborhood
of the latter might be contaminated by target components (energy spillover \cite{Spillover2011Danilo});
the same effect is also observed in the presence of range-spread targets due to an enhanced range resolution \cite{Conte-RangeSpread}. Thus,
we assume that the system performs the decision over a set of primary data
denoted by $\bz_k\in\mathds{C}^{N\times 1}$, $k=1,\ldots,K_P$, corresponding to ${K_P}$ consecutive CUTs.
Furthermore, we denote by $\bor_k\in\mathds{C}^{N\times 1}$, $k=1,\ldots,K_S$, a set of secondary data gathered from the
leading and lagging window \cite{richards2013principles} in the proximity of the CUTs.
Finally, we assume that before testing for the presence of a target the system  performs a preliminary scan of the entire region of
interest to classify the scenario and build up a kind of clutter map \cite{Skolnik,barton}. For this reason, at the design stage, all data are assumed free
of target components as well as independent and identically distributed circularly symmetric complex Gaussian random vectors with zero mean and positive
definite covariance matrix.

Moreover, we consider two situations of clutter covariance variation and accordingly formulate two classification problems, where the second one accounts for a more general scenario.  Specifically, the first scenario assumes that primary and secondary data share the same covariance matrix except for scaling factors representative of different power levels. Notice that in this case, all of the secondary data can be used in the detection stage to estimate the covariance structure. The second scenario considers the case that the background
experiences extremely serious fluctuations caused by, for example, a drastic variation of terrain type, multiple interferences,
or other extreme transitions, and the variation of the clutter power and subspace is thus
arbitrary.\footnote{When a priori information about the spatial variation of the clutter is available,
the first model is more suitable than the second one. On the other hand, the latter
represents a viable means to deal with situations where the system is completely unaware of the surrounding clutter.}

\begin{figure}[tb]
  \centering
  \includegraphics[scale=0.44]{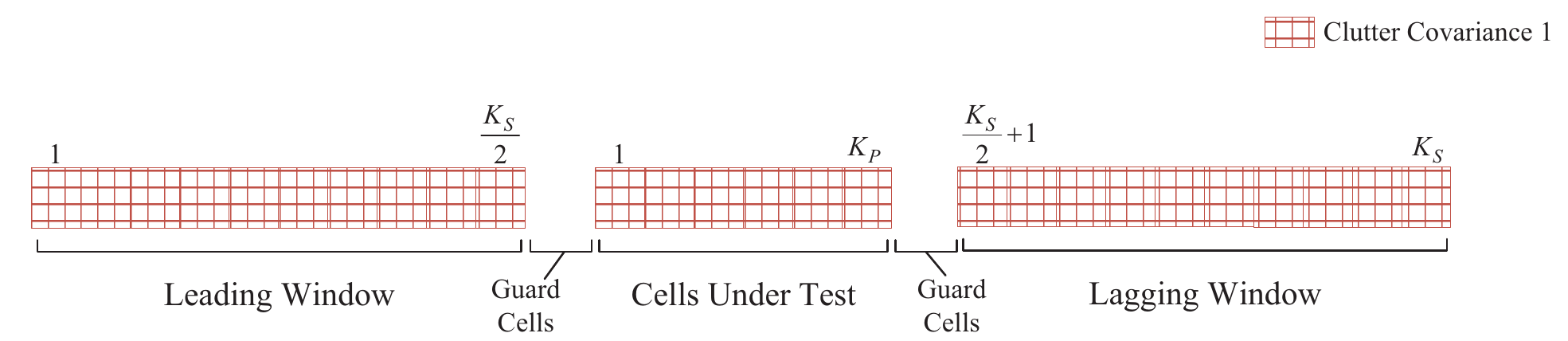}\\
  \caption{Diagrams of homogeneous environments}\label{ho}
\end{figure}

\begin{figure}[tb]
  \centering
  \includegraphics[scale=0.44]{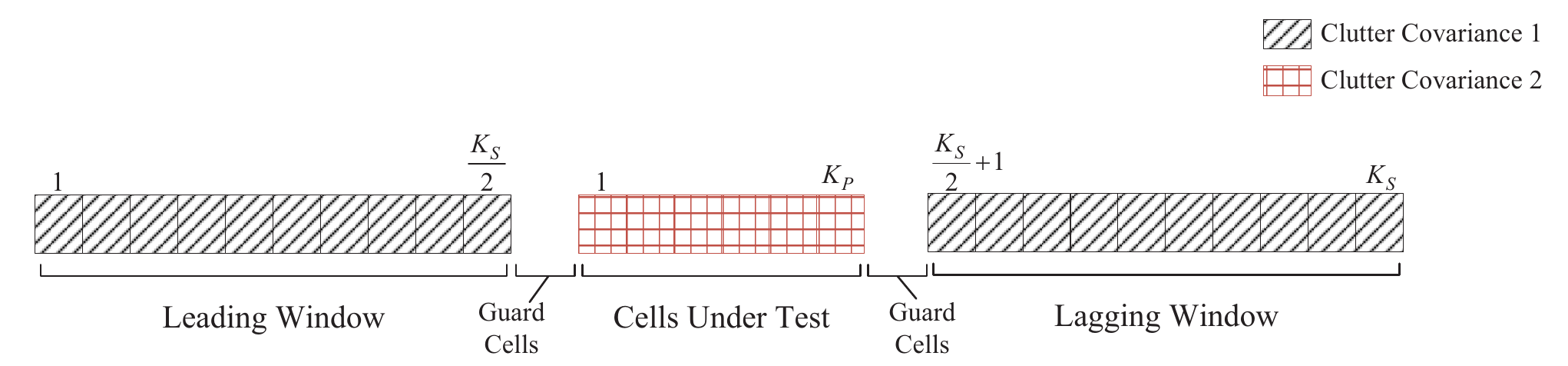}\\
  \caption{Diagrams of partially-homogeneous environments}\label{pho}
\end{figure}

\begin{figure}[tb]
  \centering
  \subfigure[Intermediate heterogeneous environment with one clutter edge]{\includegraphics[scale=0.44]{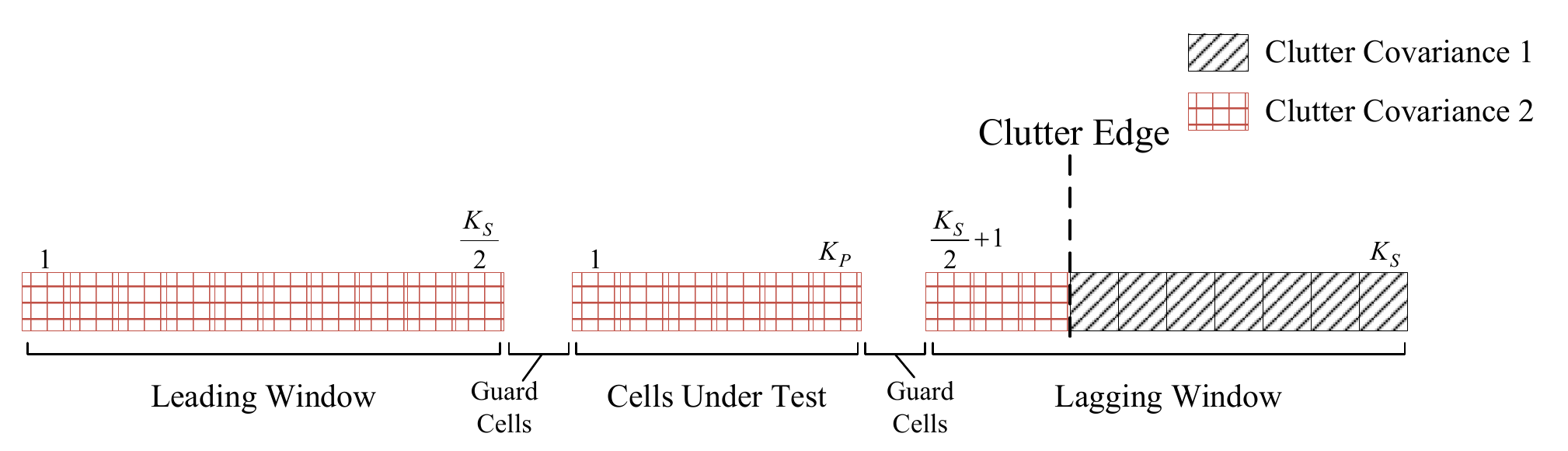}\label{f2}}
  \subfigure[Intermediate heterogeneous environment with two clutter edges]{\includegraphics[scale=0.44]{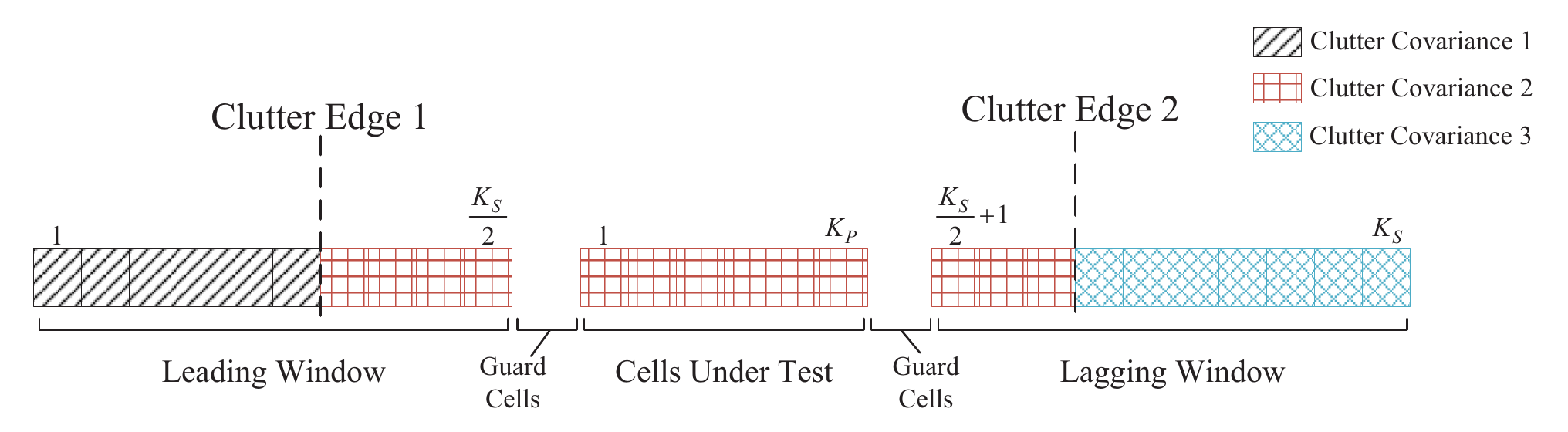}\label{f3}}
  \caption{Diagrams of intermediate heterogeneous environments}\label{F2}
\end{figure}

Let us start from the first scenario and consider the following configurations (or hypotheses):
\begin{itemize}
\item the homogeneous environment ($H_{\text{I},0}$),
where the interference component in the primary and secondary data share the same statistical characterization \cite{Kelly1986GLRT,AMF1992,Hao2016Joint}, as illustrated in Fig. \ref{ho};
\item the well-known partially-homogeneous
environment ($H_{\text{I},1}$),\footnote{Recall that the considered partially-homogeneous
environment is slightly different from the classical one since we also
consider the presence of thermal noise.} where clutter in the
primary and secondary data share the same clutter covariance structure
but different clutter power \cite{Hao2014Pers,Foglia2017joint} as illustrated in Fig. \ref{pho};
\item an intermediate heterogeneous environment where secondary data contains only one clutter edge coming from either the lagging or the leading window ($H_{\text{I},2}$) \cite{XU2021108127},
as illustrated in Fig. \ref{f2};\footnote{It is worth noticing that even though in drawing this
      figure the condition $K_{\text{I},1}<\frac{K_S}{2}$ is assumed, $K_{\text{I},1}$
      can be greater than $\frac{K_S}{2}$ as explained in what follows.}
\item an intermediate heterogeneous environment with a clutter edge present in both the leading and lagging window ($H_{\text{I},3}$)
as illustrated in Fig. \ref{f3}.
\end{itemize}

In Fig. \ref{F2}, Clutter Covariance 1, 2, and 3 differ only in terms of clutter power.
Therefore, the classification problem for radar scenarios can be formulated in terms of the
following multiple hypothesis test
\begin{equation}\label{hypothese1}
\left\{
\begin{array}{lll}
H_{\text{I},0}:
\left\{
\begin{array}{lll}
\bz_k\sim \cC\cN_N(\textbf{0},\sigma^2\bI+\bM),
\\ k=1,\ldots,{K_P},\\
\bor_{k}\sim \cC\cN_N(\textbf{0},\sigma^2\bI+\bM),
\\ k=1,\ldots,K_S,\\
\end{array}
\right.
\\
\vspace{-2mm}
\\
H_{\text{I},1}:
\left\{
\begin{array}{lll}
\bz_k\sim \cC\cN_N(\textbf{0},\sigma^2\bI+\bM),
\\ k=1,\ldots,{K_P},\\
\bor_{k}\sim \cC\cN_N(\textbf{0},\sigma^2\bI+\bM_1),
\\ k=1,\ldots,K_S,\\
\end{array}
\right.
\\
\vspace{-2mm}
\\
H_{\text{I},2}:
\left\{
\begin{array}{lll}
\bz_k\sim \cC\cN_N(\textbf{0},\sigma^2\bI+\bM),
\\ k=1,\ldots,{K_P},\\
\bor_{k}\sim \cC\cN_N(\textbf{0},\sigma^2\bI+\bM),
\\ k=1,\ldots,K_{\text{I},1},\\
\bor_{k}\sim \cC\cN_N(\textbf{0},\sigma^2\bI+\bM_2),
\\ k=K_{\text{I},1}+1,\ldots,K_S,\\
\end{array}
\right.
\\
\vspace{-2mm}
\\
H_{\text{I},3}:
\left\{
\begin{array}{lll}
\bz_k\sim \cC\cN_N(\textbf{0},\sigma^2\bI+\bM),
\\ k=1,\ldots,{K_P},\\
\bor_{k}\sim \cC\cN_N(\textbf{0},\sigma^2\bI+\bM_3),
\\ k=1,\ldots,K_{\text{I},2},\\
\bor_{k}\sim \cC\cN_N(\textbf{0},\sigma^2\bI+\bM),
\\ k=K_{\text{I},2}+1,\ldots,K_{\text{I},3},\\
\bor_{k}\sim \cC\cN_N(\textbf{0},\sigma^2\bI+\bM_4),
\\ k=K_{\text{I},3}+1,\ldots,K_S,\\
\end{array}
\right.
\end{array}
\right.
\end{equation}
where $\sigma^2\bI$ is the thermal noise component with $\sigma^2>0$ the unknown noise power level;
$\bM=\bm{U}\bm{\Lambda}\bm{U}^\dag\in\mathds{C}^{N\times N}$ is the positive semidefinite clutter covariance matrix of the primary data where $\bm{\Lambda}=diag\left(\lambda_1,\ldots,\lambda_r,0,\ldots,0\right)$, $\lambda_1\geq\ldots\geq\lambda_r>0$,
contains the eigenvalues of $\bM$ with $r<N$ the rank for the moment assumed known, $\bU$ is the unitary matrix of the corresponding eigenvectors; $K_P>r$, $K_S>r$ with $K_S$ an even number, and we denote by $K_{\text{I},1}\in\left\{r,\ldots,K_S-r\right\}$, $K_{\text{I},2}\in\left\{r,\ldots,\frac{K_S}{2}\right\}$,
and $K_{\text{I},3}\in\left\{\frac{K_S}{2}+1,\ldots,K_S-r\right\}$ the unknown
positions of the clutter power transitions within the leading and lagging windows under $H_{\text{I},2}$ and $H_{\text{I},3}$, respectively.
In order to better understand the clutter characteristics under $H_{\text{I},2}$ and $H_{\text{I},3}$, it is important to highlight that we assume that in the spatial proximity of the CUTs there exist range bins that share the same clutter power level as that in the primary data. As for $\bM_l\in\mathds{C}^{N\times N}$, $l=1,\ldots,4$, they are defined as $\bM_l=\bU\bm{\Gamma}_l\bm{\Lambda}\bm{\Gamma}_l\bU^\dag$,
where
$\bm{\Gamma}_l=diag(\sqrt{\gamma_{1,l}},\ldots,\sqrt{\gamma_{r,l}},0,\ldots,0)$, $\gamma_{1,l}\geq\ldots\geq \gamma_{r,l}>0$,
is representative of possible power variations of the clutter along different directions. It is important to notice
that these covariance structures model situations where the spatial distribution of the clutter remains approximately
unaltered over the range (clutter scatterers respond from a given set of directions) whereas clutter reflectivity might
change over the range leading to different power levels associated with the main echoes' directions.

For notation convenience, let us denote by
$\bZ_P=\left[\bz_1,\ldots,\bz_{K_P}\right]\in \mathds{C}^{N\times {K_P}}$, $\bZ_S=\left[\bor_{1},\ldots,\bor_{K_S}\right]\in \mathds{C}^{N\times K_S}$,
and $\bZ=\left[\bZ_P,\bZ_S\right]\in \mathds{C}^{N\times ({K_P}+K_S)}$ the overall data set.
Based upon the above assumptions, the probability density function (PDF) of $\bZ$ under $H_{\text{I},m}$, $m=0,\ldots,3$, namely $f_{\text{I},m}(\bZ;\bm{\theta}_m)$ can be expressed as
$f_{\text{I},m}(\bZ;\bm{\theta}_m)=f_P(\bZ_P;\sigma^2,\bM)f_{S_{\text{I},m}}(\bZ_S;\bm{\theta}_m)$,
where
\begin{equation}
\begin{cases}
f_P(\bZ_P;\sigma^2,\bM)=\left[\frac{\exp{\left\{-\frac{1}{K_P}\sum_{k=1}^{K_P}\bz_k^\dag(\sigma^2\bm{I}+\bm{M})^{-1}\bz_k\right\}}}
{\pi^N\det(\sigma^2\bm{I}+\bm{M})}\right]^{K_P},\\
f_{S_{\text{I},0}}(\bZ_S;\bm{\theta}_0)=\left[\frac{\exp{\left\{-\frac{1}{K_S}\sum_{k=1}^{K_S}\bor_k^\dag(\sigma^2\bm{I}+\bm{M})^{-1}\bor_k\right\}}}
{\pi^N\det(\sigma^2\bm{I}+\bm{M})}\right]^{K_S},\\
f_{S_{\text{I},1}}(\bZ_S;\bm{\theta}_1) = \left[\frac{\exp{\left\{-\frac{1}{K_S}\sum_{k=1}^{K_S}\bor_k^\dag(\sigma^2\bm{I}+\bm{M}_1)^{-1}\bor_k\right\}}}
{\pi^N\det(\sigma^2\bm{I}+\bm{M}_1)}\right]^{K_S},\\
f_{S_{\text{I},2}}(\bZ_S;\bm{\theta}_2) =\left[\frac{\exp{\left\{-\frac{1}{K_{\text{I},1}}\sum_{k=1}^{K_{\text{I},1}}\bor_k^\dag(\sigma^2\bm{I}+\bm{M})^{-1}\bor_k\right\}}}
{\pi^N\det(\sigma^2\bm{I}+\bm{M})}\right]^{K_{\text{I},1}}\\
\qquad \times\left[\frac{\exp{\left\{-\frac{1}{K_S-K_{\text{I},1}}\sum_{k=K_{\text{I},1}+1}^{K_S}\bor_k^\dag(\sigma^2\bm{I}+\bm{M}_2)^{-1}\bor_k\right\}}}
{\pi^N\det(\sigma^2\bm{I}+\bm{M}_2)}\right]^{K_S-K_{\text{I},1}},\\
f_{S_{\text{I},3}}(\bZ_S;\bm{\theta}_3) = \left[\frac{\exp{\left\{-\frac{1}{K_{\text{I},2}}\sum_{k=1}^{K_{\text{I},2}}\bor_k^\dag(\sigma^2\bm{I}+\bm{M}_3)^{-1}\bor_k\right\}}}
{\pi^N\det(\sigma^2\bm{I}+\bm{M}_3)}\right]^{K_{\text{I},2}}\\
\qquad\times\left[\frac{\exp{\left\{-\frac{1}{K_{\text{I},3}-K_{\text{I},2}}\sum_{k=K_{\text{I},2}+1}^{K_{\text{I},3}}\bor_k^\dag(\sigma^2\bm{I}+\bm{M})^{-1}\bor_k\right\}}}
{\pi^N\det(\sigma^2\bm{I}+\bm{M})}\right]^{K_{\text{I},3}-K_{\text{I},2}}\\
\qquad\times\left[\frac{\exp{\left\{-\frac{1}{K_S-K_{\text{I},3}}\sum_{k=K_{\text{I},3}+1}^{K_S}\bor_k^\dag(\sigma^2\bm{I}+\bm{M}_4)^{-1}\bor_k\right\}}}
{\pi^N\det(\sigma^2\bm{I}+\bm{M}_4)}\right]^{K_S-K_{\text{I},3}},
\end{cases}
\end{equation}
with
\begin{align}
\bm{\theta}_0=\left[\sigma^2,\bm{\nu}^T(\bM)\right]^T, \bm{\theta}_1=\left[\bm{\theta}_0^T,\bm{\nu}^T(\bM_1)\right]^T,\nonumber\\
\bm{\theta}_2=\left[\bm{\theta}_0^T,K_{\text{I},1},\bm{\nu}^T(\bM_2)\right]^T,\qquad\qquad\qquad\quad\nonumber\\
\bm{\theta}_3=\left[\bm{\theta}_0^T,K_{\text{I},2},K_{\text{I},3},\bm{\nu}^T(\bM_3),\bm{\nu}^T(\bM_4)\right]^T,\ \
\end{align}
and $\bm{\nu}(\bM)$ the vector-valued function selecting the distinct entries of $\bM$.

\begin{figure}[t]
  \centering
  \includegraphics[scale=0.45]{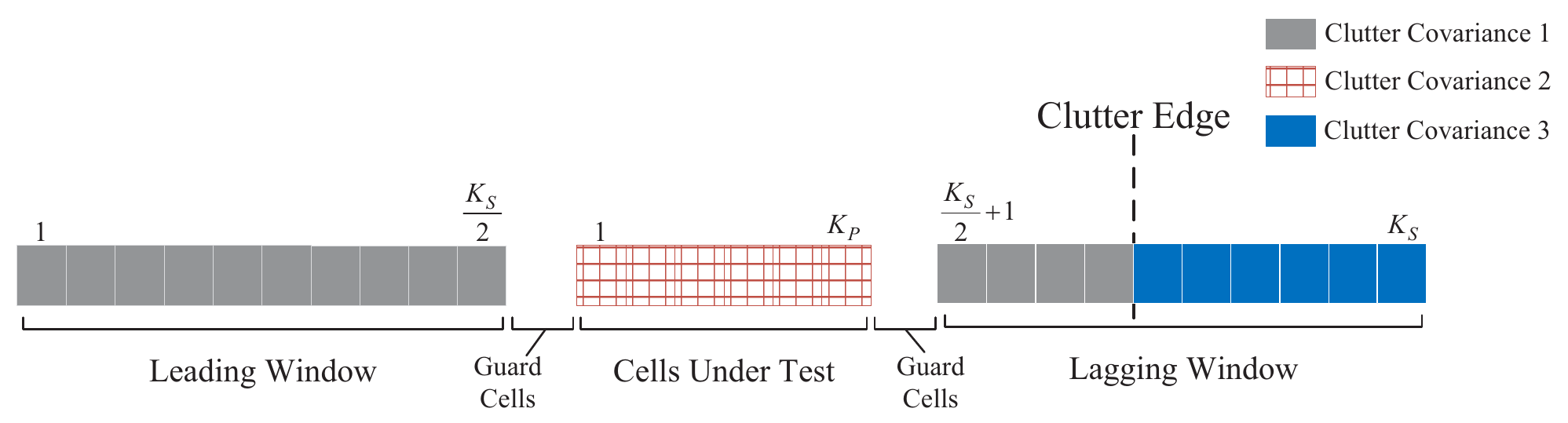}
  \caption{Heterogeneous environment with one clutter edge}\label{f5}
\end{figure}

In the second scenario, we assume that the variation of the clutter power and subspace over the range is arbitrary.
Specifically, we modify the hypothesis test \eqref{hypothese1} by adopting
a different covariance variation model that does not assume any relationship among the covariances. Moreover, we introduce an additional hypothesis, $H_{\text{II},4}$ say, where the data in lagging and leading windows belong to two separated and different clutter regions whose properties are also different from those of the clutter in the primary data, as depicted in Fig. \ref{f5}.
Notice that the situation represented by $H_{\text{II},4}$ is not considered in
the first scenario since, when secondary data are partitioned into different clutter regions, the estimation quality of some parameters can seriously degrade due to the specific constraints of the first covariance model. Specifically, under such assumptions, they can impair the estimation performance in the presence of a limited amount of data characterized by $\bM$.

With the above remarks in mind, the second classification problem can be formulated as the following multiple hypothesis testing problem
\begin{equation}\label{hypothese2}
\left\{
\begin{array}{lll}
H_{\text{II},0}:
\left\{
\begin{array}{lll}
\bz_k\sim \cC\cN_N(\textbf{0},\sigma^2\bI+\bM),
\\ k=1,\ldots,{K_P},\\
\bor_{k}\sim \cC\cN_N(\textbf{0},\sigma^2\bI+\bM),
\\ k=1,\ldots,K_S,\\
\end{array}
\right.
\\
\vspace{-2mm}
\\
H_{\text{II},1}:
\left\{
\begin{array}{lll}
\bz_k\sim \cC\cN_N(\textbf{0},\sigma^2\bI+\bM),
\\ k=1,\ldots,{K_P},\\
\bor_{k}\sim \cC\cN_N(\textbf{0},\sigma^2\bI+\bR_1),
\\ k=1,\ldots,K_S,\\
\end{array}
\right.
\\
\vspace{-2mm}
\\
H_{\text{II},2}:
\left\{
\begin{array}{lll}
\bz_k\sim \cC\cN_N(\textbf{0},\sigma^2\bI+\bM),
\\ k=1,\ldots,{K_P},\\
\bor_{k}\sim \cC\cN_N(\textbf{0},\sigma^2\bI+\bM),
\\ k=1,\ldots,K_{\text{II},1},\\
\bor_{k}\sim \cC\cN_N(\textbf{0},\sigma^2\bI+\bR_2),
\\ k=K_{\text{II},1}+1,\ldots,K_S,\\
\end{array}
\right.
\\
\vspace{-2mm}
\\
H_{\text{II},3}:
\left\{
\begin{array}{lll}
\bz_k\sim \cC\cN_N(\textbf{0},\sigma^2\bI+\bM),
\\ k=1, \ldots,{K_P},\\
\bor_{k}\sim \cC\cN_N(\textbf{0},\sigma^2\bI+\bR_3),
\\ k=1,\ldots,K_{\text{II},2},\\
\bor_{k}\sim \cC\cN_N(\textbf{0},\sigma^2\bI+\bM),
\\ k=K_{\text{II},2}+1,\ldots,K_{\text{II},3},\\
\bor_{k}\sim \cC\cN_N(\textbf{0},\sigma^2\bI+\bR_4),
\\ k=K_{\text{II},3}+1,\ldots,K_S,\\
\end{array}
\right.
\\
\vspace{-2mm}
\\
H_{\text{II},4}:
\left\{
\begin{array}{lll}
\bz_k\sim \cC\cN_N(\textbf{0},\sigma^2\bI+\bM),
\\ k=1,\ldots,{K_P},\\
\bor_{k}\sim \cC\cN_N(\textbf{0},\sigma^2\bI+\bR_5),
\\ k=1,\ldots,K_{\text{II},4},\\
\bor_{k}\sim \cC\cN_N(\textbf{0},\sigma^2\bI+\bR_6),
\\ k=K_{\text{II},4}+1,\ldots,K_S,\\
\end{array}
\right.
\end{array}
\right.
\end{equation}
where $\bM=\bm{U}\bm{\Lambda}\bm{U}^\dag$ is the clutter covariance matrix of primary data, $\bR_l=\bm{U}_l\bm{\Lambda}_l\bm{U}_l^\dag$, $l=1,\ldots,6$, with $\bm{\Lambda}_l=diag\left(\lambda_{1,l},\ldots,\lambda_{r,l},0,\ldots,0\right)$, $\lambda_{1,l}\geq\ldots\geq\lambda_{r,l}>0$,
containing the eigenvalues of $\bR_l$, and $\bU_l$ the unitary matrix of the corresponding eigenvectors.
In \eqref{hypothese2}, $H_{\text{II},0}$ is equivalent to $H_{\text{I},0}$, namely the homogeneous environment, whereas $H_{\text{II},1}$ is the totally heterogeneous environment where the covariance matrices of primary and secondary data are arbitrarily different except for the rank,
$H_{\text{II},2}$ and $H_{\text{II},3}$ are the same configurations as $H_{\text{I},2}$ and $H_{\text{I},3}$, respectively, and $H_{\text{II},4}$ is the situation that a clutter edge is  present under $H_{\text{II},1}$. $K_{\text{II},1}\in\left\{r,\ldots,K_S-r\right\}$, $K_{\text{II},2}\in\left\{r,\ldots,\frac{K_S}{2}\right\}$, $K_{\text{II},3}\in\left\{\frac{K_S}{2}+1,\ldots,K_S-r\right\}$, and $K_{\text{II},4}\in\left\{r,\ldots,K_S-r\right\}$ are the positions
of the edges under $H_{\text{II},2}$, $H_{\text{II},3}$ and $H_{\text{II},4}$, respectively. It can be observed that, by solving \eqref{hypothese2}, heterogeneous data sets are clustered into homogeneous subsets and only the subsets of secondary data which share the same clutter covariance matrix with primary data can be used for detection.

The PDF of $\bZ$ under $H_{\text{II},n}$, $n=0,\ldots,4$, namely $f_{\text{II},n}(\bZ;\bm{\delta}_n)$ can be expressed as
$f_{\text{II},n}(\bZ;\bm{\delta}_n)=f_P(\bZ_P;\sigma^2,\bM)f_{S_{\text{II},n}}(\bZ_S;\bm{\delta}_n)$,
where $f_{S_{\text{II},0}}(\bZ_S;\bm{\delta}_0)=f_{S_{\text{I},0}}(\bZ_S;\bm{\theta}_0)$,
\begin{equation}
\begin{cases}
f_{S_{\text{II},1}}(\bZ_S;\bm{\delta}_1) = \left[\frac{\exp{\left\{-\frac{1}{K_S}\sum_{k=1}^{K_S}\bor_k^\dag(\sigma^2\bm{I}+\bm{R}_1)^{-1}\bor_k\right\}}}
{\pi^N\det(\sigma^2\bm{I}+\bm{R}_1)}\right]^{K_S},\\
f_{S_{\text{II},2}}(\bZ_S;\bm{\delta}_2) =\left[\frac{\exp{\left\{-\frac{1}{K_{\text{II},1}}\sum_{k=1}^{K_{\text{II},1}}\bor_k^\dag(\sigma^2\bm{I}+\bm{M})^{-1}\bor_k\right\}}}
{\pi^N\det(\sigma^2\bm{I}+\bm{M})}\right]^{K_{\text{II},1}}\\
\qquad \times\left[\frac{\exp{\left\{-\frac{1}{K_S-K_{\text{II},1}}\sum_{k=K_{\text{II},1}+1}^{K_S}\bor_k^\dag(\sigma^2\bm{I}+\bm{R}_2)^{-1}\bor_k\right\}}}
{\pi^N\det(\sigma^2\bm{I}+\bm{R}_2)}\right]^{K_S-K_{\text{II},1}},\\
f_{S_{\text{II},3}}(\bZ_S;\bm{\delta}_3) = \left[\frac{\exp{\left\{-\frac{1}{K_{\text{II},2}}\sum_{k=1}^{K_{\text{II},2}}\bor_k^\dag(\sigma^2\bm{I}+\bm{R}_3)^{-1}\bor_k\right\}}}
{\pi^N\det(\sigma^2\bm{I}+\bm{R}_3)}\right]^{K_{\text{II},2}}\\
\qquad\times\left[\frac{\exp{\left\{-\frac{1}{K_{\text{II},3}-K_{\text{II},2}}\sum_{k=K_{\text{II},2}+1}^{K_{\text{II},3}}\bor_k^\dag(\sigma^2\bm{I}+\bm{M})^{-1}\bor_k\right\}}}
{\pi^N\det(\sigma^2\bm{I}+\bm{M})}\right]^{K_{\text{II},3}-K_{\text{II},2}}\\
\qquad\times\left[\frac{\exp{\left\{-\frac{1}{K_S-K_{\text{II},3}}\sum_{k=K_{\text{II},3}+1}^{K_S}\bor_k^\dag(\sigma^2\bm{I}+\bm{R}_4)^{-1}\bor_k\right\}}}
{\pi^N\det(\sigma^2\bm{I}+\bm{R}_4)}\right]^{K_S-K_{\text{II},3}},\\
f_{S_{\text{II},4}}(\bZ_S;\bm{\delta}_4) = \left[\frac{\exp{\left\{-\frac{1}{K_{\text{II},4}}\sum_{k=1}^{K_{\text{II},4}}\bor_k^\dag(\sigma^2\bm{I}+\bm{R}_5)^{-1}\bor_k\right\}}}
{\pi^N\det(\sigma^2\bm{I}+\bm{R}_5)}\right]^{K_{\text{II},4}}\\
\qquad\times\left[\frac{\exp{\left\{-\frac{1}{K_S-K_{\text{II},4}}\sum_{k=K_{\text{II},4}+1}^{K_S}\bor_k^\dag(\sigma^2\bm{I}+\bm{R}_6)^{-1}\bor_k\right\}}}
{\pi^N\det(\sigma^2\bm{I}+\bm{R}_6)}\right]^{K_S-K_{\text{II},4}},\\
\end{cases}
\end{equation}
with $\bm{\delta}_0=\bm{\theta}_0$, and
\begin{align}
\bm{\delta}_1=\left[\bm{\delta}_0^T,\bm{\nu}^T(\bR_1)\right]^T,
\bm{\delta}_2=\left[\bm{\delta}_0^T,K_{\text{II},1},\bm{\nu}^T(\bR_2)\right]^T,\nonumber\\
\bm{\delta}_3=\left[\bm{\delta}_0^T,K_{\text{II},2},K_{\text{II},3},\bm{\nu}^T(\bR_3),\bm{\nu}^T(\bR_4)\right]^T,\qquad
\quad\nonumber\\
\bm{\delta}_4=\left[\bm{\delta}_0^T,K_{\text{II},4},\bm{\nu}^T(\bR_5),\bm{\nu}^T(\bR_6)\right]^T.\qquad\quad\ \qquad
\end{align}

\section{Classification Architecture Design}
\label{section3}

In this section, for each multiple hypothesis testing problem, we conceive a classification architecture based on the MOS rules which consists in penalizing the compressed log-likelihood functions \cite{KL2021I}.
Such a choice is dictated by the fact that \eqref{hypothese1} or \eqref{hypothese2}
contains nested hypotheses \cite{Vincenzo2019Hetegeneous,VC2017CMC,YLJ2019ECCM}
and, hence, applying the plain MLA would lead to an overestimation of the
number of the unknown parameters. The penalty term in the general structure
of a MOS rule promotes low-dimensional models in contrast to the log-likelihood function
thus balancing the overestimation of the MLA.

Firstly, we make explicit the relationship between the hypotheses and the number of unknown parameters
by denoting it under $H_{\text{I},m}$, $m=0,\ldots,3$, and $H_{\text{II},n}$, $n=0,\ldots,4$ as follows
\begin{equation}\label{parameternumber1}
\left\{
\begin{array}{lll}
{\varrho}_{\text{I},0}=p(r)+1,&H_{\text{I},0},\\
{\varrho}_{\text{I},1}=p(r)+r+1,&H_{\text{I},1},\\
{\varrho}_{\text{I},2}=p(r)+r+2,&H_{\text{I},2},\\
{\varrho}_{\text{I},3}=p(r)+2r+3,&H_{\text{I},3},
\end{array}
\right.
\end{equation}
where $p(r)=r(2N-r)$ \cite{VT2004}, and
\begin{equation}\label{parameternumber2}
\left\{
\begin{array}{lll}
{\varrho}_{\text{II},0}=p(r)+1,&H_{\text{II},0},\\
{\varrho}_{\text{II},1}=2p(r)+1,&H_{\text{II},1},\\
{\varrho}_{\text{II},2}=2p(r)+2,&H_{\text{II},2},\\
{\varrho}_{\text{II},3}=3p(r)+3,&H_{\text{II},3},\\
{\varrho}_{\text{II},4}=3p(r)+2,&H_{\text{II},4},
\end{array}
\right.
\end{equation}
respectively.
The expression of the MOS-based architectures are given by the sum of two terms related to the log-likelihood function and the number of unknown parameters, respectively, namely
\begin{equation}\label{MOSscheme1}
H_{\text{I},\widehat{m}}=\arg\dmin_{m=0,1,2,3}{\left\{-2h_{\text{I},m}(\bZ;\widehat{\bm{\theta}}_m)+\kappa\cdot\varrho_{\text{I},m}\right\}},
\end{equation}
\begin{equation}\label{MOSscheme2}
H_{\text{II},\widehat{n}}=\arg\dmin_{n=0,1,2,3,4}{\left\{-2h_{\text{II},n}(\bZ;\widehat{\bm{\delta}}_n)+\kappa\cdot\varrho_{\text{II},n}\right\}},
\end{equation}
respectively, where $h_{\text{I},m}(\bZ;\widehat{\bm{\theta}}_m)=\log{f_{\text{I},m}(\bZ;\widehat{\bm{\theta}}_m)}$,
$h_{\text{II},n}(\bZ;\widehat{\bm{\delta}}_n)=\log{f_{\text{II},n}(\bZ;\widehat{\bm{\delta}}_n)}$, $\widehat{\bm{\theta}}_m$ and $\widehat{\bm{\delta}}_n$ are suitable
estimates\footnote{Notice that \eqref{MOSscheme1} and \eqref{MOSscheme2} are approximations
of the MOS rules since we replace the MLEs with
alternative estimates under some hypotheses.
} of $\bm{\theta}_m$ and $\bm{\delta}_n$, respectively,
and $\kappa$ is a factor depending on the specific MOS rule, namely
\begin{equation}\label{kappa}
\kappa=
\left\{
\begin{array}{lll}
2,&\text{AIC},\\
(1+\rho),\ \rho\geq 1&\text{GIC},\\
\log{[2N(K_P+K_S)]},&\text{BIC}.
\end{array}
\right.
\end{equation}

\subsection{Clutter Covariance Variation Model 1}
In this subsection, we focus on problem \eqref{hypothese1} and provide the estimates used in \eqref{MOSscheme1}.
For simplicity, we start from $H_{\text{I},1}$ since we will reuse the related estimates
under the other hypotheses.
The related optimization problem with respect to $\btheta_1$ can be written as
\begin{equation}
\max\limits_{\btheta_1}\ \ \log f_{\text{I},1}(\bZ;\btheta_1),
\label{eqn:maxLL_HI1}
\end{equation}
where
\begin{multline}
\log f_{\text{I},1}(\bZ;\btheta_1)=
-N(K_P+K_S)\log\pi
\\
-K_P\log\det\left(\sigma^2\bI+\bM\right)
-K_S\log\det\left(\sigma^2\bI+\bM_1\right)
\\
-\sum_{k=1}^{K_P}\bz_k^\dag(\sigma^2\bI+\bM)^{-1}\bz_k
-\sum_{k=1}^{K_S}\bor_k^\dag(\sigma^2\bI+\bM_1)^{-1}\bor_k.
\end{multline}
Recalling that $\bM=\bU\bLambda\bU^\dag$ and $\bM_l=\bU\bm{\Gamma}_l\bLambda\bm{\Gamma}_l\bU^\dag$, $l=1,\ldots,4$,
\eqref{eqn:maxLL_HI1} can be rewritten as

\begin{multline}\label{maxH1}
\max\limits_{\sigma^2,\bm{\Lambda},\bm{U},\bm{\Gamma}_1}
-N(K_P+K_S)\log\pi-K_P\log\det\left(\sigma^2\bI+\bm{\Lambda}\right)\\
-K_S\log\det\left(\sigma^2\bI+\bm{\Gamma}_1\bm{\Lambda}\bm{\Gamma}_1\right)
-\sum_{k=1}^{K_P}\bz_k^\dag\bm{U}(\sigma^2\bI+\bm{\Lambda})^{-1}\bm{U}^\dag\bz_k\\
-\sum_{k=1}^{K_S}\bor_k^\dag\bm{U}(\sigma^2\bI+\bm{\Gamma}_1\bm{\Lambda}\bm{\Gamma}_1)^{-1}\bm{U}^\dag\bor_k.
\end{multline}
Now, the plain maximization of the objective function in \eqref{maxH1} with respect to $\bU$ is difficult
from a mathematical point of view. For this reason, we resort to an estimation
procedure that is suboptimum with respect to the MLA and
consists in minimizing a specific residual error (as described in Appendix \ref{appendix0}). Denoting by $\widehat{\bU}$ the obtained estimate, \eqref{maxH1} can be approximated as
\begin{multline}
\max\limits_{\sigma^2,\lambda_1,\ldots,\lambda_r\atop \gamma_{1,1},\ldots,\gamma_{r,1}}
-K_P\sum_{i=1}^r\log\left(\sigma^2+\lambda_i\right)
-K_S\sum_{i=1}^r\log\left(\sigma^2+\lambda_i^{(p)}\right)\\
-(K_P+K_S)(N-r)\log\sigma^2-t_1(\bZ;\bm{\theta}_1),
\end{multline}
where $\lambda_i^{(p)}=\gamma_{i,1}\lambda_i$, and
\begin{eqnarray}
t_1(\bZ;\bm{\theta}_1)&=&\sum_{i=1}^r\frac{\bS_P(i,i)}{\sigma^2+\lambda_i}+\sum_{i=r+1}^N\frac{\bS_P(i,i)}{\sigma^2}\nonumber\\
&&+\sum_{i=1}^r\frac{\bS_S(i,i)}{\sigma^2+\lambda_i^{(p)}}+\sum_{i=r+1}^N\frac{\bS_S(i,i)}{\sigma^2},
\end{eqnarray}
with $\bS_P=\sum\limits_{k=1}^{K_P}\widehat{\bU}^\dag\bz_k\bz_k^\dag\widehat{\bU}$ and $\bS_S=\sum\limits_{k=1}^{K_S}\widehat{\bU}^\dag\bor_k\bor_k^\dag\widehat{\bU}$.
At the same time, since given $\lambda_i$, $\lambda_i^{(p)}$ is completely determined by $\gamma_{i,1}$, we first estimate $\lambda_i$ following the lead of \cite{XU2021108127} and with primary data only, specifically
\begin{eqnarray}\label{sigma}
\widehat{\sigma}^2=\frac{1}{K_P(N-r)}\sum_{i=r+1}^N\mu_{i,P},
\end{eqnarray}
\begin{eqnarray}\label{lambda}
&&\widehat{\lambda}_{i}=\max\left\{\frac{1}{K_P}\mu_{i,P}-\widehat{\sigma}^2, 0\right\}, \ i=1,\ldots,r,
\end{eqnarray}
where $\mu_{1,P}\geq\ldots\geq\mu_{N,P}\geq0$ are the eigenvalues of $\bZ_{P}\bZ_{P}^\dag$.

The estimation of $\gamma_{i,1}$, $i=1,\ldots,r$, is thus tantamount to
\begin{eqnarray}\label{H1-gamma-1}
\max\limits_{\gamma_{1,1},\ldots,\gamma_{r,1}}
-K_S\sum_{i=1}^r\log\left(\widehat{\sigma}^2+\gamma_{i,1}\widehat{\lambda}_{i}\right)
-\sum_{i=1}^r\frac{\bS_S(i,i)}{\widehat{\sigma}^2+\gamma_{i,1}\widehat{\lambda}_{i}}.
\end{eqnarray}
Under the constraint that $\gamma_{1,1}\geq\ldots\geq \gamma_{r,1}>0$, we introduce the following auxiliary variables
\begin{equation}\label{H1-gamma-2}
\left\{
\begin{array}{lll}
\gamma_{r,1}=\tau_{r,p},\\
\gamma_{r-1,1}=\tau_{r,p}+\tau_{r-1,p},\\
\gamma_{r-2,1}=\tau_{r,p}+\tau_{r-1,p}+\tau_{r-2,p},\\
\qquad \vdots\\
\gamma_{1,1}=\tau_{r,p}+\tau_{r-1,p}+\tau_{r-2,p}+\ldots+\tau_{1,p},
\end{array}
\right.
\end{equation}
where $\tau_{r,p}>0$, and $\tau_{i,p}\geq 0$, $i=1,\ldots,r-1$. It follows that \eqref{H1-gamma-1} can be recast as
\begin{multline}\label{H1-gamma-3}
\max\limits_{\gamma_{1,1},\ldots,\gamma_{r,1}}
-K_S\sum_{i=1}^r\log\left[\widehat{\sigma}^2+\widehat{\lambda}_{i}\sum\limits_{j=i}^{r}\tau_{j,p}\right]
\\
-\sum_{i=1}^r\frac{\bS_S(i,i)}{\widehat{\sigma}^2+\widehat{\lambda}_{i}\sum\limits_{j=i}^{r}\tau_{j,p}}.
\end{multline}
Solving the above problem is a difficult task (at least to the best of the authors' knowledge) and, hence,
we resort to a cyclic optimization procedure that consists in the following steps. Firstly, we
assume that $\tau_{1,p}, \ldots, \tau_{h-1,p}$, are equal to their respective estimates at the $n$th step denoted
by $\widehat{\tau}_{1,p}^{(n)}, \ldots, \widehat{\tau}^{(n)}_{h-1,p}$, and that $\tau_{h+1,p}, \ldots, \tau_{r,p}$, are set to their estimate
at the $(n-1)$th step, namely, $\widehat{\tau}^{(n-1)}_{h+1,p}, \ldots, \widehat{\tau}^{(n-1)}_{r,p}$, then setting to zero the derivative of \eqref{H1-gamma-2}
with respect to $\tau_{h,p}$, we obtain the following equation
\begin{multline}\label{algorithm2-2}
\sum\limits_{i=1}^h\frac{K_S}{\widehat{\sigma}^2+\left(\sum\limits_{j=i}^{h-1}\widehat{\tau}^{(n)}_{j,p}+\tau_{h,p}
+\sum\limits_{j=h+1}^{r}\widehat{\tau}^{(n-1)}_{j,p}\right)\widehat{\lambda}_{i}}\\
-\sum\limits_{i=1}^h\frac{\bS_S(i,i)}{\left[\widehat{\sigma}^2+\left(\sum\limits_{j=i}^{h-1}\widehat{\tau}^{(n)}_{j,p}+\tau_{h,p}
+\sum\limits_{j=h+1}^{r}\widehat{\tau}^{(n-1)}_{j,p}\right)\widehat{\lambda}_{i}\right]^2}=0,\\
 h=1,\ldots,r.
\end{multline}
If there exist several positive solutions, we select the value that maximizes the objective function,
and if the equation does not have any positive solution, we set $\tau_{h,p}=0$ including when $h=r$. The above steps are repeated
until a maximum number, $n_{\text{max}}$ say, is achieved.
The final estimates of $\widehat{\gamma}_{i,1}$, are as follows:
\begin{equation}\label{gamma1}
\widehat{\gamma}_{i,1}=\sum\limits_{j=i}^{r}\widehat{\tau}^{(n_{\text{max}})}_{j,p},\ i=1,\ldots,r.
\end{equation}
Summarizing, an approximation of \eqref{maxH1} can be written as
\begin{multline}\label{hI1}
h_{\text{I},1}(\bZ;\widehat{\bm{\theta}}_1)=-K_P\sum_{i=1}^r\log\left(\widehat{\sigma}^2+\widehat{\lambda}_{i}\right)
\\
-K_S\sum_{i=1}^r\log\left(\widehat{\sigma}^2+\widehat{\gamma}_{i,1}\widehat{\lambda}_{i}\right)
-(K_P+K_S)(N-r)\log\widehat{\sigma}^2
\\
-\sum_{i=1}^r\frac{\bS_P(i,i)}{\widehat{\sigma}^2+\widehat{\lambda}_{i}}
-\sum_{i=r+1}^N\frac{\bS_P(i,i)}{\widehat{\sigma}^2}
\\
-\sum_{i=1}^r\frac{\bS_S(i,i)}{\widehat{\sigma}^2+\widehat{\gamma}_{i,1}\widehat{\lambda}_{i}}-\sum_{i=r+1}^N\frac{\bS_S(i,i)}{\widehat{\sigma}^2}.
\end{multline}

Under $H_{\text{I},0}$, the maximization of $\log f_{\text{I},0}(\bZ;\bm{\theta}_0)$ with respect to $\bm{\theta}_0$ leads to
\begin{align}\label{maxH0}
&\max\limits_{\sigma^2,\bm{\Lambda},\bm{U}}
-N(K_P+K_S)\log\pi-(K_P+K_S)\log\det\left(\sigma^2\bI+\bm{\Lambda}\right) \nonumber
\\
&-\sum_{k=1}^{K_P}\bz_k^\dag\bm{U}(\sigma^2\bI+\bm{\Lambda})^{-1}\bm{U}^\dag\bz_k-\sum_{k=1}^{K_S}\bor_k^\dag\bm{U}(\sigma^2\bI+\bm{\Lambda})^{-1}\bm{U}^\dag\bor_k.
\end{align}
Even though it is possible to obtain the exact MLEs as shown in \cite{XU2021108127}, we approximate the compressed log-likelihood as under $H_{\text{I},1}$ so that the estimated clutter subspace $\bU$ is the same under both hypotheses. Through the same estimation procedure for $\bU$ shown in Appendix \ref{appendix0}, \eqref{maxH0} can be approximated as
\begin{eqnarray}
&\max\limits_{\sigma^2,\lambda_1,\ldots,\lambda_r}&
-(K_P+K_S)(N-r)\log\sigma^2-t_0(\bZ;\bm{\theta}_0)\nonumber\\
&&-(K_P+K_S)\sum_{i=1}^r\log\left(\sigma^2+\lambda_i\right),
\end{eqnarray}
where
\be
t_0(\bZ;\bm{\theta}_0)=\sum\limits_{i=1}^r\frac{\bS_Z(i,i)}{\sigma^2+\lambda_i}+\sum\limits_{i=r+1}^N\frac{\bS_Z(i,i)}{\sigma^2}
\ee
with $\bS_Z=\widehat{\bU}^\dag\bZ\bZ^\dag\widehat{\bU}$.
As for the estimates for $\sigma^2$ and $\lambda_i$, they are given by \eqref{sigma} and \eqref{lambda}.
Thus, \eqref{maxH0} can be approximated as
\begin{multline}\label{hI0}
h_{\text{I},0}(\bZ;\widehat{\bm{\theta}}_0)=-(K_P+K_S)\sum_{i=1}^r\log\left(\widehat{\sigma}^2+\widehat{\lambda}_{i}\right)
-\sum_{i=1}^r\frac{\bS_Z(i,i)}{\widehat{\sigma}^2+\widehat{\lambda}_{i}}\\
-\sum_{i=r+1}^N\frac{\bS_Z(i,i)}{\widehat{\sigma}^2}
-(K_P+K_S)(N-r)\log\widehat{\sigma}^2.
\end{multline}

Under $H_{\text{I},2}$, resorting to the same approximation for $\widehat{\bU}$, the optimization problem over $\log{f_{\text{I},2}(\bZ;\widehat{\bm{\theta}}_2)}$ is given by
\begin{multline}\label{likelihood_H2}
\max\limits_{K_{\text{I},1}}\left\{\max\limits_{\sigma^2,\lambda_1,\ldots,\lambda_r\atop \gamma_{1,2},\ldots,\gamma_{r,2}}\right.-(K_P+K_S)(N-r)\log\sigma^2-t_2(\bZ;\bm{\theta}_2)\\
-\left[K_P+aK_{\text{I},1}+(1-a)(K_S-K_{\text{I},1})\right]\sum_{i=1}^r\log\left(\sigma^2+\lambda_i\right)\\
\left.-\left[(1-a)K_{\text{I},1}+a(K_S-K_{\text{I},1})\right]\sum_{i=1}^r\log\left(\sigma^2+\lambda_i^{(c)}\right)\right\},
\end{multline}
where $\lambda^{(c)}_i=\gamma_{i,2}\lambda_i$, $i=1,\ldots,r$,
\begin{equation}\label{A}
a =
\begin{cases}
1,&\ K_{\text{I},1}>\frac{K_S}{2},\\
0,&\ K_{\text{I},1}\leq\frac{K_S}{2},
\end{cases}
\end{equation}
\begin{multline}
t_2(\bZ;\bm{\theta}_2)=\sum_{i=1}^r\frac{\bS_P(i,i)}{\sigma^2+\lambda_i}+\sum_{i=r+1}^N\frac{\bS_P(i,i)}{\sigma^2}+
\sum_{i=1}^r\frac{\bS_{C_1}(i,i)}{\sigma^2+\lambda_i}\\
+\sum_{i=r+1}^N\frac{\bS_{C_1}(i,i)}{\sigma^2}
+\sum_{i=1}^r\frac{\bS_{C_2}(i,i)}{\sigma^2+\lambda_i^{(c)}}+\sum_{i=r+1}^N\frac{\bS_{C_2}(i,i)}{\sigma^2},
\end{multline}
with
\begin{equation}
\bS_{C_1}=
\left\{
\begin{array}{lll}
\sum\limits_{k=1}^{K_{\text{I},1}}\widehat{\bU}^\dag\bor_k\bor_k^\dag\widehat{\bU},&\ K_{\text{I},1}>\frac{K_S}{2},\\
\sum\limits_{k=K_{\text{I},1}+1}^{K_S}\widehat{\bU}^\dag\bor_k\bor_k^\dag\widehat{\bU},&\ K_{\text{I},1}\leq\frac{K_S}{2},
\end{array}
\right.
\end{equation}
and
\be
\bS_{C_2}=\sum\limits_{k=1}^{K_S}\widehat{\bU}^\dag\bor_k\bor_k^\dag\widehat{\bU}-\bS_{C_1}.
\ee
Since given $\lambda_i$, $\lambda^{(c)}_i$ is completely determined by $\gamma_{i,2}$, we first estimate $\sigma^2$ and $\lambda_i$ by means of \eqref{sigma} and \eqref{lambda}.
As for $\gamma_{i,2}$, $i=1,\ldots,r$, under the constraint that $\gamma_{1,2}\geq\ldots\geq \gamma_{r,2}>0$, we introduce $\gamma_{i,2}=\sum\limits_{j=i}^{r}\tau_{j,c}$,
where $\tau_{r,c}>0$, and $\tau_{j,c}\geq 0$, $j=1,\ldots,r-1$. The estimates of $\tau_{j,c}$, $j=1,\ldots,r$ are obtained by means of the
cyclic optimization procedure proposed under $H_{\text{I},1}$, by replacing \eqref{algorithm2-2} with
\begin{multline}\label{tau2}
\sum\limits_{i=1}^h\frac{\left[(1-a)K_{\text{I},1}+a(K_S-K_{\text{I},1})\right]}{\widehat{\sigma}^2+\left(\sum\limits_{j=i}^{h-1}\widehat{\tau}^{(n)}_{j,c}+\tau_{h,c}
+\sum\limits_{j=h+1}^{r}\widehat{\tau}^{(n-1)}_{j,c}\right)\widehat{\lambda}_{i}}\\
-\sum\limits_{i=1}^h\frac{\bS_{C_2}(i,i)}{\left[\widehat{\sigma}^2+\left(\sum\limits_{j=i}^{h-1}\widehat{\tau}^{(n)}_{j,c}+\tau_{h,c}
+\sum\limits_{j=h+1}^{r}\widehat{\tau}^{(n-1)}_{j,c}\right)\widehat{\lambda}_{i}\right]^2}=0,\\
 h=1,\ldots,r.
\end{multline}
The estimates of $\gamma_{i,2}$,
denoted by $\widehat{\gamma}_{i,2}$, are given by
\begin{equation}\label{gamma2}
\widehat{\gamma}_{i,2}=\sum\limits_{j=i}^{r}\widehat{\tau}^{(n_{\text{max}})}_{j,c},\ i=1,\ldots,r.
\end{equation}

The compressed log-likelihood of secondary data under $H_{\text{I},2}$ can be written as
\begin{multline}\label{hI2}
h_{\text{I},2}(\bZ;\widehat{\bm{\theta}}_2)=\max\limits_{K_{\text{I},1}}\left\{
-\left[K_P+aK_{\text{I},1}+(1-a)(K_S-K_{\text{I},1})\right]\right.\\
\times\sum_{i=1}^r\log\left(\widehat{\sigma}^2+\widehat{\lambda}_{i}\right)
-\left[(1-a)K_{\text{I},1}+a(K_S-K_{\text{I},1})\right]\\
\times\sum_{i=1}^r\log\left(\widehat{\sigma}^2+\widehat{\gamma}_{i,2}\widehat{\lambda}_{i}\right)
-\sum_{i=1}^r\frac{\bS_P(i,i)}{\widehat{\sigma}^2+\widehat{\lambda}_{i}}
-\sum_{i=r+1}^N\frac{\bS_P(i,i)}{\widehat{\sigma}^2}\\
\\
-\sum_{i=1}^r\frac{\bS_{C_1}(i,i)}{\widehat{\sigma}^2+\widehat{\lambda}_{i}}-\sum_{i=r+1}^N\frac{\bS_{C_1}(i,i)}{\widehat{\sigma}^2}
-\sum_{i=1}^r\frac{\bS_{C_2}(i,i)}{\widehat{\sigma}^2+\widehat{\gamma}_{i,2}\widehat{\lambda}_{i}}\\
\left.-\sum_{i=r+1}^N\frac{\bS_{C_2}(i,i)}{\widehat{\sigma}^2}-(K_P+K_S)(N-r)\log\widehat{\sigma}^2\right\}.
\end{multline}

Finally, under $H_{\text{I},3}$, we approximate the maximization over $\log{f_{\text{I},3}(\bZ,\bm{\theta}_3)}$ exploiting again $\widehat{\bU}$ as follows
\begin{multline}
\max\limits_{K_{\text{I},2},K_{\text{I},3}}\left\{\max\limits_{\sigma^2,\lambda_1,\ldots,\lambda_r\atop {\gamma_{1,3},\ldots,\gamma_{r,3},\atop \gamma_{1,4},\ldots,\gamma_{r,4}}}\right.
-(K_P+K_{\text{I},3}-K_{\text{I},2})\sum_{i=1}^r\log\left(\sigma^2+\lambda_i\right)\\
-(K_P+K_S)(N-r)\log\sigma^2-K_{\text{I},2}\sum_{i=1}^r\log\left(\sigma^2+\lambda_i^{(f)}\right)\\
\left.-(K_S-K_{\text{I},3})\sum_{i=1}^r\log\left(\sigma^2+\lambda_i^{(g)}\right)-t_3(\bZ;\bm{\theta}_3)\right\},
\end{multline}
where $\lambda_i^{(f)}=\gamma_{i,3}\lambda_i$, $\lambda_i^{(g)}=\gamma_{i,4}\lambda_i$, $i=1,\ldots,r$, and
\begin{multline}
t_3(\bZ;\bm{\theta}_3)=\sum_{i=1}^r\frac{\bS_P(i,i)}{\sigma^2+\lambda_i}+\sum_{i=r+1}^N\frac{\bS_P(i,i)}{\sigma^2}\\
+\sum_{i=1}^r\frac{\bS_{C_3}(i,i)}{\sigma^2+\lambda_i}+\sum_{i=r+1}^N\frac{\bS_{C_3}(i,i)}{\sigma^2}
+\sum_{i=1}^r\frac{\bS_{C_4}(i,i)}{\sigma^2+\lambda_i^{(f)}}\\
+\sum_{i=r+1}^N\frac{\bS_{C_4}(i,i)}{\sigma^2}
+\sum_{i=1}^r\frac{\bS_{C_5}(i,i)}{\sigma^2+\lambda_i^{(g)}}+\sum_{i=r+1}^N\frac{\bS_{C_5}(i,i)}{\sigma^2},
\end{multline}
with
\be
\bS_{C_3}=\sum\limits_{k=K_{\text{I},2}+1}^{K_{\text{I},3}}\widehat{\bU}^\dag\bor_k\bor_k^\dag\widehat{\bU}, \
\bS_{C_4}=\sum\limits_{k=1}^{K_{\text{I},2}}\widehat{\bU}^\dag\bor_k\bor_k^\dag\widehat{\bU},
\ee
and
\be
\bS_{C_5}=\sum\limits_{k=K_{\text{I},3}+1}^{K_S}\widehat{\bU}^\dag\bor_k\bor_k^\dag\widehat{\bU}.
\ee
Since given $\lambda_i$, $\lambda^{(f)}_i$ and $\lambda^{(g)}_i$ are completely determined by $\gamma_{i,3}$ and $\gamma_{i,4}$, we first estimate $\sigma^2$ and $\lambda_i$ by means of \eqref{sigma} and \eqref{lambda}.
As for $\gamma_{i,3}$, $\gamma_{i,4}$, $i=1,\ldots,r$, we introduce $\gamma_{i,3}=\sum\limits_{j=i}^{r}\tau_{j,f}$, $\gamma_{i,4}=\sum\limits_{j=i}^{r}\tau_{j,g}$,
where $\tau_{r,f},\tau_{r,g}>0$, and $\tau_{i,f},\tau_{i,g}\geq 0$, $i=1,\ldots,r-1$.
The estimates of $\tau_{j,f}$, $j=1,\ldots,r$, are obtained through the cyclic optimization procedure proposed under $H_{\text{I},1}$ by
replacing \eqref{algorithm2-2} with
\begin{multline}\label{tau3}
\sum\limits_{i=1}^h\frac{K_{\text{I},2}}{\widehat{\sigma}^2+\left(\sum\limits_{j=i}^{h-1}\widehat{\tau}^{(n)}_{j,f}+\tau_{h,f}
+\sum\limits_{j=h+1}^{r}\widehat{\tau}^{(n-1)}_{j,f}\right)\widehat{\lambda}_{i}}\\
-\sum\limits_{i=1}^h\frac{\bS_{C_4}(i,i)}{\left[\widehat{\sigma}^2+\left(\sum\limits_{j=i}^{h-1}\widehat{\tau}^{(n)}_{j,f}+\tau_{h,f}
+\sum\limits_{j=h+1}^{r}\widehat{\tau}^{(n-1)}_{j,f}\right)\widehat{\lambda}_{i}\right]^2}=0,\\
h=1,\ldots,r.
\end{multline}

The estimates of $\tau_{j,g}$, $j=1,\ldots,r$, are obtained through the same approach  by
replacing \eqref{algorithm2-2} with
\begin{multline}\label{tau4}
\sum\limits_{i=1}^h\frac{K_S-K_{\text{I},3}}{\widehat{\sigma}^2+\left(\sum\limits_{j=i}^{h-1}\widehat{\tau}^{(n)}_{j,g}+\tau_{h,g}
+\sum\limits_{j=h+1}^{r}\widehat{\tau}^{(n-1)}_{j,g}\right)\widehat{\lambda}_{i}}\\
-\sum\limits_{i=1}^h\frac{\bS_{C_5}(i,i)}{\left[\widehat{\sigma}^2+\left(\sum\limits_{j=i}^{h-1}\widehat{\tau}^{(n)}_{j,g}+\tau_{h,g}
+\sum\limits_{j=h+1}^{r}\widehat{\tau}^{(n-1)}_{j,g}\right)\widehat{\lambda}_{i}\right]^2}=0,\\
 h=1,\ldots,r.
\end{multline}
Thus the estimates of $\gamma_{i,3}$, $\gamma_{i,4}$,
denoted by $\widehat{\gamma}_{i,3}$, $\widehat{\gamma}_{i,4}$, respectively, are given by
\begin{equation}\label{gamma3}
\widehat{\gamma}_{i,3}=\sum\limits_{j=i}^{r}\widehat{\tau}^{(n_{\text{max}})}_{j,f},\ i=1,\ldots,r,
\end{equation}
\begin{equation}\label{gamma4}
\widehat{\gamma}_{i,4}=\sum\limits_{j=i}^{r}\widehat{\tau}^{(n_{\text{max}})}_{j,g},\ i=1,\ldots,r.
\end{equation}

Finally, $h_{\text{I},3}(\bZ;\widehat{\bm{\theta}}_3)$ is written as
\begin{multline}\label{hI3}
h_{\text{I},3}(\bZ;\widehat{\bm{\theta}}_3)
\\
=\max\limits_{K_{\text{I},2},K_{\text{I},3}}\left\{
-(K_P+K_{\text{I},3}-K_{\text{I},2})\sum_{i=1}^r\log\left(\widehat{\sigma}^2+\widehat{\lambda}_{i}\right)\right.
\\
-K_{\text{I},2}\sum_{i=1}^r\log\left(\widehat{\sigma}^2+\widehat{\gamma}_{i,3}\widehat{\lambda}_{i}\right)-(K_P+K_S)(N-r)\log\widehat{\sigma}^2
\\
-(K_S-K_{\text{I},3})\sum_{i=1}^r\log\left(\widehat{\sigma}^2+\widehat{\gamma}_{i,4}\widehat{\lambda}_{i}\right)
-\sum_{i=1}^r\frac{\bS_P(i,i)}{\widehat{\sigma}^2+\widehat{\lambda}_{i}}
\\
-\sum_{i=r+1}^N\frac{\bS_P(i,i)}{\widehat{\sigma}^2}
-\sum_{i=1}^r\frac{\bS_{C_3}(i,i)}{\widehat{\sigma}^2+\widehat{\lambda}_{i}}
-\sum_{i=r+1}^N\frac{\bS_{C_3}(i,i)}{\widehat{\sigma}^2}
\\
-\sum_{i=1}^r\frac{\bS_{C_4}(i,i)}{\widehat{\sigma}^2+\widehat{\gamma}_{i,3}\widehat{\lambda}_{i}}
-\sum_{i=r+1}^N\frac{\bS_{C_4}(i,i)}{\widehat{\sigma}^2}
-\sum_{i=1}^r\frac{\bS_{C_5}(i,i)}{\widehat{\sigma}^2+\widehat{\gamma}_{i,4}\widehat{\lambda}_{i}}
\\
\left.-\sum_{i=r+1}^N\frac{\bS_{C_5}(i,i)}{\widehat{\sigma}^2}
\right\}.\
\end{multline}

Plugging \eqref{parameternumber1}, \eqref{hI1}, \eqref{hI0}, \eqref{hI2}, \eqref{hI3} into \eqref{MOSscheme1} leads to the final classification results for multiple hypothesis testing problem \eqref{hypothese1}.

\subsection{Clutter Covariance Variation Model 2}
In this subsection we provide the expressions of the compressed log-likelihood functions that are used to compute \eqref{MOSscheme2}. To this end, we derive the MLEs of $\bm{\delta}_n$, $n=1,\ldots,4$.
Again, for the reader ease, we start with the optimization over $\bm{\delta}_1$ under $H_{\text{II},1}$, which is tantamount to
\begin{multline}\label{maxHII1}
\max\limits_{\sigma^2,\bm{\Lambda},\bm{U} \atop \bm{\Lambda}_1,\bm{U}_1}-N(K_P+K_S) \log\pi-K_P\sum_{i=1}^r\log{\left(\sigma^2+\lambda_i\right)}\\
-K_S\sum_{i=1}^r\log{\left(\sigma^2+\lambda_{i,1}\right)}
-(K_P+K_S)(N-r)\log\sigma^2\\
-\mbox{Tr}\left[\left(\sigma^2\bI+\bm{\Lambda}\right)^{-1}\bU^\dag\bV_P\bm{\Theta}_P\bV_P^\dag\bU\right]\\
-\mbox{Tr}\left[\left(\sigma^2\bI+\bm{\Lambda}_1\right)^{-1}\bU_1^\dag\bV_S\bm{\Theta}_S\bV_S^\dag\bU_1\right],
\end{multline}
where $\bm{\Theta}_P=diag(\mu_{1,P},\ldots,\mu_{N,P})$ with $\mu_{1,P}\geq\ldots\geq\mu_{N,P}\geq0$ and $\bm{\Theta}_S=diag(\mu_{1,S},\ldots,\mu_{N,S})$ with $\mu_{1,S}\geq\ldots\geq\mu_{N,S}\geq0$ being the eigenvalues of $\bZ_P\bZ_P^\dag$ and $\bZ_S\bZ_S^\dag$, respectively, $\bV_P\in\mathds{C}^{N\times N}$ and $\bV_S\in\mathds{C}^{N\times N}$ contains the corresponding eigenvectors.
It is important to observe that under $H_{\text{II},1}$, $\bU_1$ is different from $\bU$, thus exploiting {\em Theorem 1} in \cite{Mirsky} yields
\begin{multline}
\max\limits_{\bm{U}}-\mbox{Tr}\left[\left(\sigma^2\bI+\bm{\Lambda}\right)^{-1}\bU^\dag\bV_P\bm{\Theta}_P\bV_P^\dag\bU\right]\\
=-\mbox{Tr}\left[\left(\sigma^2\bI+\bm{\Lambda}\right)^{-1}\bm{\Theta}_P\right],
\end{multline}
and
\begin{multline}
\max\limits_{\bm{U}_1}
-\mbox{Tr}\left[\left(\sigma^2\bI+\bm{\Lambda}_1\right)^{-1}\bU_1^\dag\bV_S\bm{\Theta}_S\bV_S^\dag\bU_1\right]\\
=-\mbox{Tr}\left[\left(\sigma^2\bI+\bm{\Lambda}_1\right)^{-1}\bm{\Theta}_S\right].
\end{multline}
As a consequence, following the lead of \cite{XU2021108127} and \cite{YLJ2019ECCM}, It can be shown that the estimates of, $\sigma^2$, $\lambda_i$, and $\lambda_{i,1}$ under $H_{\text{II},1}$ are given by
\begin{equation}\label{estHII1}
\begin{cases}
\widehat{\sigma}_{(1)}^2=\frac{1}{(K_P+K_S)(N-r)}\sum_{i=r+1}^N\left(\mu_{i,P}+\mu_{i,S}\right),
\\
\\
\widehat{\lambda}_{(1),i}=\max\left\{\frac{1}{K_P}\mu_{i,P}-\widehat{\sigma}_{(1)}^2, 0\right\}, \ i=1,\ldots,r,
\\
\\
\widehat{\lambda}_{i,1}=\max\left\{\frac{1}{K_S}\mu_{i,S}-\widehat{\sigma}_{(1)}^2, 0\right\}, \ i=1,\ldots,r.
\end{cases}
\end{equation}
Plugging \eqref{estHII1} into \eqref{maxHII1} leads to
\begin{multline}\label{hII1}
h_{\text{II},1}(\bZ;\widehat{\bm{\delta}}_0)=-N(K_P+K_S)\log\pi
\\
-(K_P+K_S)(N-r)\log{\widehat{\sigma}_{(1)}^2}
-K_P\sum_{i=1}^r\log(\widehat{\sigma}^2_{(1)}+\widehat{\lambda}_{(1),i})
\\
-K_S\sum_{i=1}^r\log(\widehat{\sigma}^2_{(1)}+\widehat{\lambda}_{i,1})
-\sum_{i=1}^r\frac{\mu_{i,P}}{\widehat{\sigma}_{(1)}^2+\widehat{\lambda}_{(1),i}}
\\
-\sum_{i=1}^r\frac{\mu_{i,S}}{\widehat{\sigma}_{(1)}^2+\widehat{\lambda}_{i,1}}
-\frac{1}{\widehat{\sigma}_{(1)}^2}\sum_{i=r+1}^N(\mu_{i,P}+\mu_{i,S}).
\end{multline}

Under $H_{\text{II},0}$, following the lead of \cite{XU2021108127}, it can be shown that the estimation of $\bm{\delta}_0$ leads to the following maximization problem
\begin{multline}\label{maxHII0}
\max_{\sigma^2,\bm{\Lambda},\bm{U}}-N(K_P+K_S)\log\pi-(K_P+K_S)(N-r)\log\sigma^2\\
-(K_P+K_S)\sum_{i=1}^r\log{\left(\sigma^2+\lambda_i\right)}
-\mbox{Tr}\left[\left(\sigma^2\bI+\bm{\Lambda}\right)^{-1}\bU^\dag\bV\bm{\Theta}\bV^\dag\bU\right],\nonumber
\end{multline}
where $\bm{\Theta}=diag\left(\mu_{1},\ldots,\mu_{N}\right)$ with $\mu_{1}\geq\ldots\geq\mu_{N}\geq0$ being the eigenvalues of $\bZ\bZ^\dag$ and $\bV\in\mathds{C}^{N\times N}$ is the unitary matrix of corresponding eigenvectors.
The estimates of $\sigma^2$, and $\lambda_i$ under $H_{\text{II},0}$ are given by
\begin{equation}
\begin{cases}
\widehat{\sigma}_{(0)}^2=\frac{1}{(K_P+K_S)(N-r)}\sum_{i=r+1}^N\mu_{i},
\\
\\
\widehat{\lambda}_{(0),i}=\max\left\{\frac{1}{K_P+K_S}\mu_{i}-\widehat{\sigma}_{(0)}^2, 0\right\}, \ i=1,\ldots,r.
\end{cases}
\end{equation}

Gathering the above results, the compressed log-likelihood under $H_{\text{II},0}$ can be written as
\begin{multline}\label{hII0}
h_{\text{II},0}(\bZ;\widehat{\bm{\delta}}_0)=- N(K_P+K_S)\log\pi
\\
-(K_P+K_S)(N-r)\log{\widehat{\sigma}_{(0)}^2}
\\
-(K_P+K_S)\sum_{i=1}^r\log(\widehat{\sigma}^2_{(0)}+\widehat{\lambda}_{(0),i})
\\
-\sum_{i=1}^r\frac{\mu_{i}}{\widehat{\sigma}_{(0)}^2+\widehat{\lambda}_{(0),i}}
-\sum_{i=r+1}^N\frac{\mu_{i}}{\widehat{\sigma}_{(0)}^2}.
\end{multline}

Similarly, it can be shown that the compressed log-likelihood functions under $H_{\text{II},2}$, $H_{\text{II},3}$, and $H_{\text{II},4}$ are given by
\begin{multline}
h_{\text{II},2}(\bZ;\widehat{\bm{\delta}}_2)=\max\limits_{K_{\text{II},1}}\Bigg\{-N(K_P+K_S)\log\pi
\\
-\left[K_P+bK_{\text{II},1}+(1-b)(K_S-K_{\text{II},1})\right]\sum\limits_{i=1}^r\log(\widehat{\sigma}^2_{(2)}+\widehat{\lambda}_{(2),i})
\\
-\left[(1-b)K_{\text{II},1}+b(K_S-K_{\text{II},1})\right]\sum\limits_{i=1}^r\log(\widehat{\sigma}^2_{(2)}+\widehat{\lambda}_{i,2})
\\
-(K_P+K_S)(N-r)\log\widehat{\sigma}_{(2)}^2
-\sum_{i=1}^r\frac{\mu_{i,1}}{\widehat{\sigma}_{(2)}^2+\widehat{\lambda}_{(2),i}}
\\
-\sum_{i=1}^r\frac{\mu_{i,2}}{\widehat{\sigma}_{(2)}^2+\widehat{\lambda}_{i,2}}
-\frac{1}{\widehat{\sigma}_{(2)}^2}\sum_{i=r+1}^N(\mu_{i,1}+\mu_{i,2})\Bigg\},
\end{multline}
\begin{multline}
h_{\text{II},3}(\bZ;\widehat{\bm{\delta}}_3)=\max\limits_{K_{\text{II},2},K_{\text{II},3}}\Bigg\{-N(K_P+K_S)\log\pi
\\
-(K_P+K_{\text{II},3}-K_{\text{II},2})\sum\limits_{i=1}^r\log(\widehat{\sigma}^2_{(3)}+\widehat{\lambda}_{(3),i})
\\
-K_{\text{II},2}\sum\limits_{i=1}^r\log(\widehat{\sigma}^2_{(3)}+\widehat{\lambda}_{i,3})
-(K_S-K_{\text{II},3})\sum\limits_{i=1}^r\log(\widehat{\sigma}^2_{(3)}+\widehat{\lambda}_{i,4})
\\
-(K_P+K_S)(N-r)\log\widehat{\sigma}_{(3)}^2
-\sum_{i=1}^r\frac{\mu_{i,3}}{\widehat{\sigma}_{(3)}^2+\widehat{\lambda}_{(3),i}}
\\
-\sum_{i=1}^r\frac{\mu_{i,4}}{\widehat{\sigma}_{(3)}^2+\widehat{\lambda}_{i,3}}-\sum_{i=1}^r\frac{\mu_{i,5}}{\widehat{\sigma}_{(3)}^2+\widehat{\lambda}_{i,4}}
\\
\left.-\frac{1}{\widehat{\sigma}_{(3)}^2}\sum_{i=r+1}^N(\mu_{i,3}+\mu_{i,4}+\mu_{i,5})\right\},
\end{multline}

\begin{multline}
h_{\text{II},4}(\bZ;\widehat{\bm{\delta}}_4)=\max\limits_{K_{\text{II},4}}\left\{-N(K_P+K_S)\log\pi\right.
\\
-K_P\sum\limits_{i=1}^r\log(\widehat{\sigma}^2_{(4)}+\widehat{\lambda}_{(4),i})-K_{\text{II},4}\sum\limits_{i=1}^r\log(\widehat{\sigma}^2_{(4)}+\widehat{\lambda}_{i,5})\\
-(K_S-K_{\text{II},4})\sum\limits_{i=1}^r\log(\widehat{\sigma}^2_{(4)}+\widehat{\lambda}_{i,6})
-(K_P+K_S)(N-r)\log\widehat{\sigma}_{(4)}^2\\
-\sum_{i=1}^r\frac{\mu_{i,P}}{\widehat{\sigma}_{(4)}^2+\widehat{\lambda}_{(4),i}}
-\sum_{i=1}^r\frac{\mu_{i,6}}{\widehat{\sigma}_{(4)}^2+\widehat{\lambda}_{i,5}}-\sum_{i=1}^r\frac{\mu_{i,7}}{\widehat{\sigma}_{(4)}^2+\widehat{\lambda}_{i,6}}\\
\left.-\frac{1}{\widehat{\sigma}_{(4)}^2}\sum_{i=r+1}^N(\mu_{i,P}+\mu_{i,6}+\mu_{i,7})\right\},
\end{multline}
where
\begin{itemize}
\item $\widehat{\sigma}_{(2)}^2=\frac{1}{(K_P+K_S)(N-r)}\sum\limits_{i=r+1}^N\left(\mu_{i,1}+\mu_{i,2}\right)$, $\widehat{\lambda}_{(2),i}=\max\left\{\mu_{i,1}/\left[K_P+bK_{\text{II},1}+(1-b)(K_S-K_{\text{II},1})\right]-\widehat{\sigma}_{(2)}^2, 0\right\}$,
$\widehat{\lambda}_{i,2}=\max\left\{\mu_{i,2}/\left[(1-b)K_{\text{II},1}+b(K_S-K_{\text{II},1})\right]-\widehat{\sigma}_{(2)}^2, 0\right\}$, $i=1,\ldots,r$,
with
\begin{equation}\label{B}
b =
\left\{
\begin{array}{lll}
1,&\ K_{\text{II},1}>\frac{K_S}{2},
\\
\\
0,&\ K_{\text{II},1}\leq\frac{K_S}{2};
\end{array}
\right.
\end{equation}
$\mu_{1,1}\geq\ldots\geq\mu_{N,1}\geq0$ and $\mu_{1,2}\geq\ldots\geq\mu_{N,2}\geq0$ are the eigenvalues of $\bZ_{P}\bZ_{P}^\dag+\bZ_{S_1}\bZ_{S_1}^\dag$
    and $\bZ_{S_2}\bZ_{S_2}^\dag$, respectively, with
\begin{equation}
\bZ_{S_1}=
\left\{
\begin{array}{lll}
\left[\bor_1,\ldots,\bor_{K_{\text{II},1}}\right],&\ K_{\text{II},1}>\frac{K_S}{2},
\\
\\
\left[\bor_{K_{\text{II},1}+1},\ldots,\bor_{K_S}\right],&\ K_{\text{II},1}\leq\frac{K_S}{2},
\end{array}
\right.
\end{equation}
and $\bZ_{S_2}\bZ_{S_2}^\dag=\bZ_{S}\bZ_{S}^\dag-\bZ_{S_1}\bZ_{S_1}^\dag$;
\item $\widehat{\sigma}_{(3)}^2=\frac{1}{(K_P+K_S)(N-r)}\sum\limits_{i=r+1}^N\left(\mu_{i,3}+\mu_{i,4}+\mu_{i,5}\right)$,
 $\widehat{\lambda}_{(3),i}=\max\left\{\frac{1}{K_P+K_{\text{II},3}-K_{\text{II},2}}\mu_{i,3}-\widehat{\sigma}_{(3)}^2, 0\right\}$,
 $\widehat{\lambda}_{i,3}=\max\left\{\frac{1}{K_{\text{II},2}}\mu_{i,4}-\widehat{\sigma}_{(3)}^2, 0\right\}$,
$\widehat{\lambda}_{i,4}=\max\left\{\frac{1}{K_S-K_{\text{II},3}}\mu_{i,5}-\widehat{\sigma}_{(3)}^2, 0\right\}$, $i=1,\ldots,r$;
$\mu_{1,3}\geq\ldots\geq\mu_{N,3}\geq0$, $\mu_{1,4}\geq\ldots\geq\mu_{N,4}\geq0$, and $\mu_{1,5}\geq\ldots\geq\mu_{N,5}\geq0$ are the eigenvalues of $\bZ_{P}\bZ_{P}^\dag+\bZ_{S_3}\bZ_{S_3}^\dag$, $\bZ_{S_4}\bZ_{S_4}^\dag$,
    and $\bZ_{S_5}\bZ_{S_5}^\dag$, respectively, with $\bZ_{S_3}=\left[\bor_{K_{\text{II},2}+1},\ldots,\bor_{K_{\text{II},3}}\right]$, $\bZ_{S_4}=\left[\bor_1,\ldots,\bor_{K_{\text{II},2}}\right]$ and
    $\bZ_{S_5}=\left[\bor_{K_{\text{II},3}+1},\ldots,\bor_{K_S}\right]$;
\item $\widehat{\sigma}_{(4)}^2=\frac{1}{(K_P+K_S)(N-r)}\sum\limits_{i=r+1}^N\left(\mu_{i,P}+\mu_{i,6}+\mu_{i,7}\right)$,
 $\widehat{\lambda}_{(4),i}=\max\left\{\frac{1}{K_P}\mu_{i,P}-\widehat{\sigma}_{(4)}^2, 0\right\}$,
 $\widehat{\lambda}_{i,5}=\max\left\{\frac{1}{K_{\text{II},4}}\mu_{i,6}-\widehat{\sigma}_{(4)}^2, 0\right\}$,
$\widehat{\lambda}_{i,6}=\max\left\{\frac{1}{K_S-K_{\text{II},4}}\mu_{i,7}-\widehat{\sigma}_{(4)}^2, 0\right\}$, $i=1,\ldots,r$;
$\mu_{1,6}\geq\ldots\geq\mu_{N,6}\geq0$, and $\mu_{1,7}\geq\ldots\geq\mu_{N,7}\geq0$ are the eigenvalues
of $\bZ_{S_6}\bZ_{S_6}^\dag$ and $\bZ_{S_7}\bZ_{S_7}^\dag$,
    respectively, with $\bZ_{S_6}=\left[\bor_1,\ldots,\bor_{K_{\text{II},4}}\right]$, and
    $\bZ_{S_7}=\left[\bor_{K_{\text{II},4}+1},\ldots,\bor_{K_S}\right]$.
\end{itemize}
\subsection{Implementation Issue for Unknown $r$}
In the case that $r$ is unknown, we have to estimate it from data.
To this end, we devise a preliminary stage that provides the estimate of $r$ by means of another MOS-based architecture. The estimate of $r$ exploiting
the AIC, GIC, and BIC is given by
$\widehat{r}=\arg\min_{r=1,\ldots,M}{\left\{-2h_{\text{II},0}(\bZ;\widehat{\bm{\delta}}_0)+\kappa\cdot p(r)\right\}}$,
where $M<N$ is an upper bound on $r$, $h_{\text{II},0}(\bZ;\widehat{\bm{\delta}}_0)$ is given by \eqref{hII0},
and $\kappa\cdot p(r)$ is the penalty term.

\section{Illustrative Examples}
\label{section4}
In this section, we investigate the classification performance of the proposed architectures in terms of the
Probability of Correct Classification ($P_{cc}$), and the Root
Mean Square (RMS) estimation errors of clutter edge positions, by means of standard Monte Carlo
counting techniques since closed-form expressions for the
considered performance metrics are not available.\footnote{In fact, as further claimed in \cite{MOS1}, we can
only provide a general view of the ranking of the considered criteria, but this ranking will
not necessarily hold in every application.}
Specifically, we estimate these metrics over  $L=1000$ independent trials.

The numerical examples assume that $N=9$, $K_P=8$, $K_S=32$, and $\sigma^2=1$. The clutter covariance matrix of primary data is defined as
$\bM = \sigma_c^2\sum_{\theta_i\in\Theta}\bv(\theta_i)\bv(\theta_i)^\dag$, where $\sigma_c^2$ is the clutter power of primary data set according to the Clutter to Noise Ratio (CNR) defined as
$\text{CNR}=10\log_{10}(\sigma_c^2/\sigma^2)=30\ \text{dB}$,
and $\Theta=\left\{-20^\circ, 0^\circ, 10^\circ\right\}$, which implies that the true rank of the clutter covariance matrix is $r=3$, with
$\bv(\theta_i)$ the steering vector given by
\begin{equation}\label{v}
\bv(\theta_i)=\frac{1}{\sqrt{N}}\left[1,e^{\jmath\pi\sin{\theta_i}},\ldots,e^{\jmath\pi(N-1)\sin{\theta_i}}\right].
\end{equation}
As for the GIC parameter,
since a clear guideline for choosing the values of $\rho$ does not exist \cite{MOS1},
we assess the performance of the GIC-based
architectures for different values of $\rho$.
Results not reported here for brevity show that
the GIC-based rule with $\rho\geq 5$ returns poor classification performance.
Thus, we select values lower than $5$ and in the specific case $\rho=2$ and $\rho=4$.
In what follows, we use GIC2 and GIC4 to denote the GIC-based classifier with
$\rho=2$ and $\rho=4$, respectively.

In the next subsections, we first investigate the classification and estimation performance for model 1 and then discuss the performance of the more general model 2.

\subsection{Clutter Covariance Variation Model 1}
\begin{figure}[t]
  \centering
  \subfigure[$H_{\text{I},0}$]{\includegraphics[width=4.2cm]{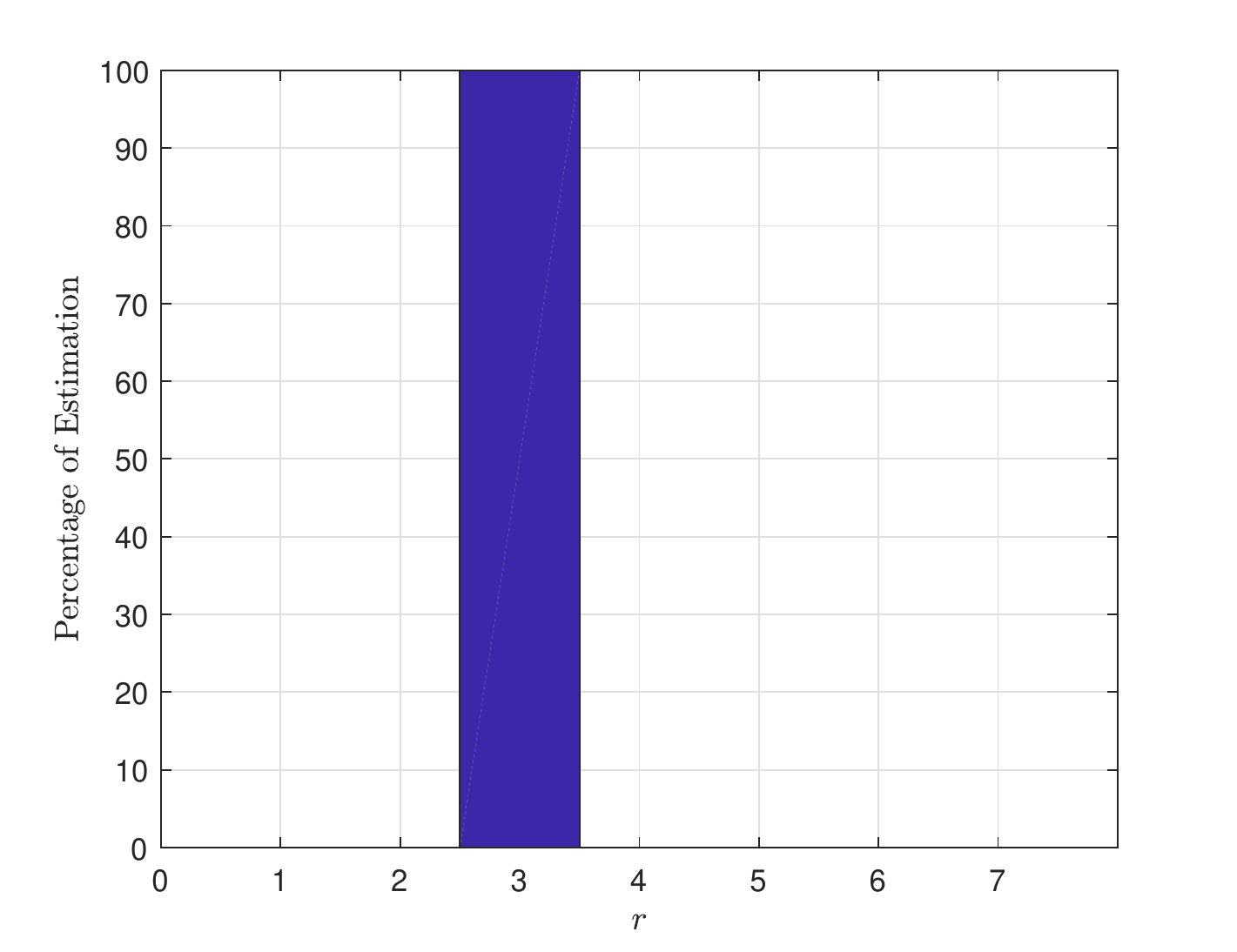}\label{BIC0}}
  \subfigure[$H_{\text{I},1}$]{\includegraphics[width=4.2cm]{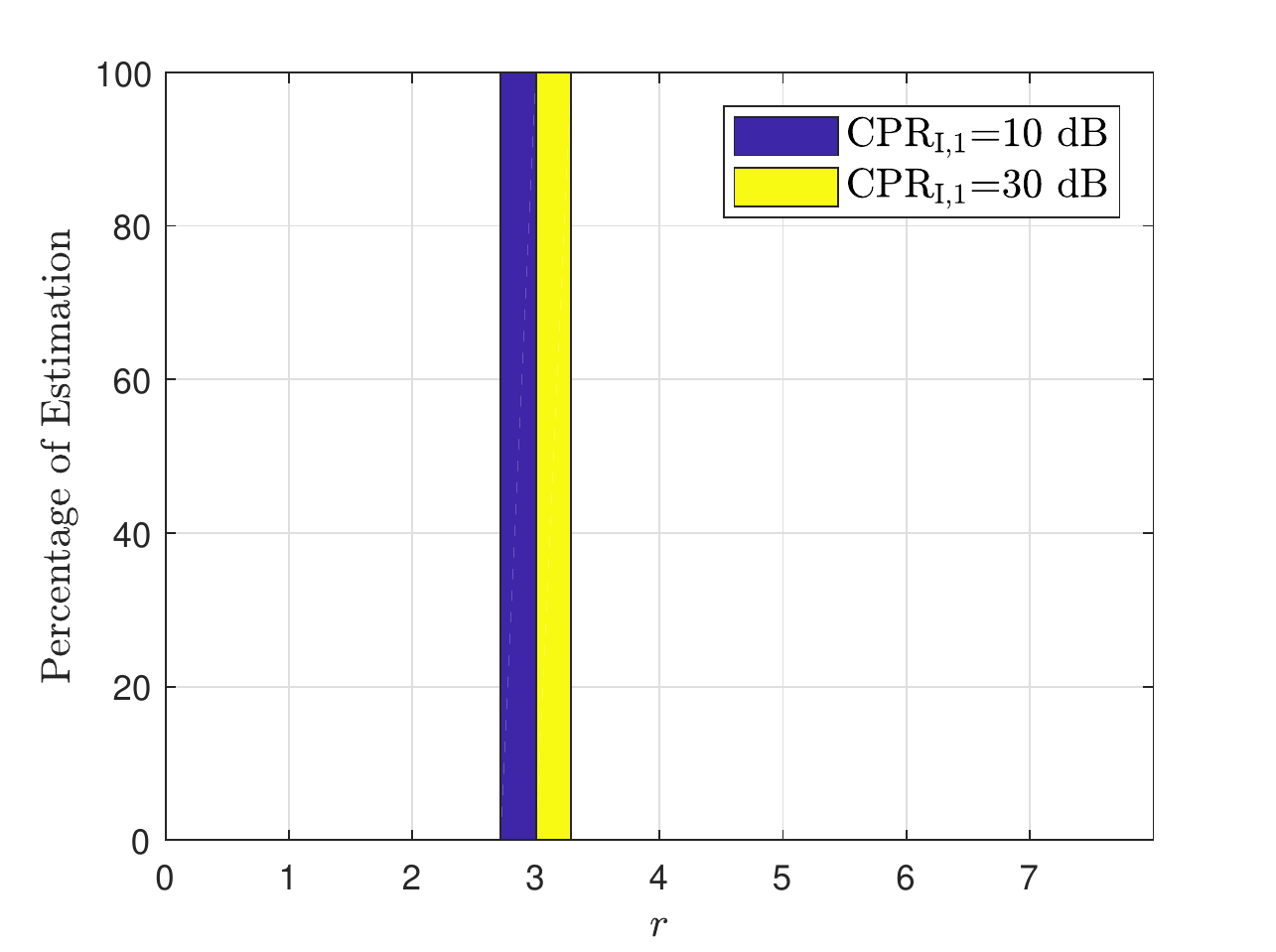}\label{BIC1}}
  \subfigure[$H_{\text{I},2}$, $K_{\text{I},1}=10$]{\includegraphics[width=4.2cm]{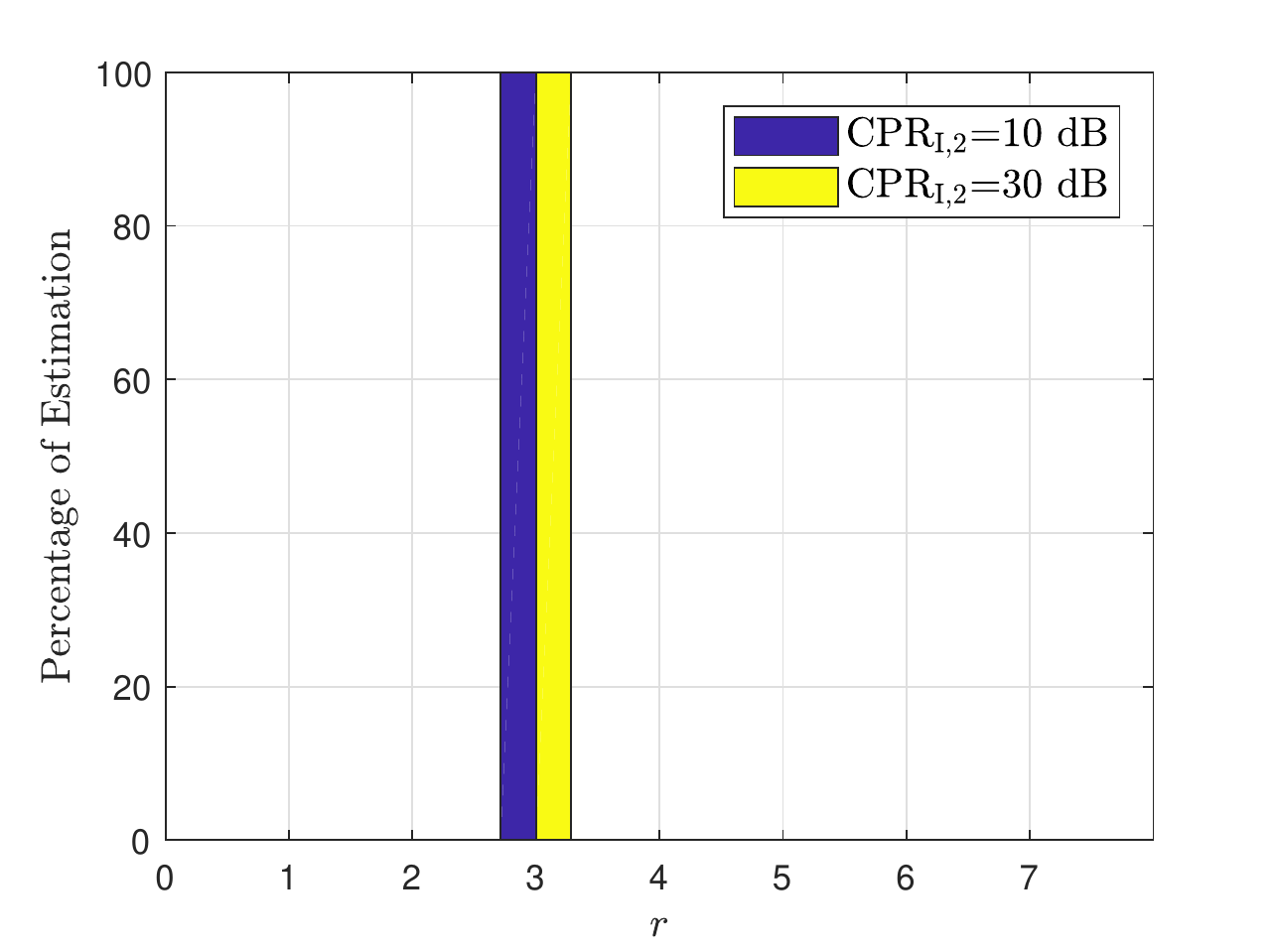}\label{BIC21}}
  \subfigure[$H_{\text{I},2}$, $K_{\text{I},1}=25$]{\includegraphics[width=4.2cm]{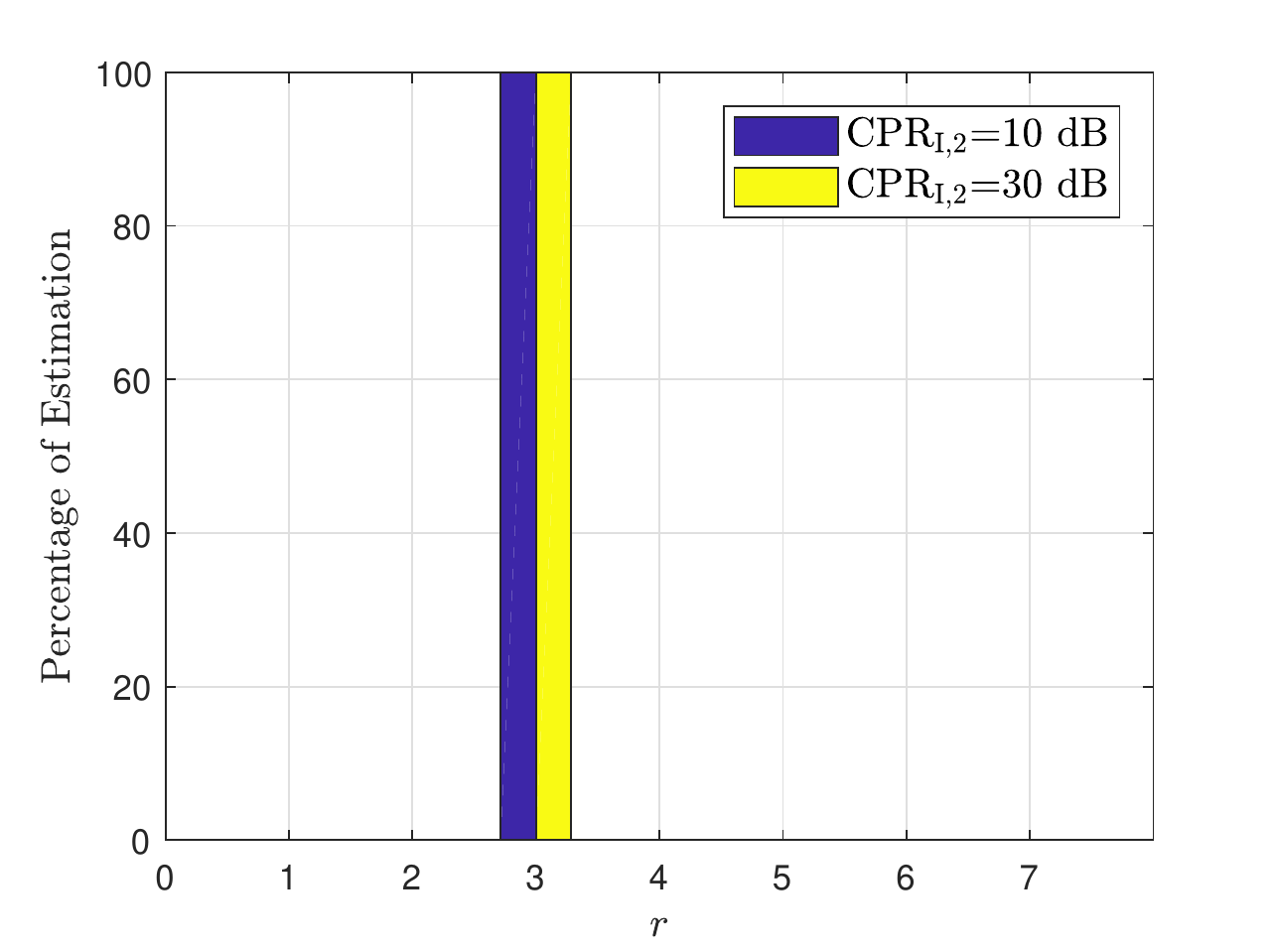}\label{BIC22}}
  \subfigure[$H_{\text{I},3}$, $\alpha=1$, $K_{\text{I},2}=4$, $K_{\text{I},3}=20$]{\includegraphics[width=4.2cm]{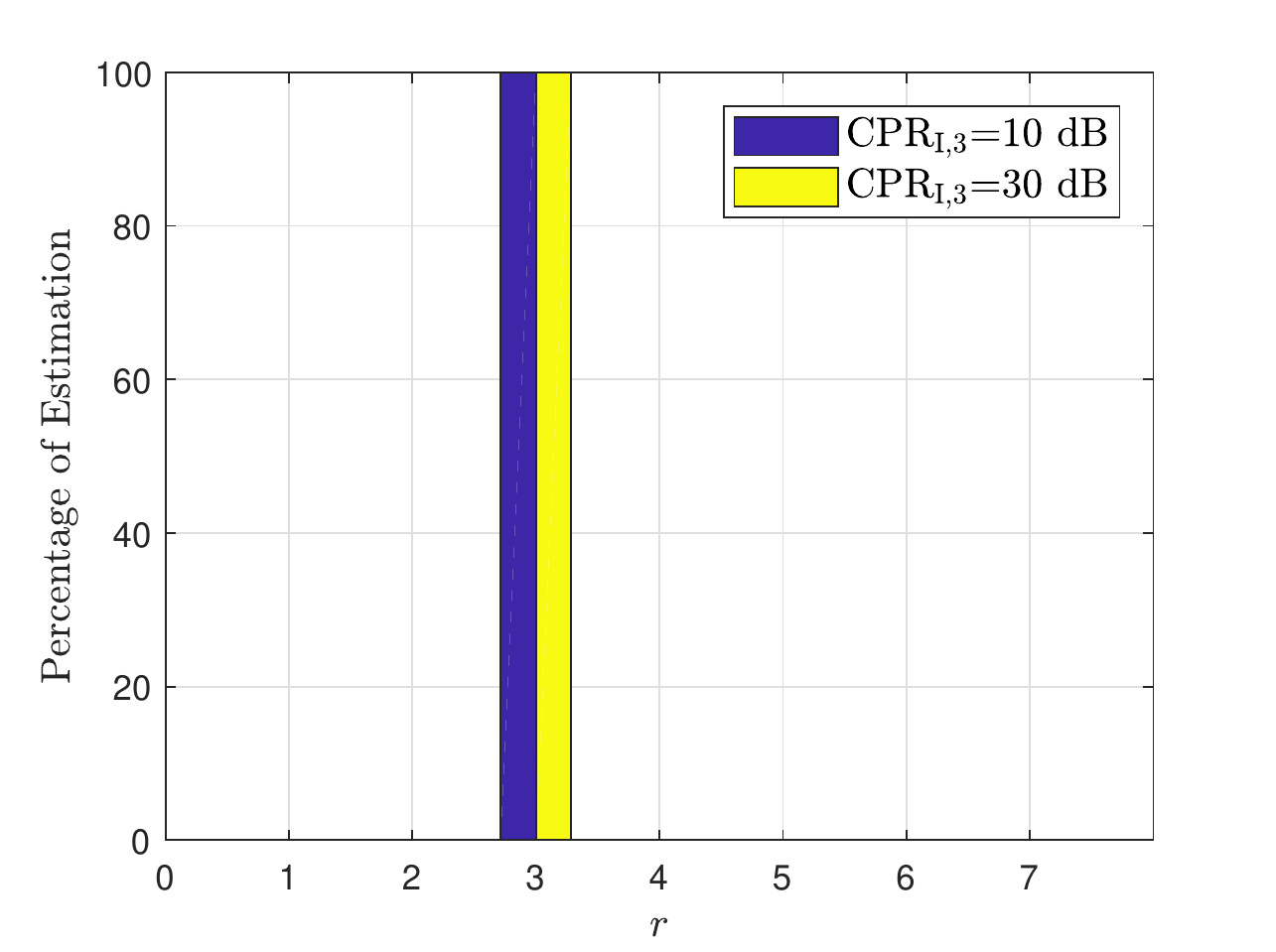}\label{BIC31}}
  \subfigure[$H_{\text{I},3}$, $\alpha=0.1$, $K_{\text{I},2}=10$, $K_{\text{I},3}=25$]{\includegraphics[width=4.2cm]{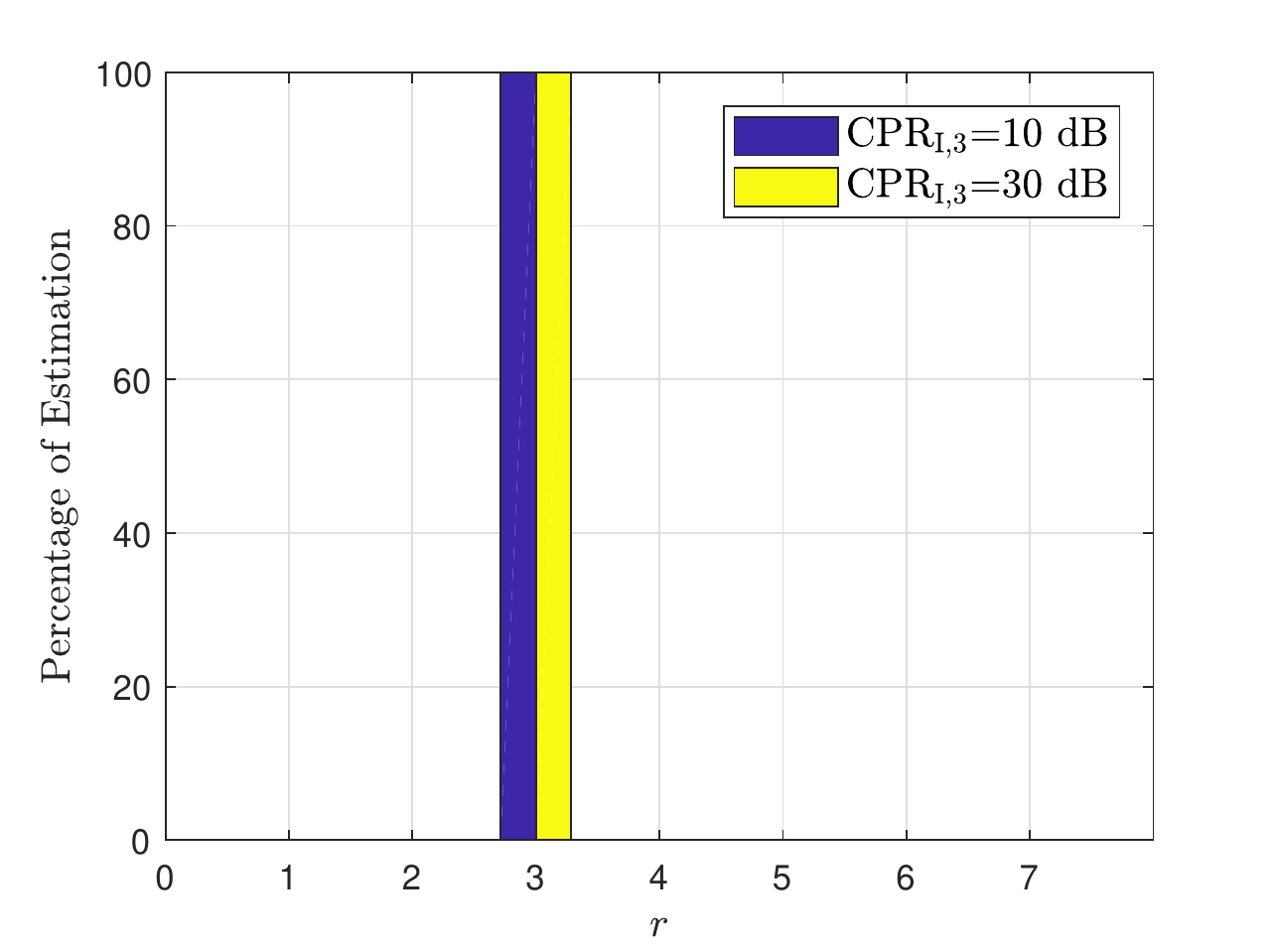}\label{BIC34}}
  \caption{Rank estimation performance under $H_{\text{I},l}$, CNR=30 dB}\label{rest}
\end{figure}

In this subsection, we estimate the classification performance of \eqref{MOSscheme1}. As for the $\bM_l$, $l=1,\ldots,4$, in order to form $\bm{\Gamma}_l$, we assume that $\omega_{1,l},\ldots,\omega_{r,l}$, are independent uniformly distributed random variables with $\omega_{i,l}\sim U(0,1)$, $i=1,\ldots,r$, then, we sort $\omega_{i,l}$ to obtain $\omega_{(1),l}\geq\ldots\geq\omega_{(r),l}$, and define
$\gamma_{i,l}=10^{\Delta_l} \omega_{(i),l}$, $i=1,\ldots,r,$ $l=1,\ldots,4$,
with $\Delta_l$, $l=1,2,3$, determined according to the Clutter Power Ratio (CPR) under $H_{\text{I},l}$, $l=1,\ldots,3$, given by
$\text{CPR}_{\text{I},l}=10\log_{10}10^{\Delta_l}=10\Delta_l$, $l=1,2,3$.
Otherwise stated, the $\gamma_{i,l}s$ are modeled as uniformly distributed random variables taking on values
in $[0,10^{\Delta_l}]$.
Notice also that the CPR indicates the ratio between the clutter power level in the region not including the
CUTs and the clutter power in the primary data.
Factors $\gamma_{i,l}$s under clutter covariance model 1 are related to the CPR through $\Delta_l$s.
At the same time, we assume that $\Delta_4=\alpha\Delta_3$ under $H_{\text{I},3}$ with $\alpha>0$ a scaling factor.
First of all, we focus on the performance of rank estimation.
In Fig. \ref{rest}, we show the percentage of estimation of the BIC rule under
each hypothesis with different configurations.\footnote{Results not shown
here indicate that the BIC rule returns better performance than AIC and GIC for rank estimation.
Such a behavior can be due to the fact that BIC is an
asymptotic approximation of the optimal Maximum a Posteriori rule unlike the other
two rules.} As can be observed, the preliminary stage returns a correct estimation probability no less than 99.7 percent for all the considered situations. In this respect, in the ensuing classification performance analysis we assume that the true value of $r$ is known.

\begin{figure}[t]
  \centering
  \includegraphics[width=7.7cm]{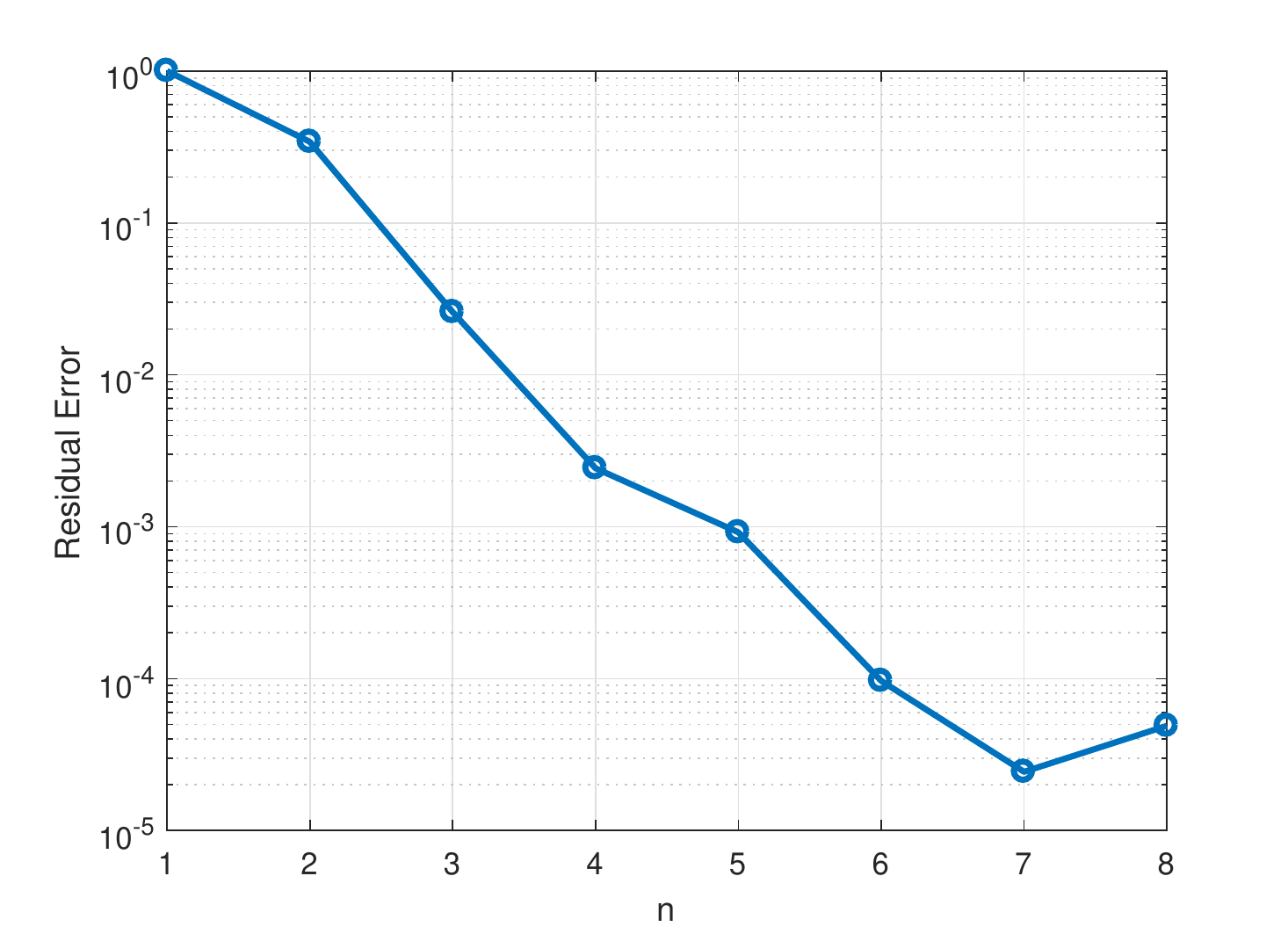}\\
  \caption{Residual errors versus iteration numbers of the heuristic approach}\label{nmax}
\end{figure}

Next, we investigate the behavior of the cyclic procedures for the purposes of $\tau$s estimations in terms of the iteration number $n$.
To this end, Fig. \ref{nmax} shows the average residual errors of
\be
\Delta\Psi(n)=\left|\frac{\left[\Psi(\bZ_S|\widehat{\bm{\theta}}^{(n)}_1)-\Psi(\bZ_S|\widehat{\bm{\theta}}^{(n-1)}_1)\right]}{\Psi(\bZ_S|\widehat{\bm{\theta}}^{(n)}_1)}\right|
\ee
over 1000 trials against $n$, where $\Psi(\bZ_S|\widehat{\bm{\theta}}^{(n)}_1)$ is the objective function in \eqref{H1-gamma-2} with the results of $n$th iteration plugged in.
It can be observed that, for the considered parameters, the variation is lower than $10^{-4}$ when $n=6$. Similar results can be found under the other hypotheses and not reported here
for brevity. As a consequence, we set $n_{\text{max}}=6$ in what follows.

Table \ref{table:H0} contains the classification results of 1000 trials when data are generated under $H_{\text{I},0}$. Inspection of
the Table highlights that the classification procedure correctly
decides $H_{\text{I},0}$ with a $P_{cc}=1$ for all the considered criteria.
\begin{table}[t]
\caption{Classification results under $H_{\text{I},0}$, CNR=30 dB}
\label{table:H0}
\centering
\begin{tabular}{c|ccccc}
  \toprule
   &$H_{\text{I},0}$ & $H_{\text{I},1}$ & $H_{\text{I},2}$ & $H_{\text{I},3}$  \\
  \midrule
   AIC & 1000     & 0   & 0 & 0\\
   GIC2 & 1000     & 0   & 0 & 0\\
   GIC4 & 1000     & 0   & 0 & 0\\
   BIC & 1000     & 0   & 0 & 0\\
  \bottomrule
\end{tabular}
\end{table}

\begin{figure}[t]
  \centering
  \includegraphics[width=7.7cm]{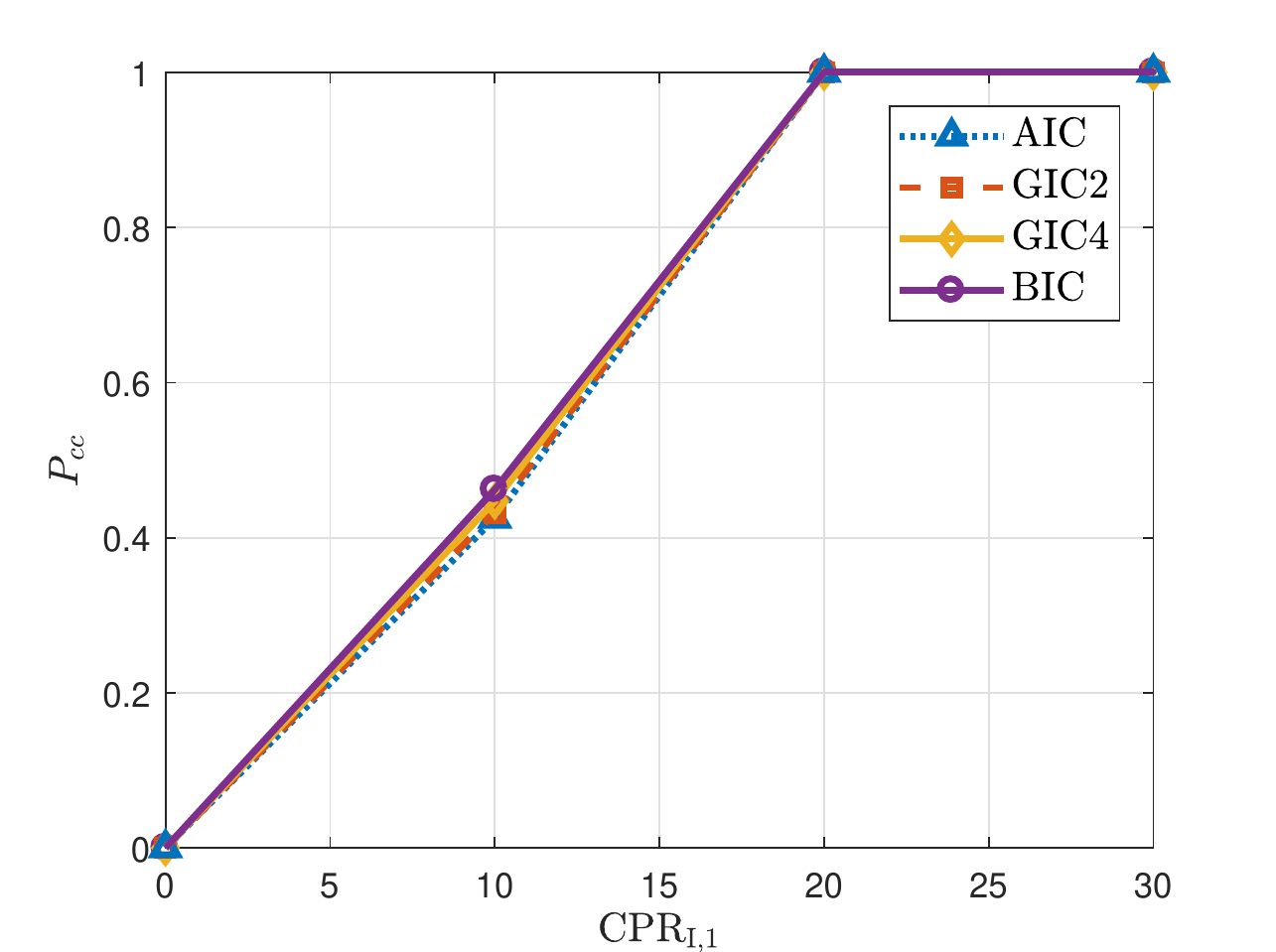}\\
  \caption{$P_{cc}$ versus $\text{CPR}_{\text{I},1}$ under $H_{\text{I},1}$, CNR=30 dB}\label{MOS-H1}
\end{figure}

In Fig. \ref{MOS-H1}, we proceed with the performance under $H_{\text{I},1}$ and plot the $P_{cc}$
versus the $\text{CPR}_{\text{I},1}$.
It is observed that when $\text{CPR}_{\text{I},1}$ is small, the scenario cannot be correctly classified since the clutter transition between
primary data and secondary data is minor. Moreover, the figure shows that
all four criteria exhibit similar classification performance and
achieve $P_{cc}\geq 0.75$ when $\text{CPR}_{\text{I},1}\geq15.7$ dB.
More specifically, the BIC slightly outperforms the other criteria and AIC has the worst performance.

\begin{figure}[t]
  \centering
  \subfigure[$K_{\text{I},1}=4$]{\includegraphics[width=4.2cm]{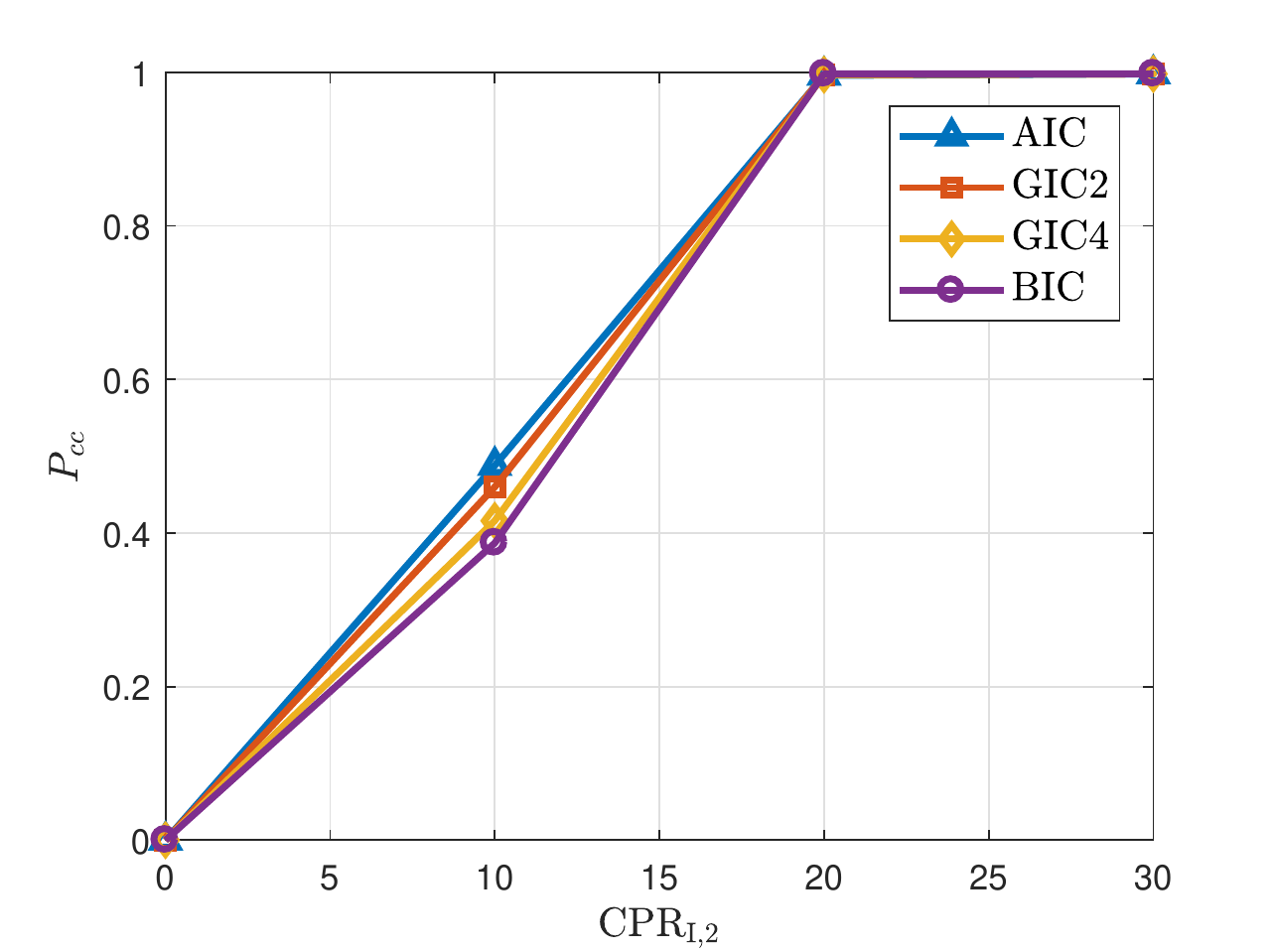}\label{MOS-H2-K4}}
  \subfigure[$K_{\text{I},1}=10$]{\includegraphics[width=4.2cm]{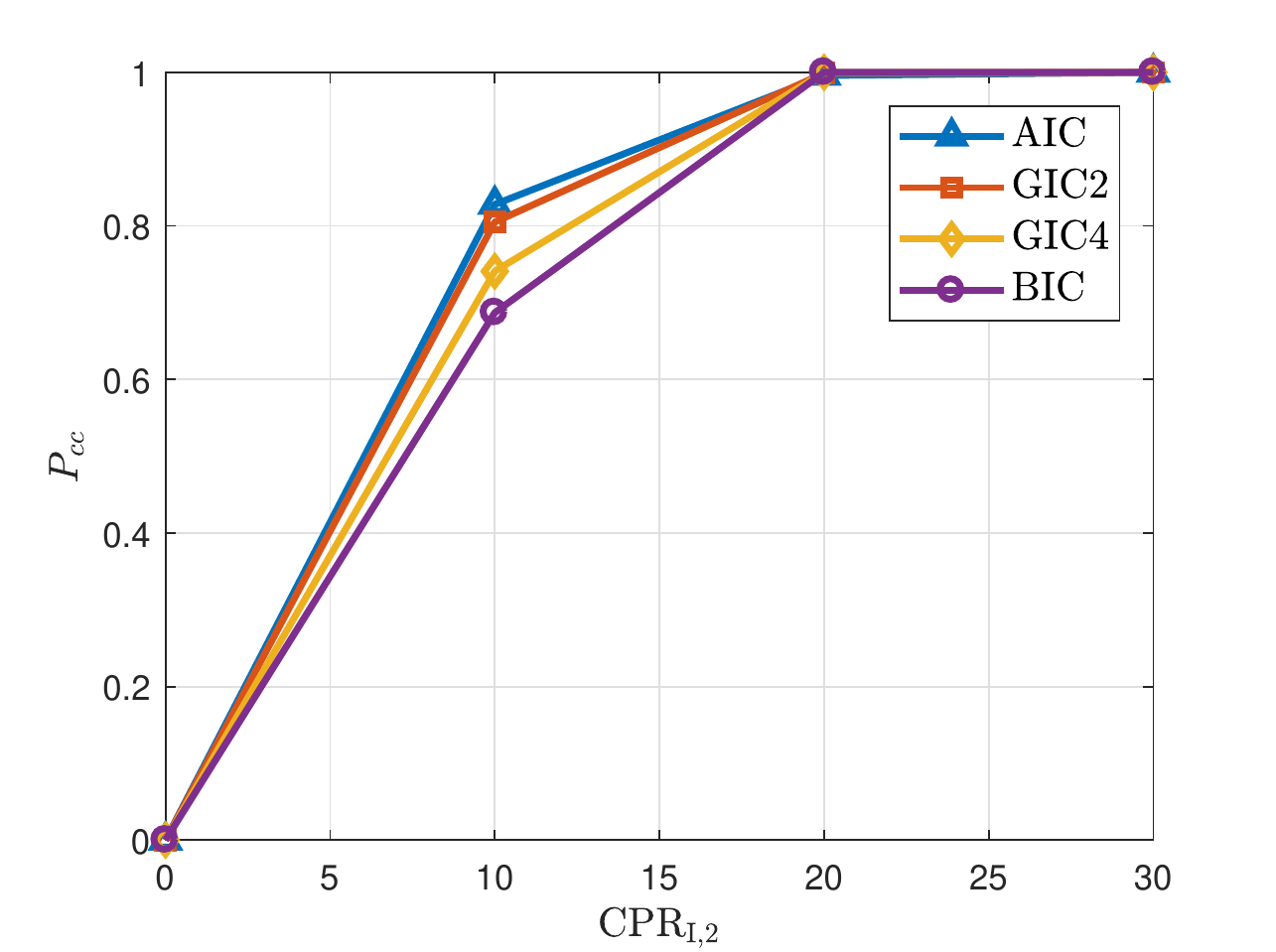}\label{MOS-H2-K10}}
  \subfigure[$K_{\text{I},1}=25$]{\includegraphics[width=4.2cm]{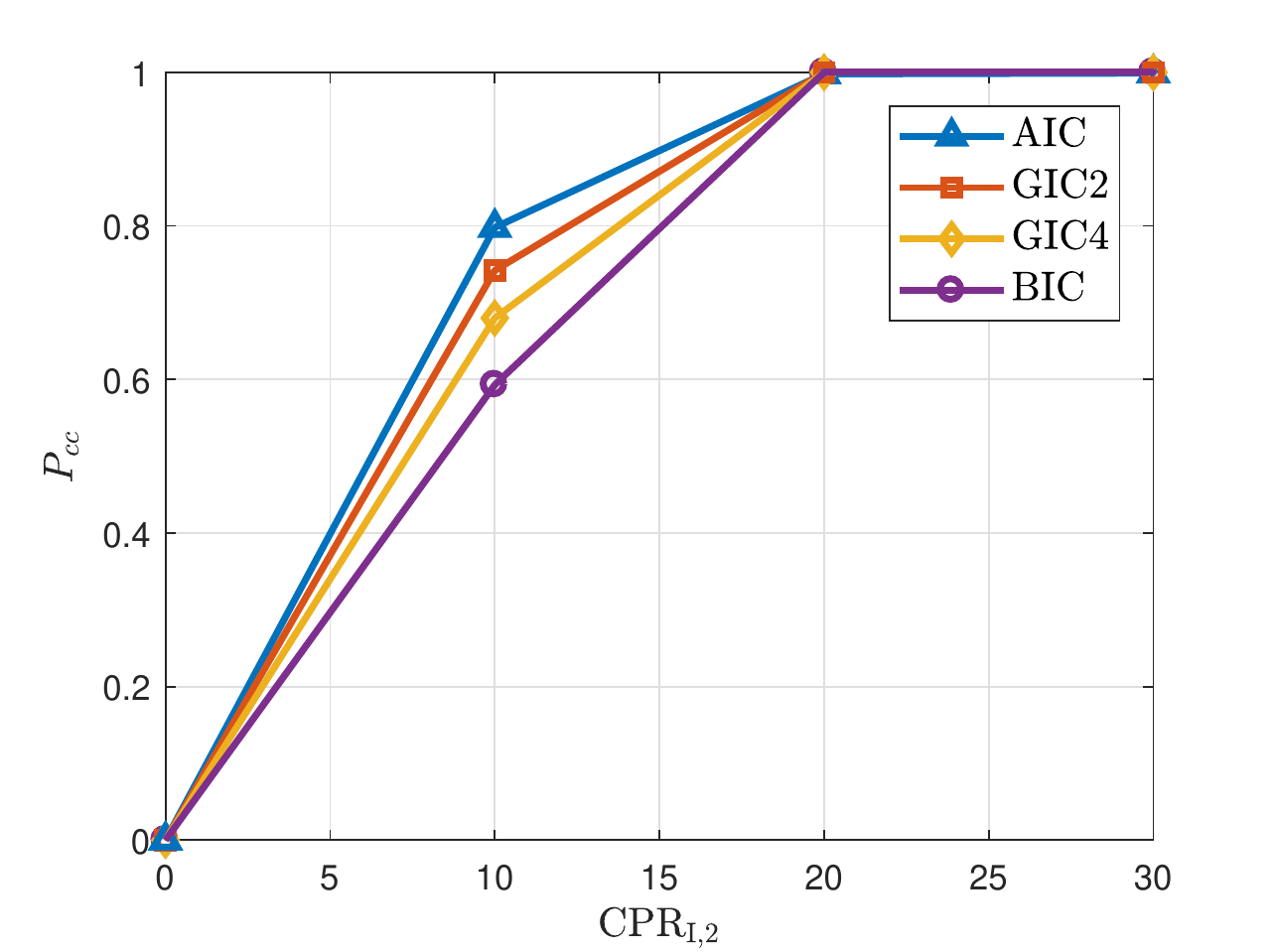}\label{MOS-H2-K25}}
  \caption{$P_{cc}$ versus $\text{CPR}_{\text{I},2}$ under $H_{\text{I},2}$, CNR=30 dB}\label{MOS-H2}
\end{figure}

Fig. \ref{MOS-H2} deals with the case where $H_{\text{I},2}$ is true and shows the classification performance with $K_{\text{I},1}=4,10,25$. Under $H_{\text{I},2}$, for all of the illustrated values of $K_{\text{I},1}$, it turns
out that AIC exhibits the best performance with remarkable improvement with respect to GIC2, GIC4 and BIC, especially when $K_{\text{I},1}=25$.
Further inspections of the figures highlight that
the detection capability of the clutter edge at $K_{\text{I},1}=10$ is significantly improved compared with $K_{\text{I},1}=4$, and $K_{\text{I},1}=25$. For example, when $K_{\text{I},1}=10$, the $P_{cc}$ of AIC is greater than 0.75 when $\text{CPR}_{\text{I},2}\geq9.1$ dB, and this value increases to 15.1 dB and 9.4 dB when $K_{\text{I},1}=4$ and $K_{\text{I},1}=25$, respectively. In other words, the classification architectures are much more sensitive with
the power transition located far from the beginning or the end of the reference window. Such result is confirmed by the fact that the clutter edge at
$K_{\text{I},1}=4$ exhibits the worst classification performance at least for the considered parameter values.

\begin{figure}[t]
  \centering
  \subfigure[$K_{\text{I},2}=4$, $K_{\text{I},3}=20$]{\includegraphics[width=4.2cm]{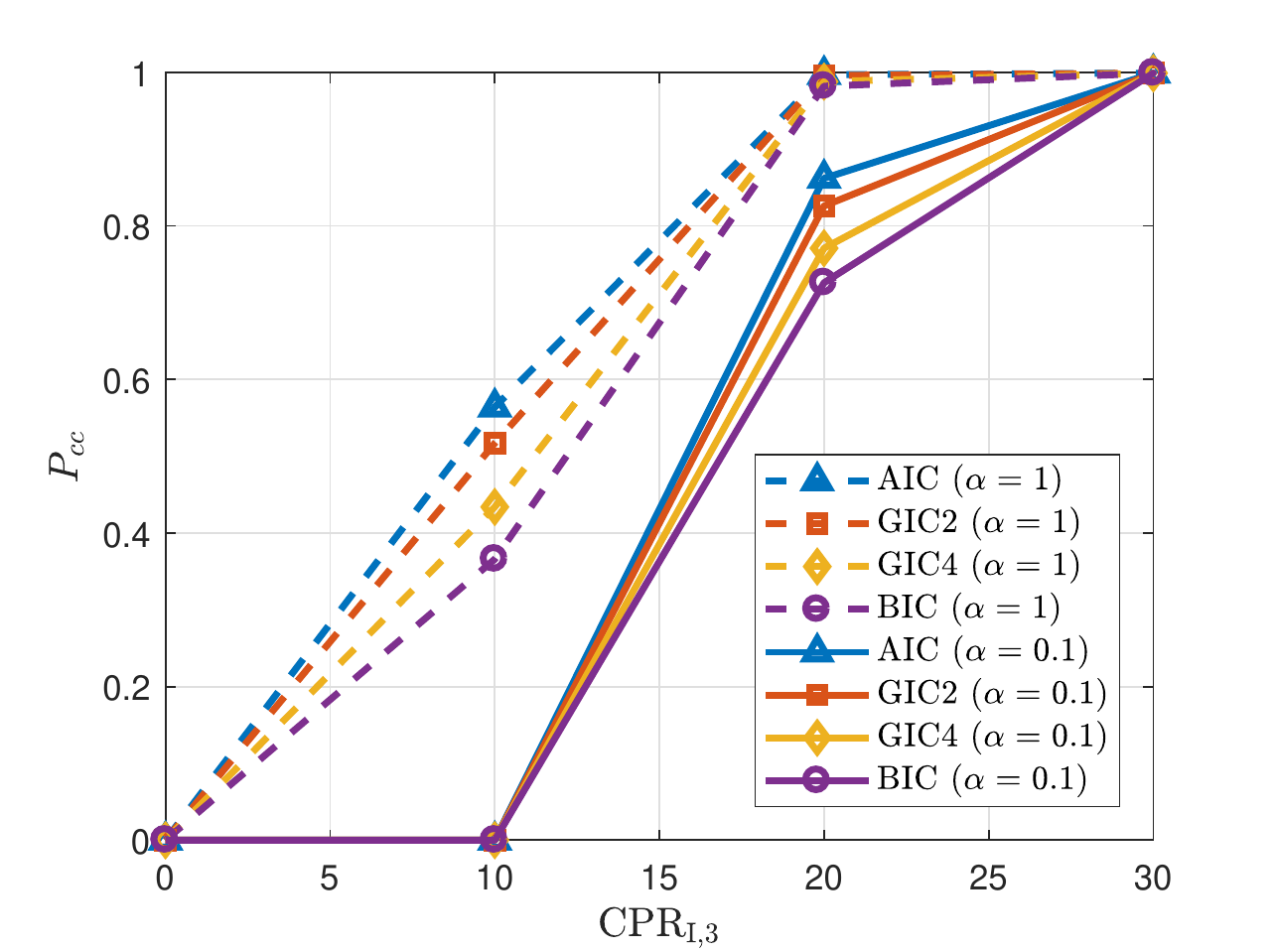}\label{MOS-H3-4-10}}
  \subfigure[$K_{\text{I},2}=10$, $K_{\text{I},3}=22$]{\includegraphics[width=4.2cm]{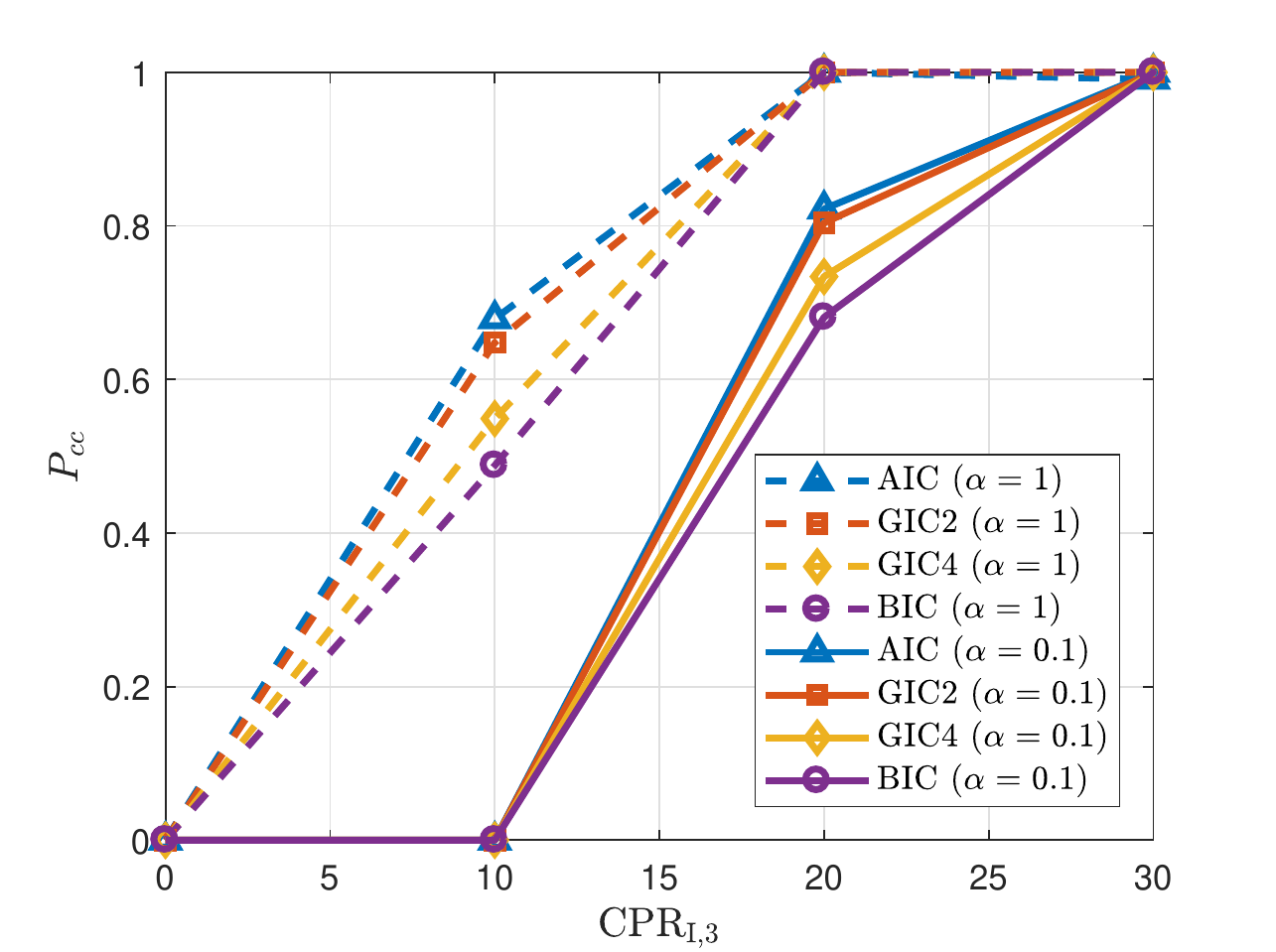}\label{MOS-H3-10-20}}
  \subfigure[$K_{\text{I},2}=12$, $K_{\text{I},3}=25$]{\includegraphics[width=4.2cm]{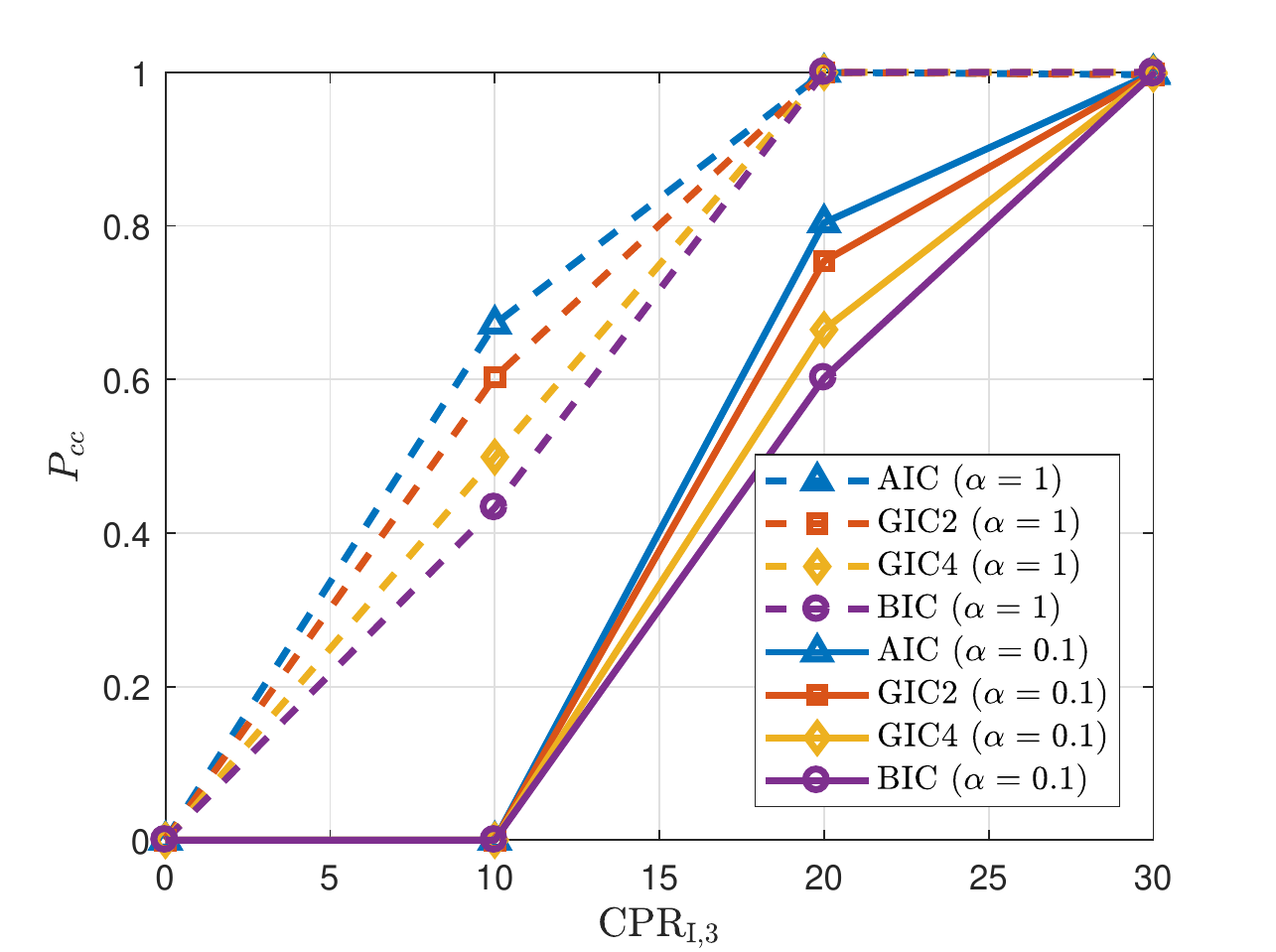}\label{MOS-H3-12-25}}
  \caption{$P_{cc}$ versus $\text{CPR}_{\text{I},3}$ under $H_{\text{I},3}$, CNR=30 dB}\label{MOS-H3}
\end{figure}

In Fig. \ref{MOS-H3}, we plot the $P_{cc}$ under $H_{\text{I},3}$ against $\text{CPR}_{\text{I},3}$ for different clutter edge positions and two different values of $\alpha$, namely $\alpha=1$ and $\alpha=0.1$. It can be seen that under all of the situations of Fig. \ref{MOS-H3}, the AIC ensures the best performance and GIC rules are better than BIC, which confirms what observed in Fig. \ref{MOS-H2}.
The figures also show that the classification architectures identify $H_{\text{I},3}$ when $\alpha=1$ much more effectively than when $\alpha=0.1$. For example, when $K_{\text{I},2}=4$, $K_{\text{I},3}=20$, $P_{cc}$ of AIC for $\alpha=1$ is 0.57 at $\text{CPR}_{\text{I},3}=10$ dB and can achieve $P_{cc}=1$ when $\text{CPR}_{\text{I},3}$ increases to 20 dB. On the other hand, for $\alpha=0.1$, the $P_{cc}$ is zero at $\text{CPR}_{\text{I},3}=10$ dB and less than 0.9 even when it increases to $\text{CPR}_{\text{I},3}=20$ dB. This is not a surprising result since the former situation contains greater clutter power.
Moreover, we can see that for $\alpha=1$, the classification performance increases when $K_{\text{I},3}-K_{\text{I},2}$ decreases, whereas for $\alpha=0.1$,
since the clutter region characterized by $\Delta_3$ maintains relatively enough power, the classification performance is more impacted by the significance of the region
characterized by $\Delta_4$, thus when $\alpha=0.1$, the classification performance increases when $K_{\text{I},3}$ decreases.

\begin{figure}[t]
  \centering
  \subfigure[$\text{RMS}_{\text{I},1}$ versus $\text{CPR}_{\text{I},2}$]{\includegraphics[width=4.2cm]{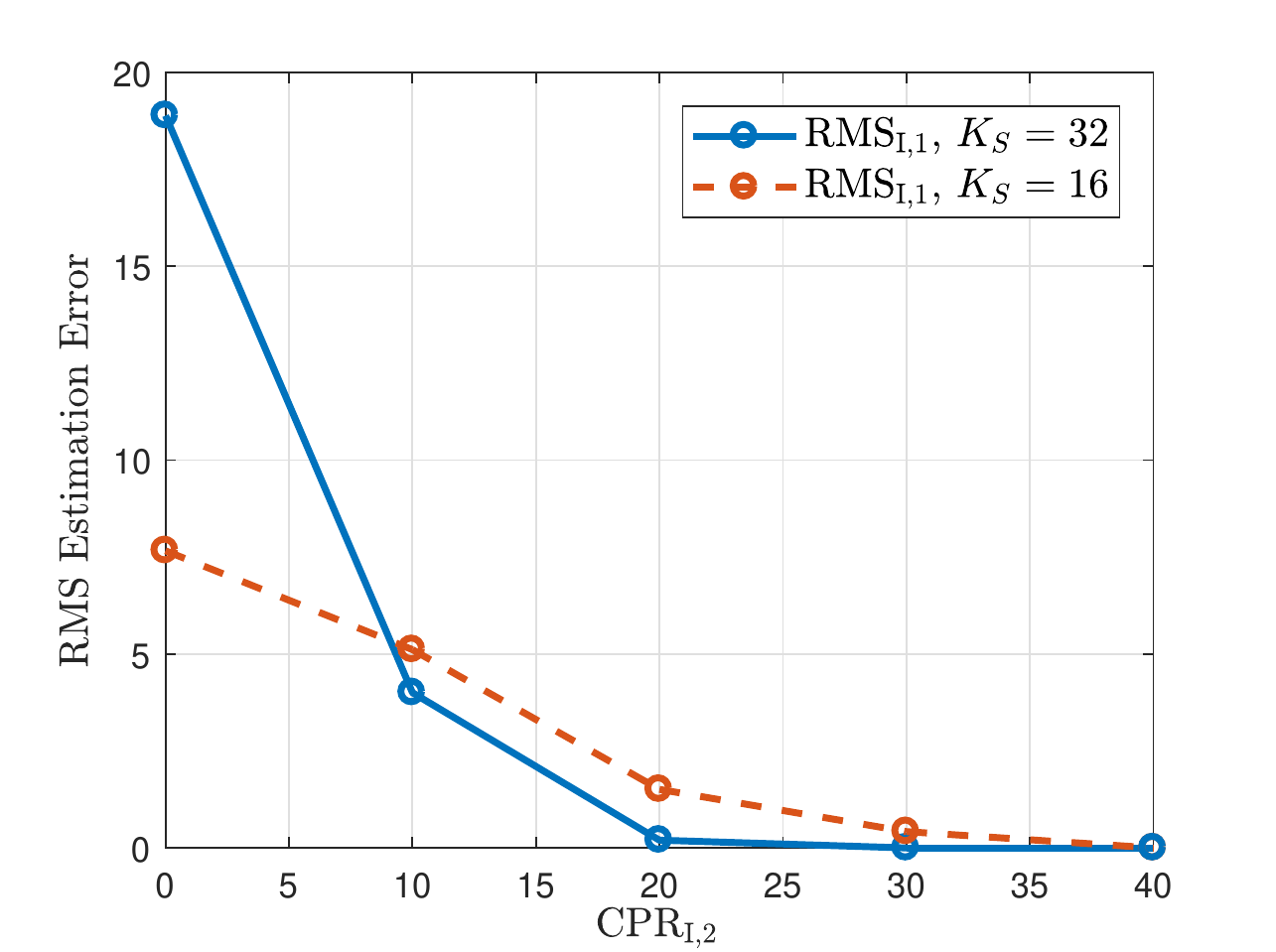}\label{MOS-RMS1}}
  \subfigure[$\text{RMS}_{\text{I},2}$ and $\text{RMS}_{\text{I},3}$ versus $\text{CPR}_{\text{I},3}$, $\alpha=1$]{\includegraphics[width=4.2cm]{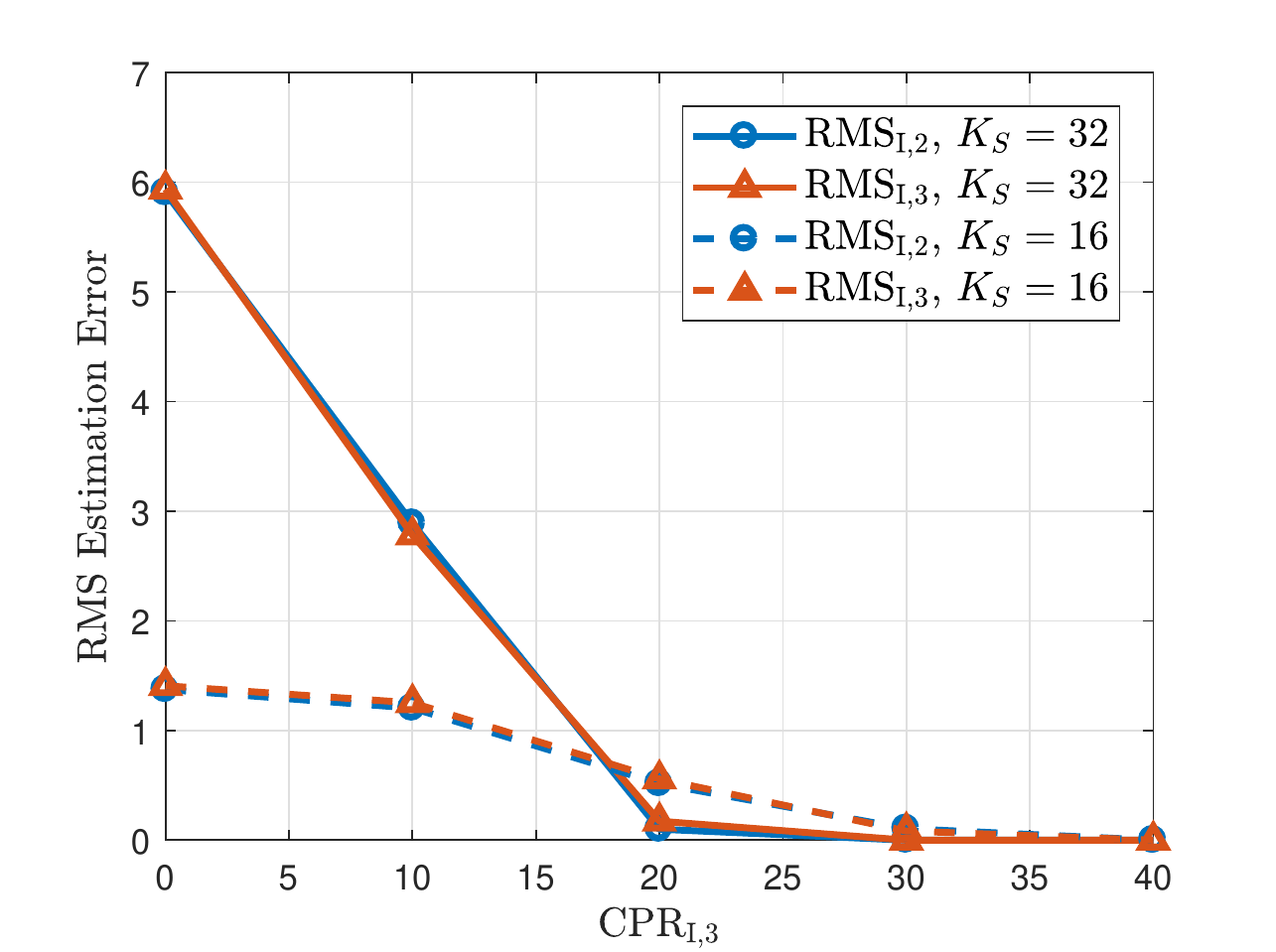}\label{MOS-RMS2}}
  \caption{RMS estimation errors, CNR=30 dB}\label{MOS-RMS}
\end{figure}

The final analysis in this subsection focuses on the estimation capabilities of the proposed approach for what concerns the position of the clutter edge. It is clear that this estimate depends on the extent of power transition between the consecutive homogeneous regions and the size of the reference window. In addition, the edge indices $K_1$, $K_2$ and $K_3$ are generated as discrete uniform random variables taking on values in $K_{\text{I},1}\in\left\{r,\ldots,K_S-r\right\}$, $K_{\text{I},2}\in\left\{r,\ldots,\frac{K_S}{2}\right\}$, and $K_{\text{I},3}\in\left\{\frac{K_S}{2}+1,\ldots,K_S-r\right\}$, respectively. In Fig. \ref{MOS-RMS1}, we plot the RMS estimation errors assuming $K_S=16,32$, where the estimates of the RMS errors for $K_{\text{I},i}, i=1,\ldots,3$ are defined as
\begin{equation}
\text{RMS}_{\text{I},i}=\sqrt{\frac{1}{L}\sum\limits_{l=1}^{L}|\widehat{K}^{(l)}_{\text{I},i}-K_{\text{I},i}|^2},
\end{equation}
with $\widehat{K}^{(l)}_{\text{I},i}$ the estimate of $K_{\text{I},i}$ at the $l$th trial.
 The figures highlight that
for low CPR values, the error is greater for the configurations associated with larger $K_S$, but the curves related to larger $K_S$ drop with faster rates than the curves obtained for smaller $K_S$.
Observe that for CPR values greater than 20 dB, the proposed architecture can return RMS$_{\text{I},1}$ values less than 2 and RMS$_{\text{I},l}$, $l=2,3$, values less than 1.
Under $H_{\text{I},3}$, the case such that $\alpha\neq1$ returns similar results except for the larger RMS estimation error values occurring in the presence of low-power clutter regions, and not reported here for brevity.

\subsection{Clutter Covariance Variation Model 2}
The classification and estimation performance of architecture \eqref{MOSscheme2} for the more general model of clutter covariance variation is assessed in this subsection. It is of primary importance to underline that when the covariance matrices of each clutter region are totally different, for example, when the angular sectors contributing to the clutter component are not the same (even though maintaining $r$), the proposed architecture returns excellent classification performance (such results are not reported here for brevity). Thus, we assume for simplicity that
$\bR_l = \sigma_{c,l}^2\sum_{\theta_i\in\Theta}\bv(\theta_i)\bv(\theta_i)^\dag$, $l=1,\ldots,6$, where $\sigma_{c,l}^2$ is the clutter power of the $l$th region and $\bv(\theta_i)$ is given by \eqref{v}. Furthermore, $\sigma_{c,l}^2$, $l=1,2,3,5$ are computed according to
\begin{equation}
\begin{cases}
\text{CPR}_{\text{II},l}=10\log_{10}(\sigma_{c,l}^2/\sigma_{c}^2),\ l=1,2,3,\\
\text{CPR}_{\text{II},4}=10\log_{10}(\sigma_{c,5}^2/\sigma_{c}^2),
\end{cases}
\end{equation}
where $\sigma_c^2$ is the clutter power of primary data, CPR$_{\text{II},i}$, $i=1,\ldots,4$ is the CPR value under $H_{\text{II},i}$, $i=1,\ldots,4$. As for $\sigma_{c,4}^2$ and $\sigma_{c,6}^2$, we define $\sigma_{c,4}^2=\beta\sigma_{c,3}^2$ with $\beta>0$ a scaling factor, and $\sigma_{c,6}^2=1.5\sigma_{c,5}^2$.

\begin{figure}[t]
  \centering
  \subfigure[$H_{\text{II},1}$]{\includegraphics[width=4.2cm]{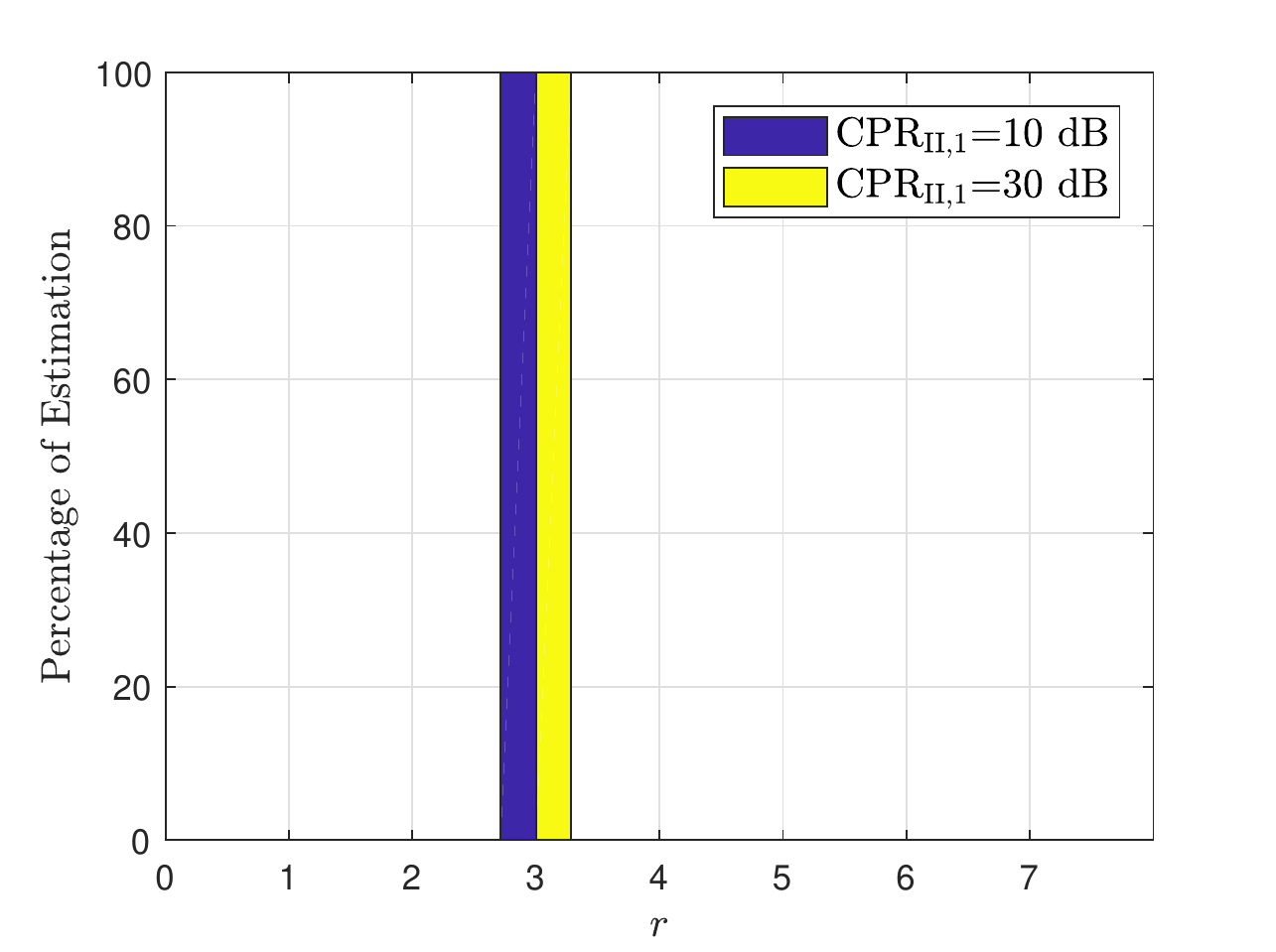}\label{BIC1II}}
  \subfigure[$H_{\text{II},2}$, $K_{\text{II},1}=4$]{\includegraphics[width=4.2cm]{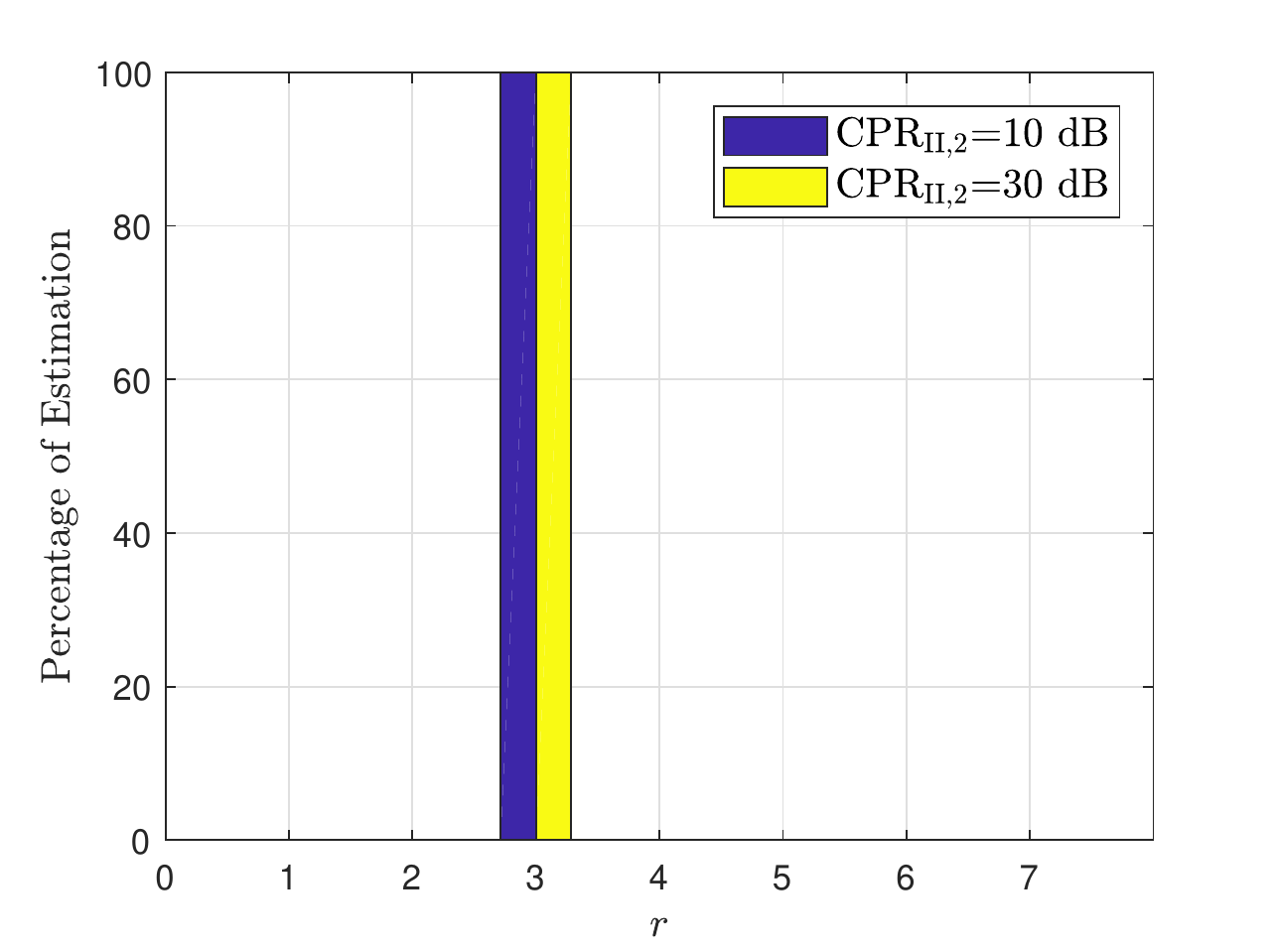}\label{BIC2II}}
  \subfigure[$H_{\text{II},3}$, $\beta=1$, $K_{\text{II},2}=12$, $K_{\text{II},3}=25$]{\includegraphics[width=4.2cm]{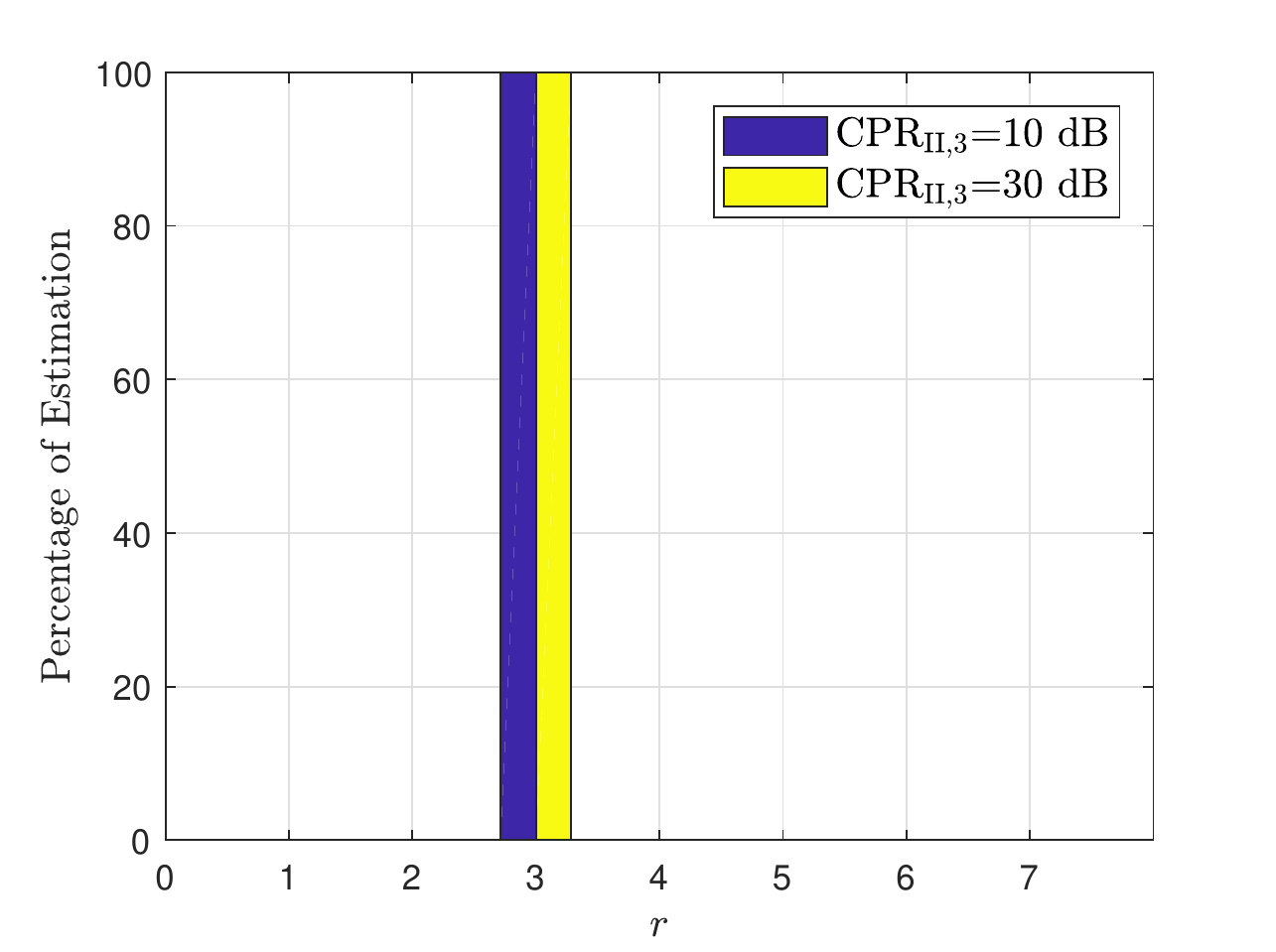}\label{BIC3II}}
  \subfigure[$H_{\text{II},4}$, $K_{\text{II},4}=10$]{\includegraphics[width=4.2cm]{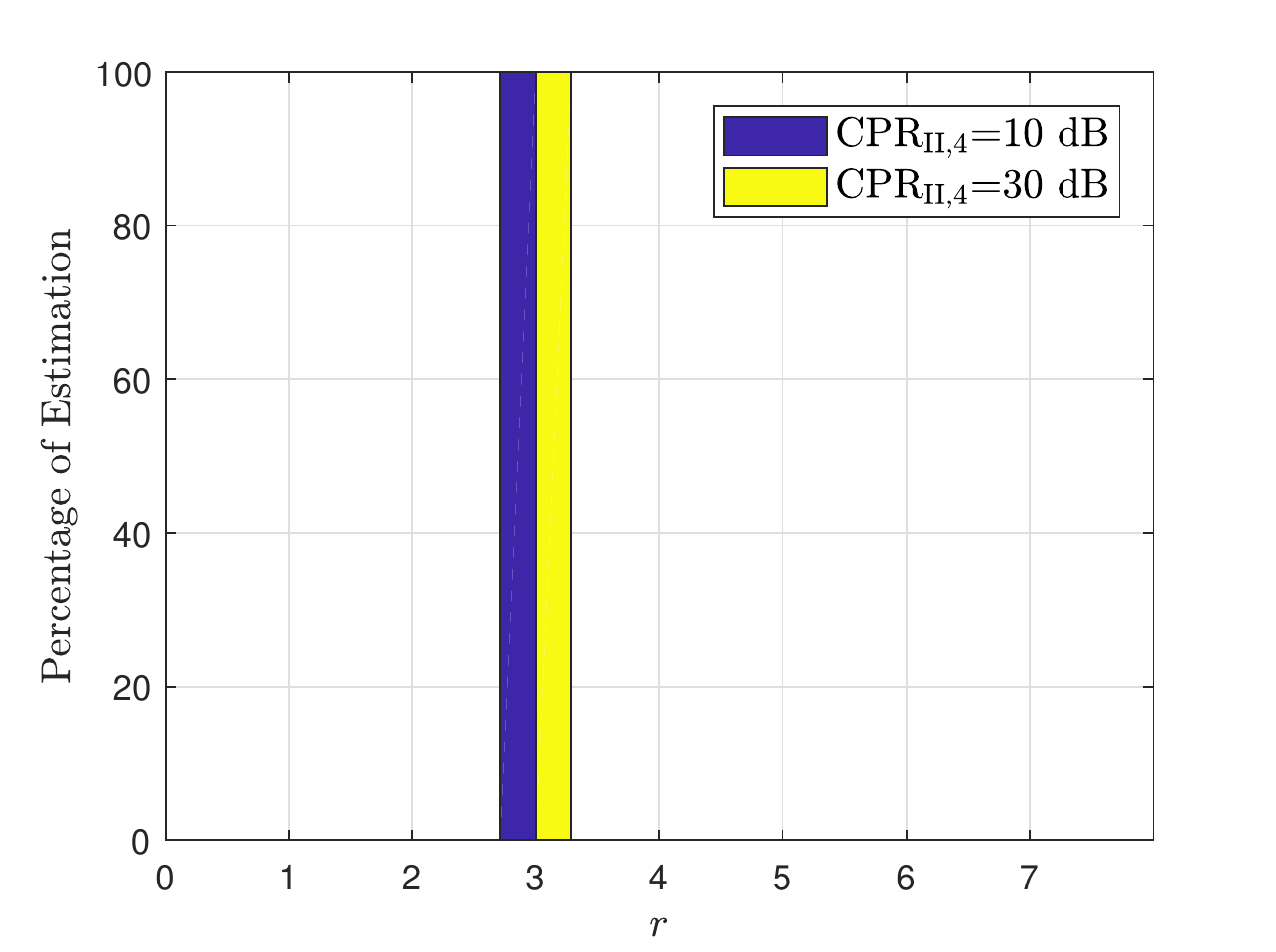}\label{BIC4II}}
  \caption{Rank estimation performance under $H_{\text{II},i}$, CNR=30 dB}\label{restII}
\end{figure}

In Fig. \ref{restII}, we show the rank estimation performance of BIC rule under $H_{\text{II},i}$, $i=1,\ldots,4$.
As for $H_{\text{II},0}$, it is shown in Fig. \ref{BIC0}.
Since correct rank estimation can be ensured as shown in the figures
(at least for the illustrated configurations), in what follows we assume that $r$ is known.

Let us start with assuming that $H_{\text{II},0}$ is in force. Table \ref{table:HII0} contains the specific classification results of the AIC rule for this hypothesis. Inspection of the table highlights that the proposed architecture correctly decides for $H_{\text{II},0}$ with a percentage greater than 0.98.
\begin{table}[t]
\caption{Classification results under $H_{\text{II},0}$, CNR=30 dB}
\label{table:HII0}
\centering
\begin{tabular}{c|ccccc}
  \toprule
   &$H_{\text{II},0}$ & $H_{\text{II},1}$ & $H_{\text{II},2}$ & $H_{\text{II},3}$ & $H_{\text{II},4}$  \\
  \midrule
   AIC & 983     & 1   & 15 & 1& 0\\
   GIC2 & 1000     & 0   & 0 & 0& 0\\
   GIC4 & 1000     & 0   & 0 & 0& 0\\
   BIC & 1000     & 0   & 0 & 0& 0\\
  \bottomrule
\end{tabular}
\end{table}

\begin{figure}[t]
  \centering
  \includegraphics[width=7.7cm]{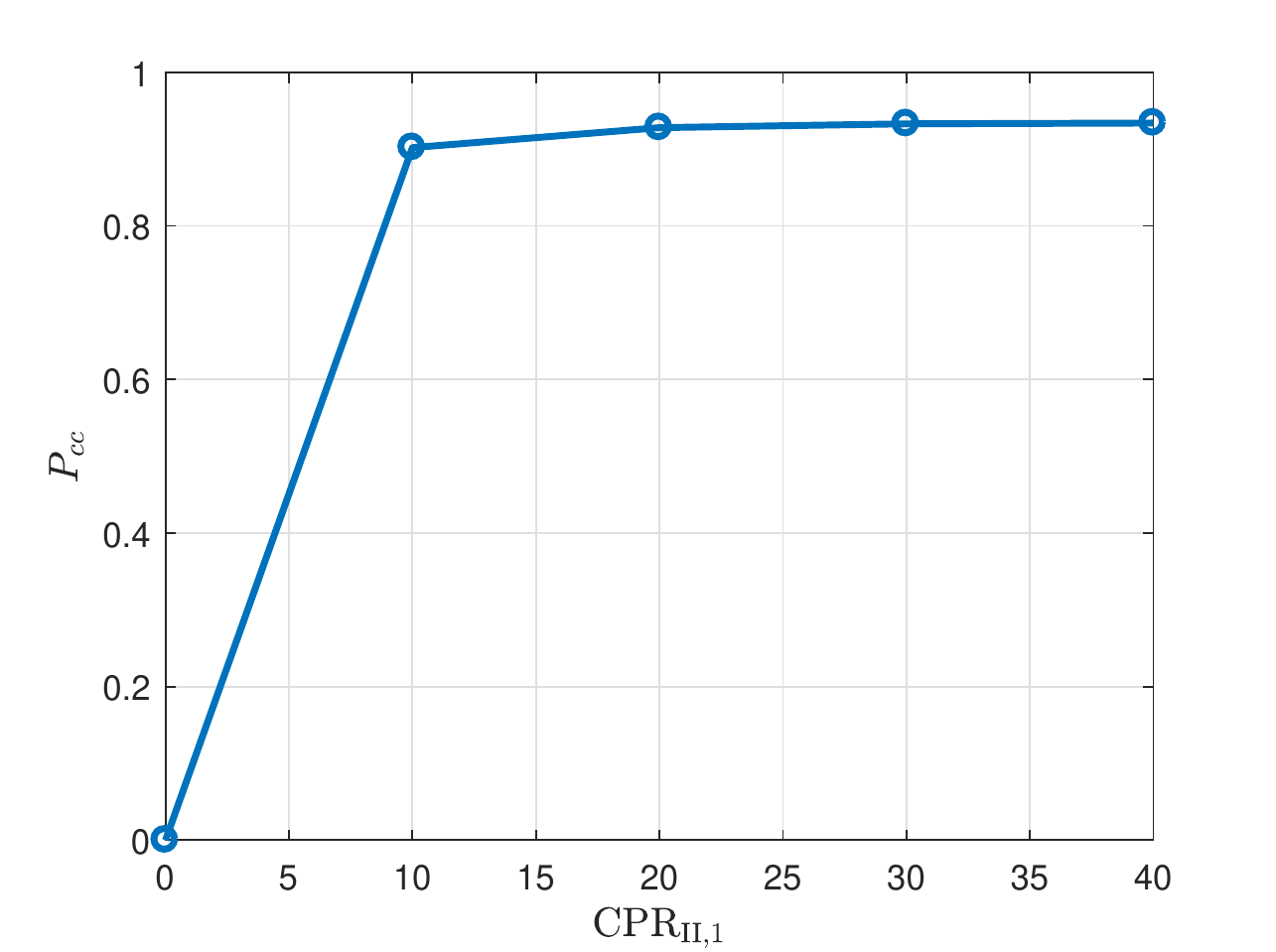}\\
  \caption{$P_{cc}$ versus $\text{CPR}_{\text{II},1}$ under $H_{\text{II},1}$, CNR=30 dB}\label{MOS-HII1}
\end{figure}

\begin{figure}[t]
  \centering
  \includegraphics[width=7.7cm]{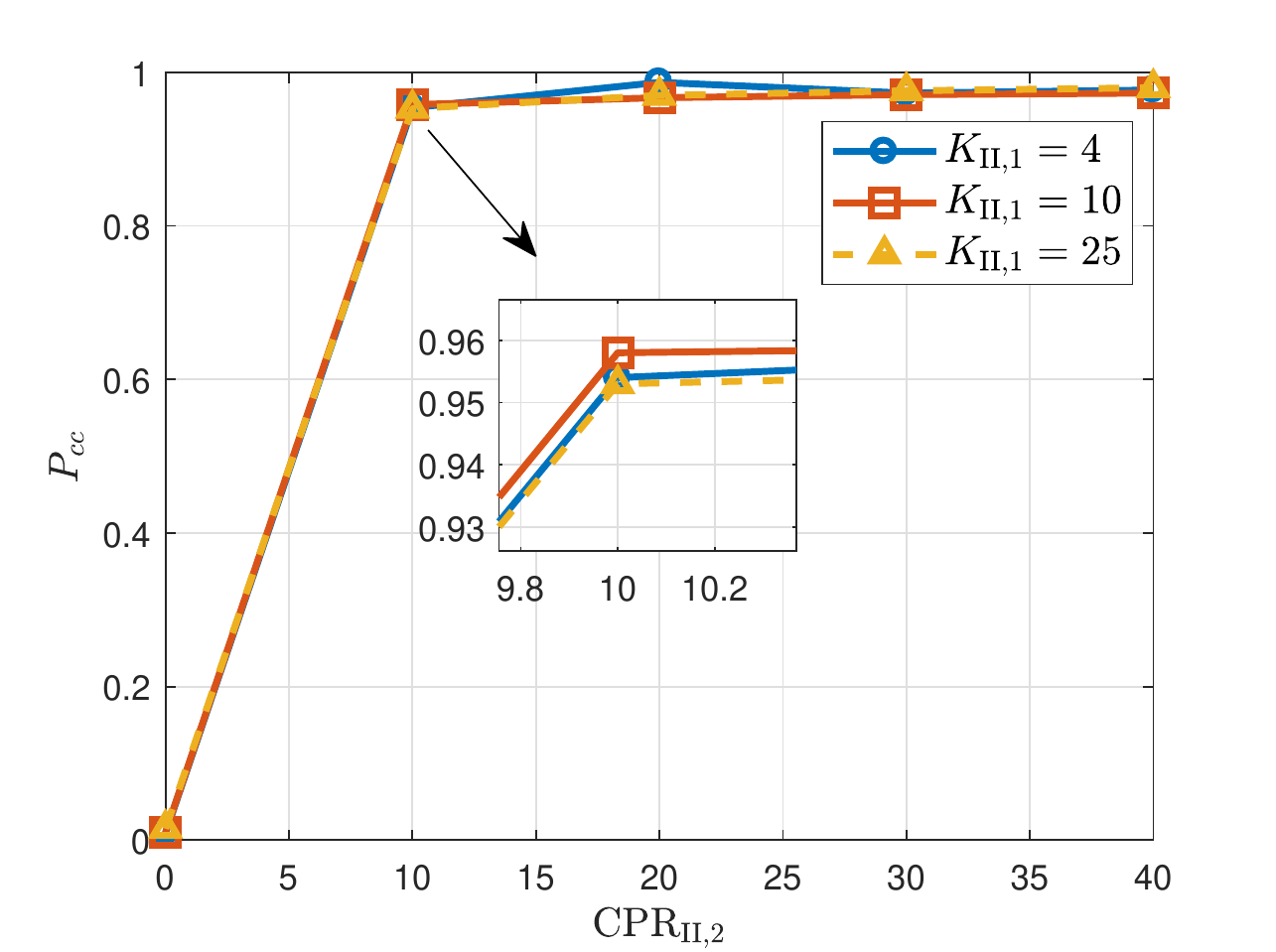}\\
  \caption{$P_{cc}$ versus $\text{CPR}_{\text{II},2}$ under $H_{\text{II},2}$, CNR=30 dB}\label{MOS-HII2}
\end{figure}

For $H_{\text{II},i}$, $i=1,\ldots,4$, results not reported here for brevity
indicate that in this case, the AIC criterion overwhelmingly
outperforms GIC and BIC. For the specific case, this behavior can be explained by the fact
that AIC has the lowest penalty term which decreases the inclination to underfit.
For this reason, in what follows, only the
performance of AIC are shown. In Fig. \ref{MOS-HII1}, we plot the $P_{cc}$ when $H_{\text{II},1}$ is true against
$\text{CPR}_{\text{II},1}$. It can be seen that $H_{\text{II},1}$ is correctly identified by the proposed classification architecture with $P_{cc}$ greater than 0.75
when $\text{CPR}_{\text{II},1}\geq 8.3$ dB, and this correct classification probability increases to 0.9 when $\text{CPR}_{\text{II},1}\geq10$ dB.
Fig. \ref{MOS-HII2} deals with the case that $H_{\text{II},2}$ is in force and shows the resulting $P_{cc}$ curves versus $\text{CPR}_{\text{II},2}$
under different values of $K_{\text{II},1}$. As can be observed, in this case, the identification ability of one clutter edge of the proposed architecture is somehow
robust with respect to the position of clutter transition and can ensure $P_{cc}>0.75$ when $\text{CPR}_{\text{II},2}\geq7.85$ dB for all three illustrated configurations.

In Fig. \ref{MOS-HII3}, we plot the classification performance against $\text{CPR}_{\text{II},3}$ under $H_{\text{II},3}$ with different positions of clutter edges for $\beta=1$ and $\beta=0.5$ in Fig. \ref{MOS-HII3-beta1} and Fig. \ref{MOS-HII3-beta05}, respectively. It turns out that, as observed under $H_{\text{II},2}$, the classification architecture exhibits similar performance for the three illustrated clutter edge positions. More specifically, $H_{\text{II},3}$ can be correctly identified with $P_{cc}>0.75$ when $\text{CPR}_{\text{II},3}$ greater than 8.2 dB and 16.7 dB for $\beta=1$ and $\beta=0.5$, respectively. Further inspections of the two figures highlight that $P_{cc}$ for $\beta=0.5$ significantly decreases compared with that for $\beta=1$ due to the lower power of the clutter region characterized by smaller $\sigma_{c,4}^2$.
Fig. \ref{MOS-HII4} shows that the proposed classification architecture can correctly classify $H_{\text{II},4}$ with $P_{cc}>0.75$ at $\text{CPR}_{\text{II},4}$ greater than 18.4 dB.

\begin{figure}[t]
  \centering
  \subfigure[$H_{\text{II},3}$, $\text{CPR}_{\text{II},3}=10$ dB, $\beta=0.5$]{\includegraphics[width=4.2cm]{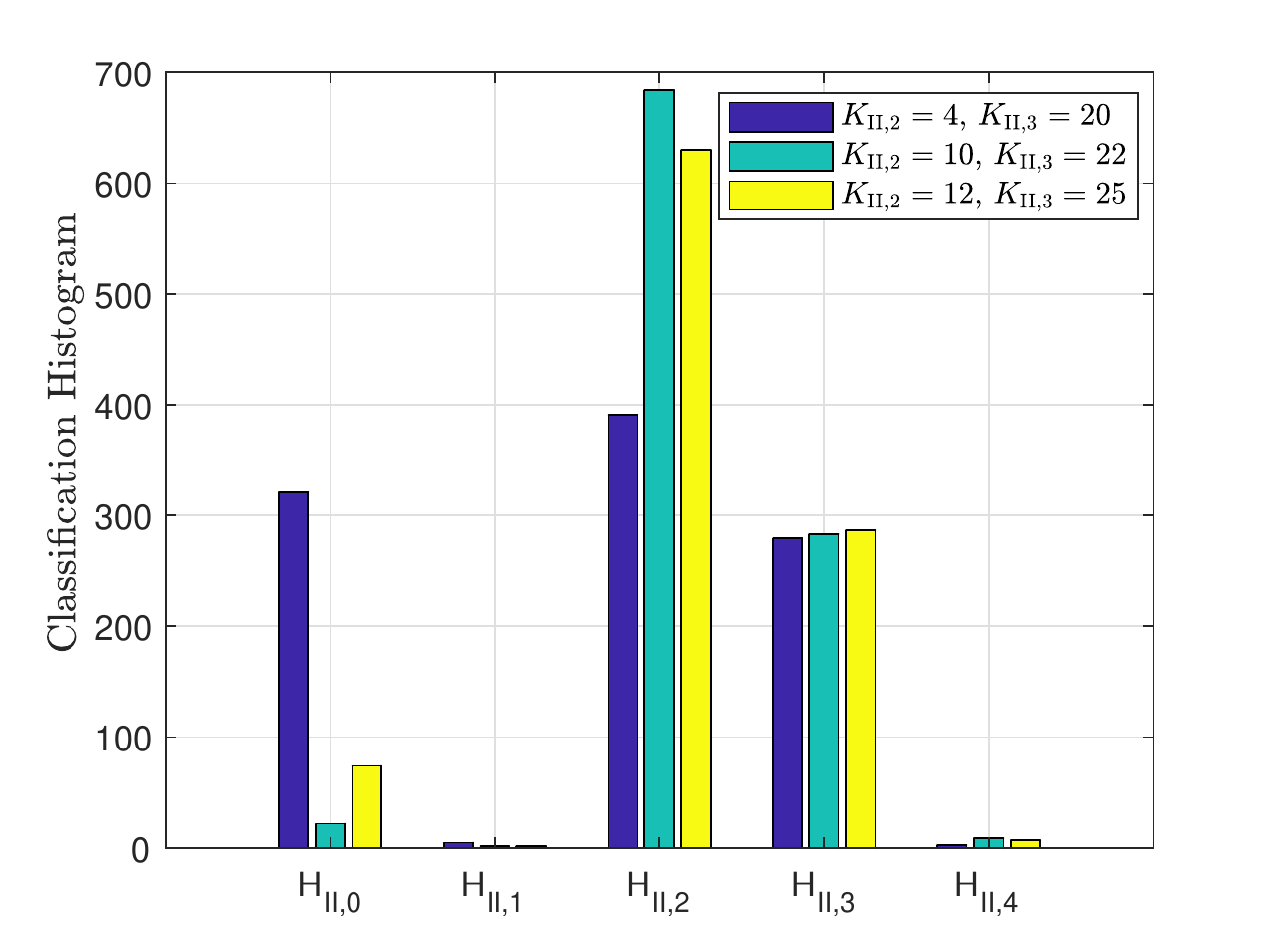}\label{MOS-HII3-stem}}
  \subfigure[$H_{\text{II},4}$, $\text{CPR}_{\text{II},4}=10$ dB]{\includegraphics[width=4.2cm]{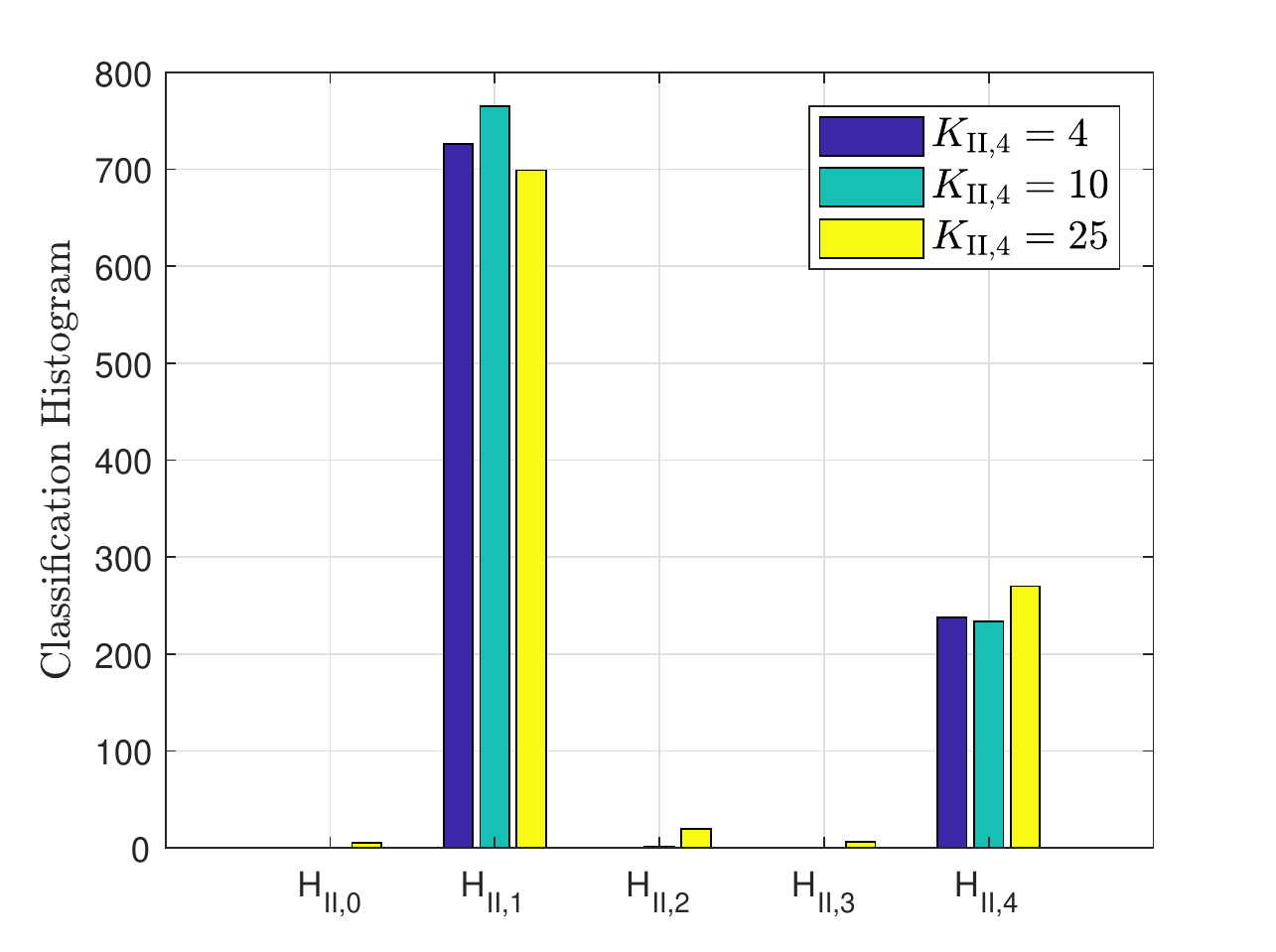}\label{MOS-HII4-stem}}
  \caption{Classification results under $H_{\text{II},3}$ and $H_{\text{II},4}$, CNR=30 dB}\label{MOS-HII-stem}
\end{figure}

Furthermore, to depict a more complete picture of the classification capabilities, Fig. \ref{MOS-HII-stem} contains the specific results under $H_{\text{II},3}$ with $\text{CPR}_{\text{II},3}=10$ dB, $\beta=0.5$, and $H_{\text{II},4}$ with $\text{CPR}_{\text{II},4}=10$ dB, respectively. The figures show that for relatively low $\text{CPR}_{\text{II},3}$, $H_{\text{II},3}$ is mainly misidentified as $H_{\text{II},2}$ when $\beta\neq1$, whereas for the cases that $\text{CPR}_{\text{II},4}$ is not large, $H_{\text{II},4}$ is inclined to be misclassified as $H_{\text{II},1}$ since the extent of clutter transition within the secondary data is less significant when $\sigma_{c,6}^2=1.5\sigma_{c,5}^2$.

\begin{figure}[t]
  \centering
  \subfigure[$\text{RMS}_{\text{II},1}$ versus $\text{CPR}_{\text{II},2}$]{\includegraphics[width=4.2cm]{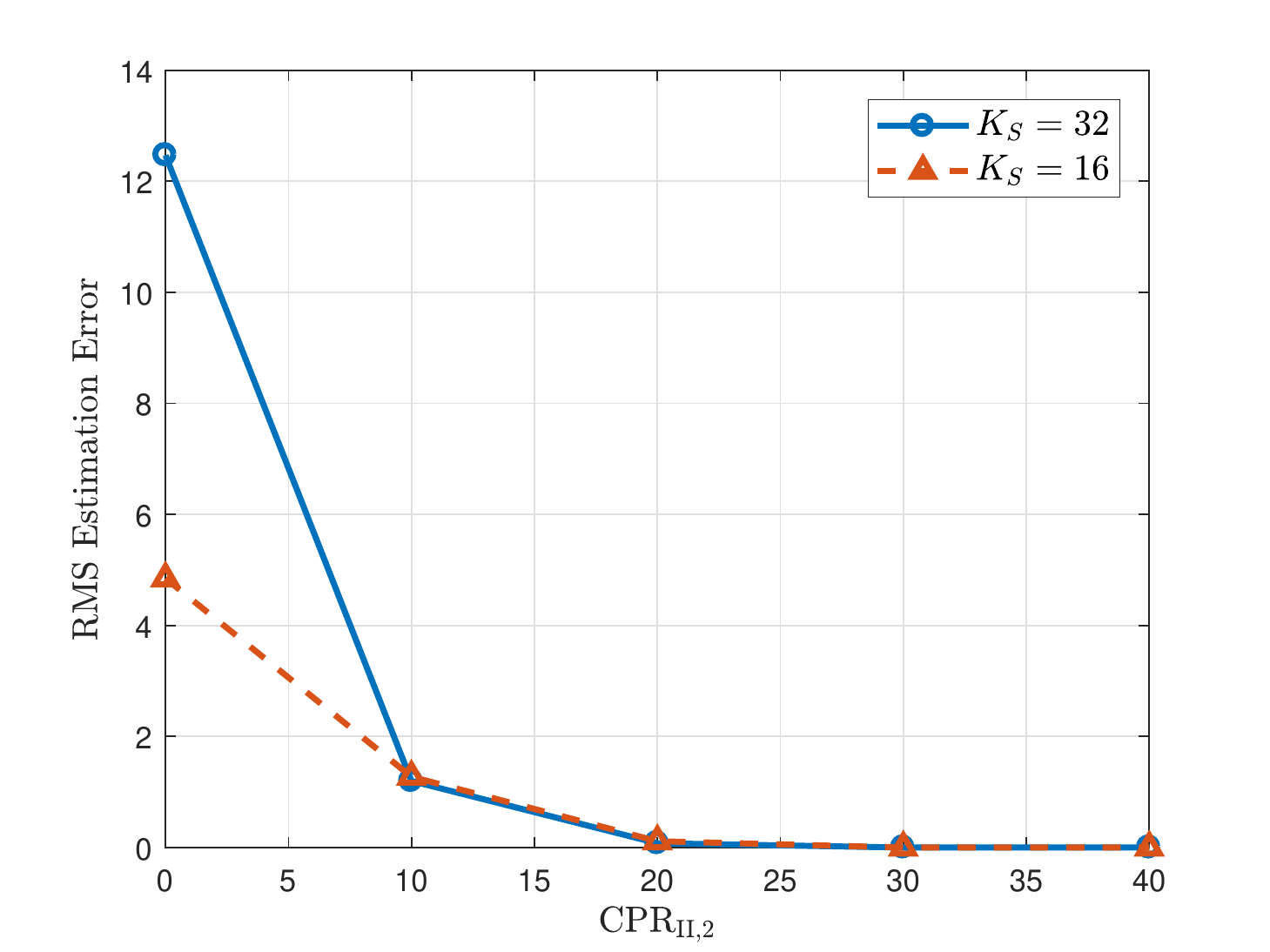}\label{MOS-RMSII1}}
  \subfigure[$\text{RMS}_{\text{II},2}$ and $\text{RMS}_{\text{II},3}$ versus $\text{CPR}_{\text{II},3}$, $\beta=1$]{\includegraphics[width=4.2cm]{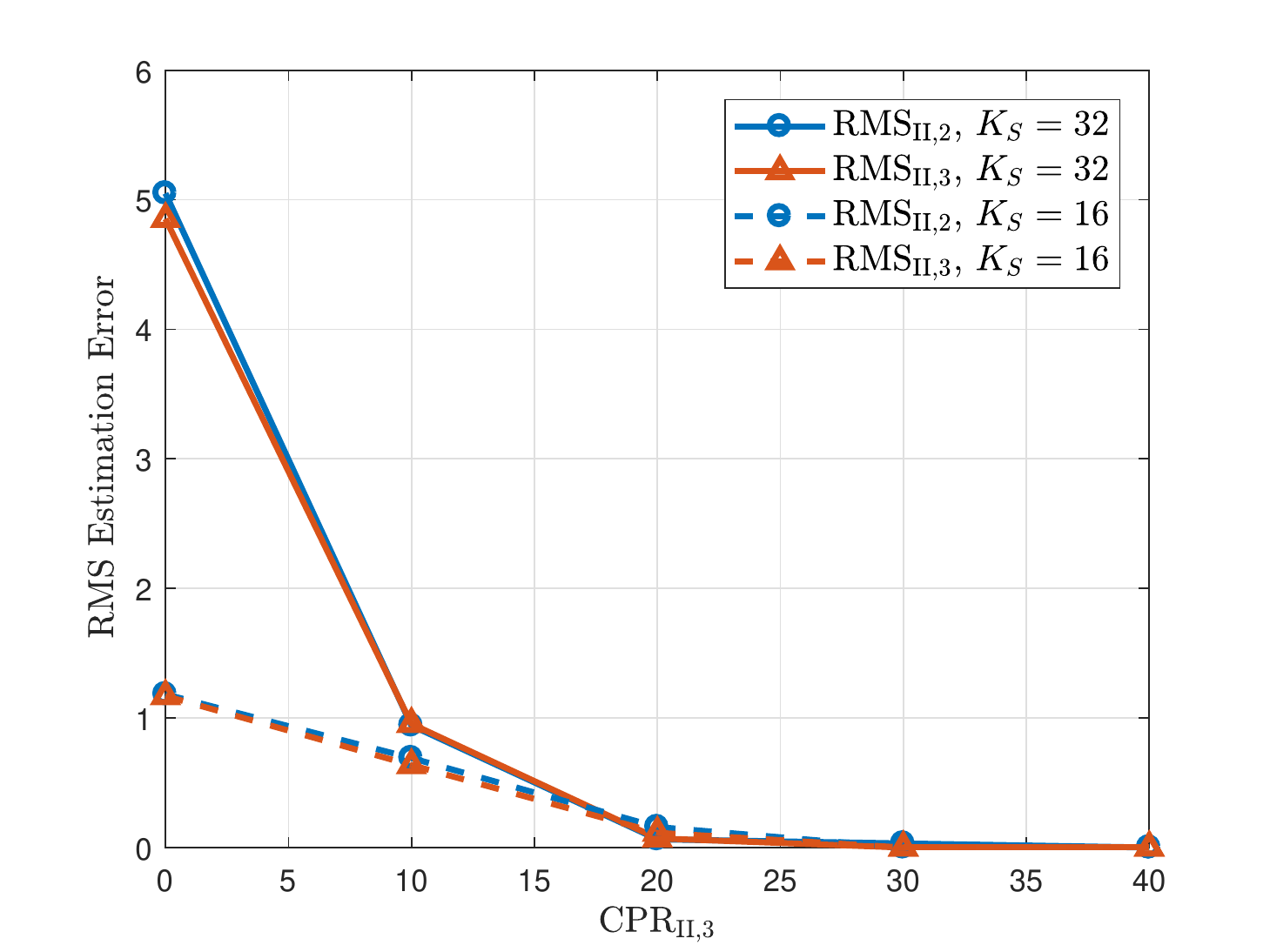}\label{MOS-RMSII2}}
  \subfigure[$\text{RMS}_{\text{II},4}$ versus $\text{CPR}_{\text{II},4}$]{\includegraphics[width=4.2cm]{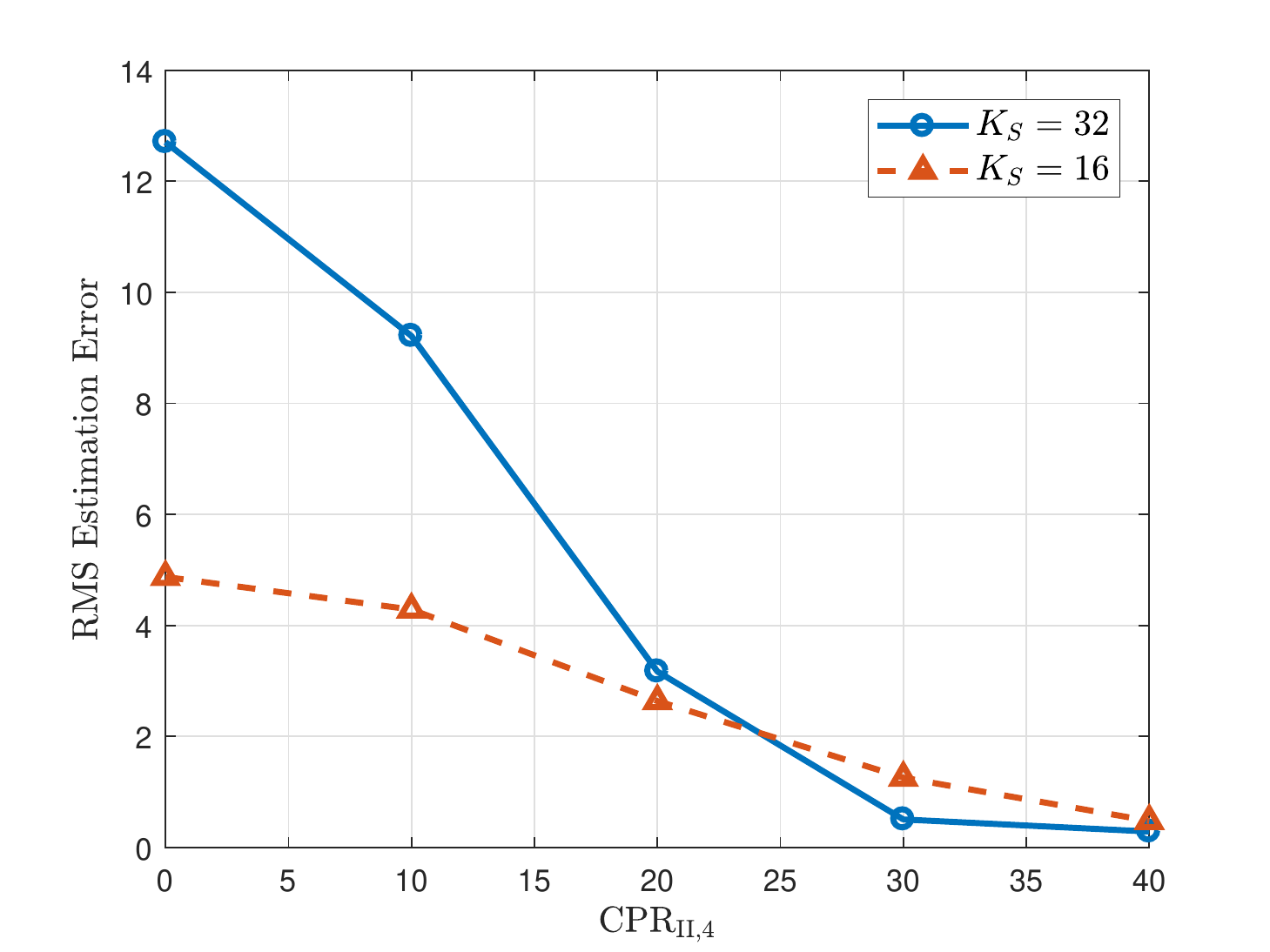}\label{MOS-RMSII3}}
  \caption{RMS estimation errors, CNR=30 dB}\label{MOS-RMSII}
\end{figure}

The final analysis concerns the estimation capability of the clutter edge positions in terms of the RMS estimation errors of $K_{\text{II},i}, i=1,\ldots,4$, defined as
\begin{equation}
\text{RMS}_{\text{II},i}=\sqrt{\frac{1}{L}\sum\limits_{l=1}^{L}|\widehat{K}^{(l)}_{\text{II},i}-K_{\text{II},i}|^2},
\end{equation}
with $\widehat{K}^{(l)}_{\text{II},i}$ the estimate of $K_{\text{II},i}$ at the $l$th trial.
The results are shown in Fig. \ref{MOS-RMSII} for different values of parameters. In Fig. \ref{MOS-RMSII}, the edge indices are generated as discrete uniform random variables taking on values in $K_{\text{II},1}\in\left\{r,\ldots,K_S-r\right\}$, $K_{\text{II},2}\in\left\{r,\ldots,\frac{K_S}{2}\right\}$, $K_{\text{II},3}\in\left\{\frac{K_S}{2}+1,\ldots,K_S-r\right\}$, and $K_{\text{II},4}\in\left\{r,\ldots,K_S-r\right\}$ respectively.
From the figures, what has been shown in Fig. \ref{MOS-RMS} is confirmed that at low CPR values, the estimation procedure associated smaller $K_S$ exhibits lower RMS estimation errors.

\section{Conclusion}
\label{section5}

In this paper, we have provided a solution to the problem of classifying the
radar operating scenario in terms of clutter statistical properties within the radar reference
window. The newly conceived architectures are also capable of effectively
estimating the clutter rank.
Unlike existing classifiers, at the design stage we have
considered two models for the clutter variation over the range and several configurations, including
the traditional homogeneous environment, a new formulation of the
partially-homogeneous environment, as well as the
cases where one or two clutter edges are present.
The classification problems have been formulated in terms of multiple hypothesis
testing problems and solved by means of suitable approximations of the MOS rules due to the
intractable mathematics associated with maximum likelihood estimation of some unknown parameters.
The performance analysis conducted on simulated data has confirmed
the effectiveness of the proposed architectures in deciding
for the correct operating scenario (hypothesis). More importantly,
the classification results provided by the proposed algorithms represent clutter maps
that allow the system to select the most adequate set of secondary data for the current scenario
and, hence, to improve the detection performance.

Future research tracks may include the extension of this work to the
case where clutter discretes are included in the region of interest or the
design of joint classification and detection architectures.
Another related issue may concern the design of classification architectures that
exploit possible knowledge about the background and/or systems leading to specific clutter covariance
structures and, at the same time, more reliable parameter estimation performances.

\appendices

\section{Estimation of \bU in \eqref{maxH1}}
\label{appendix0}
In this appendix, we describe the optimization criterion leading to the estimate of $\bm{U}$ under $H_{\text{I},1}$. To this end, notice that the optimization problem with respect to $\bm{U}$ is tantamount to
\begin{eqnarray}
&\max\limits_{\bm{U}}&
-\sum_{k=1}^{K_P}\bz_k^\dag\bm{U}(\sigma^2\bI+\bm{\Lambda})^{-1}\bm{U}^\dag\bz_k\nonumber\\
&&\qquad-\sum_{k=1}^{K_S}\bor_k^\dag\bm{U}(\sigma^2\bI+\bm{\Gamma}_1\bm{\Lambda}\bm{\Gamma}_1)^{-1}\bm{U}^\dag\bor_k\nonumber\\
\Rightarrow&\max\limits_{\bm{U}}&
-\sum_{k=1}^{K_P}||\bz_k||^2\bm{\alpha}_k^\dag\bm{U}(\sigma^2\bI+\bm{\Lambda})^{-1}\bm{U}^\dag\bm{\alpha}_k\nonumber\\
&&\qquad-\sum_{k=1}^{K_S}||\bor_k||^2\bm{\beta}_k^\dag\bm{U}(\sigma^2\bI+\bm{\Gamma}_1\bm{\Lambda}\bm{\Gamma}_1)^{-1}\bm{U}^\dag\bm{\beta}_k\nonumber\\
\Rightarrow&\max\limits_{\bm{U}}&
-\sum_{k=1}^{K_P+K_S}\bm{\xi}_k^\dag\bm{U}(\sigma^2\bI+\bQ_{1,k}\bm{\Lambda}\bQ_{1,k})^{-1}\bm{U}^\dag\bm{\xi}_k\cdot b_k\nonumber\\
\Rightarrow&\min\limits_{\bm{U}}&
\sum_{k=1}^{K_P+K_S}\bm{\xi}_k^\dag\bm{U}\bD_{1,k}\bm{U}^\dag\bm{\xi}_k,
\end{eqnarray}
where $\bm{\alpha}_k=\bz_k/||\bz_k||$, $k=1,\ldots,K_P$, $\bm{\beta}_k=\bor_k/||\bor_k||$, $k=1,\ldots,K_S$\footnote{Notice that $||\bm{\alpha}_k||=||\bm{\beta}_k||=1$.},
\begin{eqnarray}
\bm{\xi}_k=
\left\{
\begin{array}{lll}
\bm{\alpha}_k,&\ k=1,\ldots,K_P,\\
\bm{\beta}_{k-K_P},&\ k=K_P+1,\ldots,K_P+K_S,
\end{array}
\right.
\end{eqnarray}
and $\bD_{1,k}=b_k(\sigma^2\bI+\bQ_{1,k}\bm{\Lambda}\bQ_{1,k})^{-1}$ with
\begin{eqnarray}
&&b_k=
\left\{
\begin{array}{lll}
||\bz_k||^2,&\ k=1,\ldots,K_P,\\
||\bor_{k-K_P}||^2,&\ k=K_P+1,\ldots,K_P+K_S,
\end{array}
\right.\nonumber\\
&&\bQ_{1,k}=
\left\{
\begin{array}{lll}
\bI,&\ \ \ k=1,\ldots,K_P,\\
\bm{\Gamma}_1,&\ \ \ k=K_P+1,\ldots,K_P+K_S.
\end{array}
\right.
\end{eqnarray}
Now, let us observe that for all ${k=1,\ldots,K_P+K_S}$ the minimum eigenvalue of $\bD_{1,k}$ is $\delta_{1,k}=\frac{b_k}{\sigma^2+q_{1,k}\lambda_1}$ and the corresponding eigenvector is $\bm{e}_1=[1,0,\ldots,0]^T$, where $q_{1,k}$ is the square of $\bQ_{1,k}(1,1)$.
Therefore, for a given $k\in \left\{1,\ldots,K_P+K_S\right\}$, by the {\em Rayleigh-Ritz Theorem} \cite{Horn_Johnson1991matrixanalysis}, the unitary matrix $\bm{U}$ satisfying $\bm{U}^\dag\bm{\xi}_k=\bm{e}_1$ minimizes $\bm{\xi}_k^\dag\bm{U}\bD_{1,k}\bm{U}^\dag\bm{\xi}_k$, and the minimum is $\delta_{1,k}$.
However, it is not possible to obtain a $\bm{U}$ such that $\bm{U}^\dag\bm{\xi}_k=\bm{e}_1$, $\forall{k}\in\left\{1,\ldots,K_P+K_S\right\}$. For this reason, we resort to a suboptimum approach which
consists in minimizing the norm of the residual error between $\bm{U}^\dag\bm{\xi}_k$ and $\bm{e}_1$ for all $k = 1,\ldots,K_P+K_S$. Specifically, we estimate $\bm{U}$ by solving
\begin{eqnarray}\label{hatU}
&\min\limits_{\bm{U}\atop{\bm{U}^\dag\bm{U}=\bm{U}\bm{U}^\dag=\bm{I}}}
\sum\limits_{k=1}^{K_P+K_S}||\bm{U}^\dag\bm{\xi}_k-\bm{e}_1||^2,
\end{eqnarray}

The above problem can be written as
\begin{eqnarray}
&\min\limits_{\bm{U}\atop{\bm{U}^\dag\bm{U}=\bm{U}\bm{U}^\dag=\bm{I}}}&\sum_{k=1}^{K_P+K_S}\left[\bm{\xi}_k^\dag\bm{U}\bm{U}^\dag\bm{\xi}_k
-\bm{\xi}_k^\dag\bm{U}\bm{e}_1\right.\nonumber\\
&&\left.-\bm{e}_1^\dag\bm{U}^\dag\bm{\xi}_k+\bm{e}_1^\dag\bm{e}_1\right]\nonumber\\
\Rightarrow&\min\limits_{\bm{U}\atop{\bm{U}^\dag\bm{U}=\bm{U}\bm{U}^\dag=\bm{I}}}&\sum_{k=1}^{K_P+K_S}
\left[2-\bm{\xi}_k^\dag\bm{U}\bm{e}_1-\bm{e}_1^\dag\bm{U}^\dag\bm{\xi}_k\right]\nonumber\\
\Rightarrow&\min\limits_{\bm{d}:||\bm{d}||^2=1}&\left[2-\bm{\mu}^\dag\bm{d}-\bm{d}^\dag\bm{\mu}\right]\nonumber\\
\Rightarrow&\min\limits_{\bm{d}:||\bm{d}||^2=1}&\left[||\bm{d}-\bm{\mu}||^2+C\right],
\end{eqnarray}
where $\bm{d}=\bm{U}\bm{e}_1$, $\bm{\mu}=\sum_{k=1}^{K_P+K_S}\bm{\xi}_k$ and $C$ is a constant. The last problem can be further recast as follows
\begin{equation}
\left\{
\begin{array}{lll}
\min\limits_{\bm{d}}||\bm{d}-\bm{\mu}||^2,\\
\text{subject \ to} \ ||\bm{d}||^2=1.
\end{array}
\right.
\end{equation}
Using the Lagrange multipliers method, the Lagrangian is given by
\begin{equation}\label{L}
L(\bm{d},\lambda)=||\bm{d}-\bm{\mu}||^2-\lambda(||\bm{d}||^2-1)
\end{equation}
and setting to zero the first derivative of \eqref{L} with respect to $\bm{d}$ and $\lambda$, we come up with
\begin{equation}
\left\{
\begin{array}{lll}
\widehat{\lambda}=1-||\bm{\mu}||,\\
\widehat{\bm{d}}=\frac{\bm{\mu}}{||\bm{\mu}||}.
\end{array}
\right.
\end{equation}

It follows that $\bm{U}$ is any unitary matrix which rotates $\bm{e}_1$ to $\widehat{\bm{d}}$ and is given, for instance, by resorting to the singular value decomposition of $\widehat{\bm{d}}$. Notice that such a choice ensures that for all $k=1,\ldots,K_P+K_S$, the first component of $\bm{\xi}_k$ is greater than the remaining.

\bibliographystyle{IEEEtran}
\bibliography{group_bib_dan}

\begin{IEEEbiography}
[{\includegraphics[width=1in,height=1.3in]{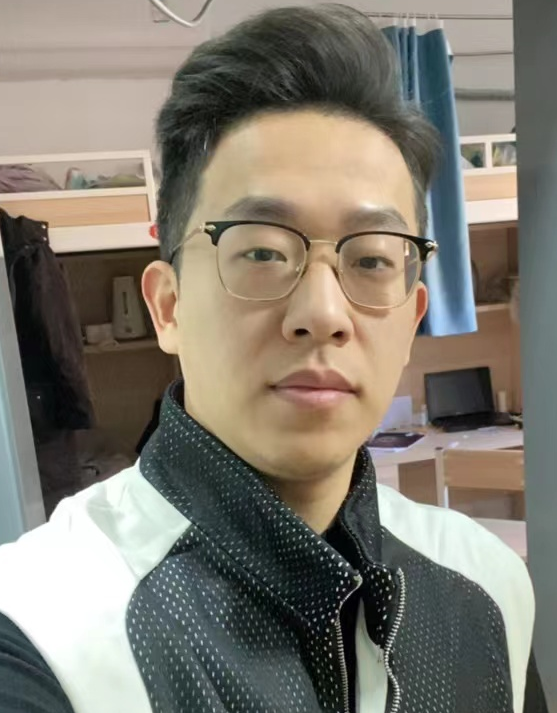}}]
{Chaoran Yin}
received the B.S. degree in
electromagnetic fields and wireless technology from Northwestern Polytechnical University,
Xi'an, Shaanxi, China, in 2018. He is currently working
toward the Ph.D. degree in signal and information
processing with Institute of Acoustics,
Chinese Academy of Sciences, and University of
Chinese Academy of Sciences, Beijing, China.
\end{IEEEbiography}

\begin{IEEEbiography}
[{\includegraphics[width=1in,height=1.3in]{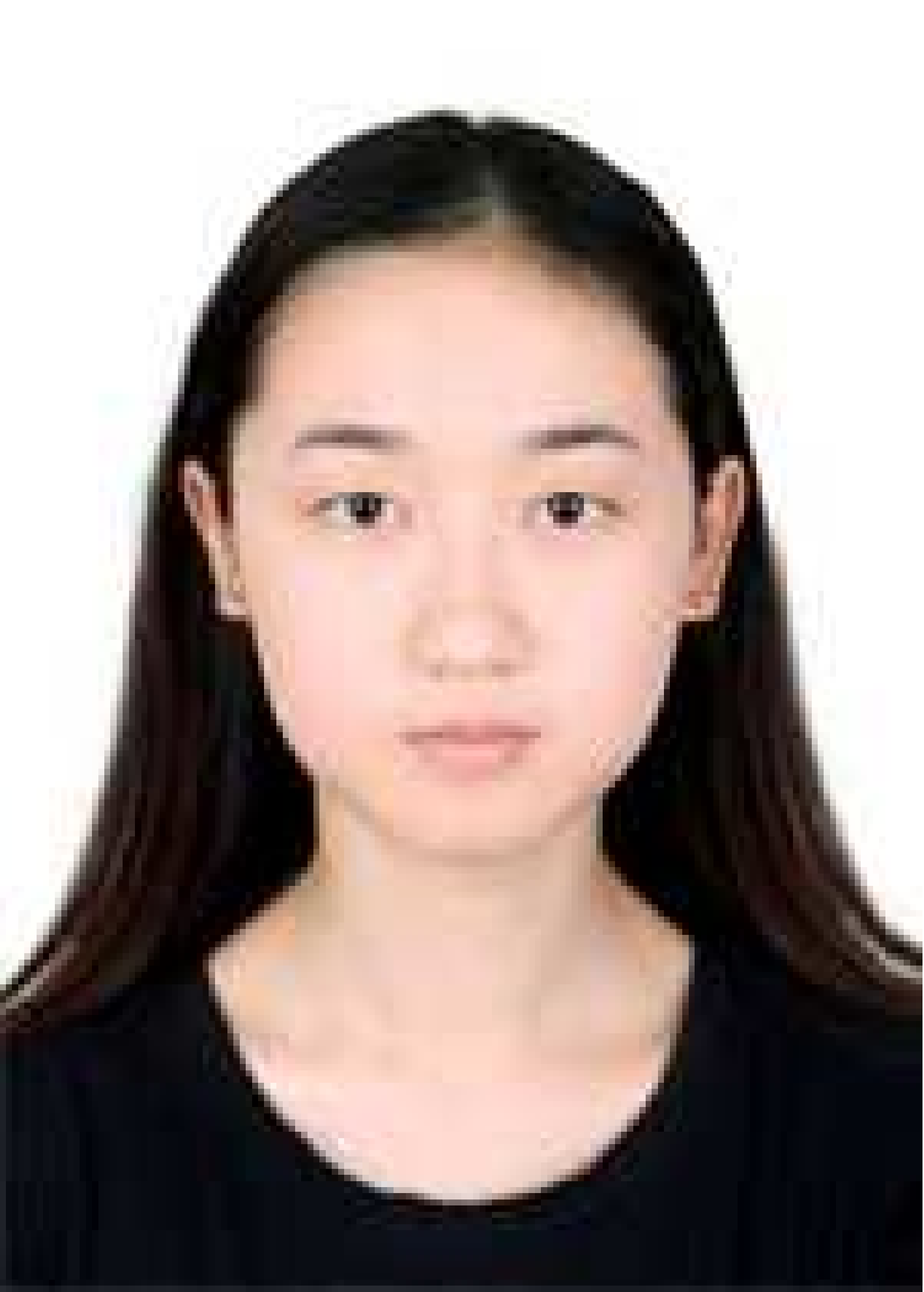}}]
{Linjie Yan}
received the B.E. degree in communication engineering from Shandong University of Science and Technology, and the Ph.D. degree in signal and information processing from the Institute of Acoustics, Chinese Academy of Sciences, Beijing, China, in 2016 and 2021, respectively. She is currently a Post-doctor with the Institute of Acoustics, Chinese Academy of Sciences, Beijing, China. Her research interests include statistical signal processing with more emphasis on adaptive sonar and radar signal processing.
\end{IEEEbiography}

\begin{IEEEbiography}
[{\includegraphics[width=1in,height=1.25in]{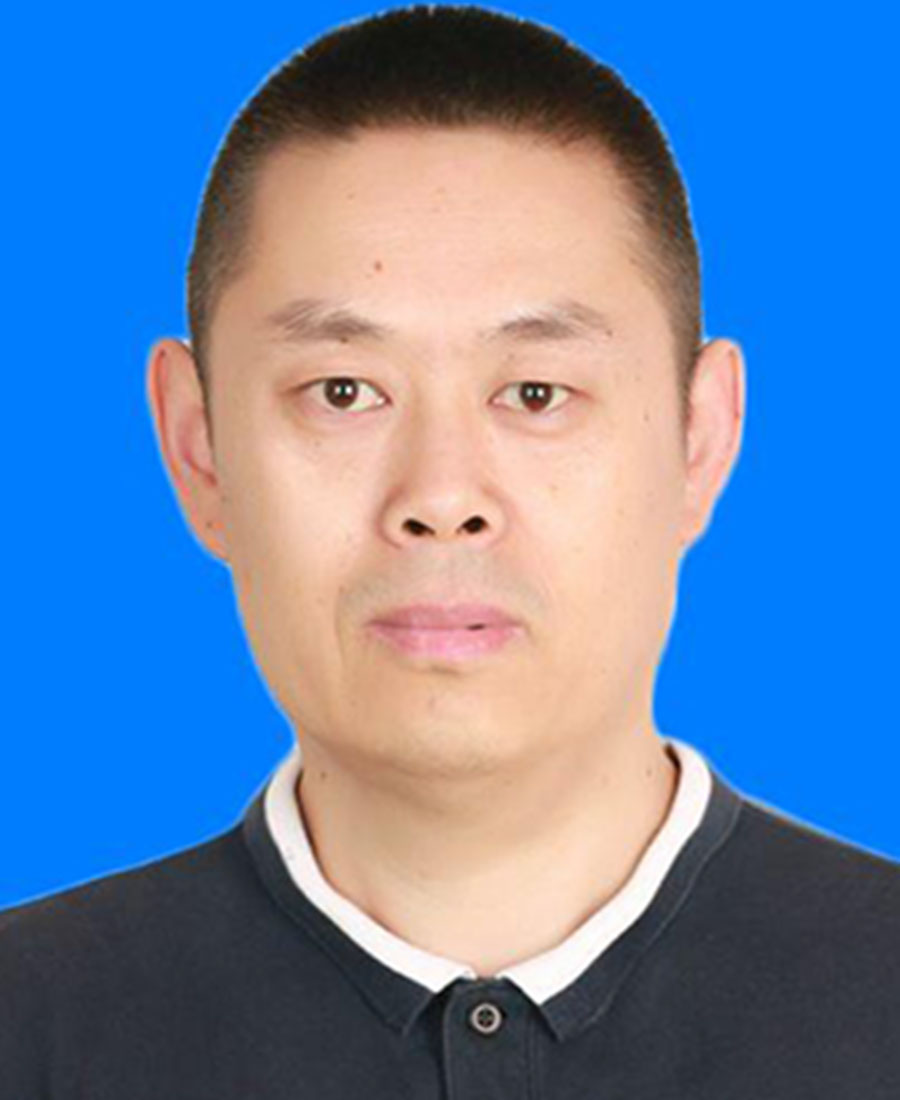}}]
{Chengpeng Hao}
(Senior Member, IEEE) received
the B.S. and M.S. degrees in electronic
engineering from Beijing Broadcasting Institute,
Beijing, China, in 1998 and 2001, respectively,
and the Ph.D. degree in signal and information
processing from the Institute of Acoustics, Chinese
Academy of Sciences, Beijing (IACAS),
China, in 2004.

He is currently a full professor with IACAS. He has held a visiting position with the Electrical and Computer Engineering Department, Queens University, Kingston, ON, Canada, from July 2013 to July 2014. He has authored or coauthored more than 160 journal and conference papers. His research interests are in the fields of statistical signal processing and array signal processing. Dr. Hao is currently serving as an Associate Editor for several international journals, including IEEE ACCESS, {\em Signal, Image and Video Processing} (Springer), and the {\em Journal of Engineering} (IET).
\end{IEEEbiography}

\begin{IEEEbiography}
[{\includegraphics[width=1in,height=1.25in]{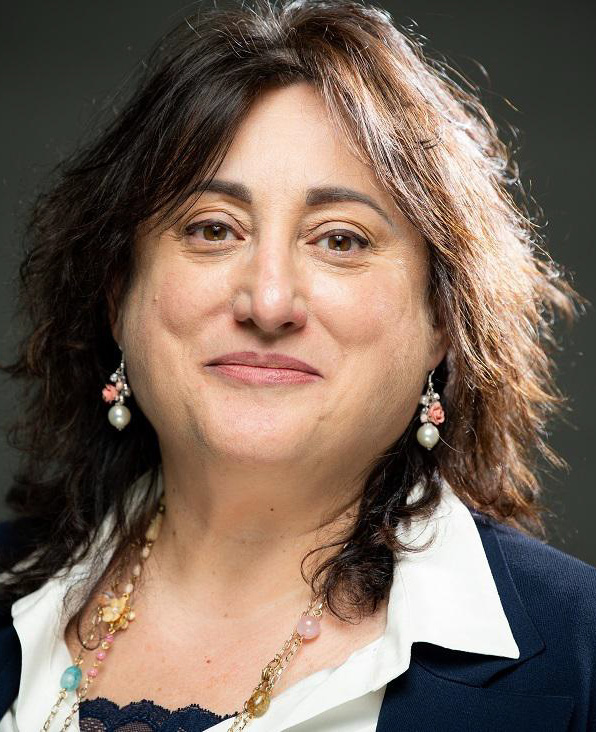}}]
{Silvia Liberata Ullo} (Senior Member, IEEE)
received the degree (cum laude) in electronic
engineering from the Faculty of Engineering,
Federico II University, Naples, Italy, in 1989, and
the M.Sc. degree from the Sloan Business School
of Boston, Massachusetts Institute of Technology
(MIT), Cambridge, MA, USA, in June 1992.

She is an Industry Liaison for the IEEE Joint
ComSoc/Vehicular Technology Membership (VTS)
Italy Chapter. She was the National Referent for
International Federation of Business and Professional
Women (FIDAPA BPW) Italy Science and Technology Task Force
from 2019 to 2021. She has been a researcher with the Department of Engineering,
University of Sannio, Benevento, Italy, since 2004. She is a Member
of the Academic Senate and the Ph.D. Professors Board. She
is teaching signal theory and elaboration, and telecommunication networks
for electronic engineering, and optical and radar remote sensing for the
Ph.D. course. She has authored 80+ research articles, coauthored many book
chapters and served as the Editor of two books, and many special issues in
reputed journals of her research sectors. Her main interests include signal
processing, remote sensing, satellite data analysis, machine learning and
quantum ML, radar systems, sensor networks, and smart grids. She has
worked in the private and public sector from 1992 to 2004, before joining the
University of Sannio.
\end{IEEEbiography}

\begin{IEEEbiography}
[{\includegraphics[width=1in,height=1.25in]{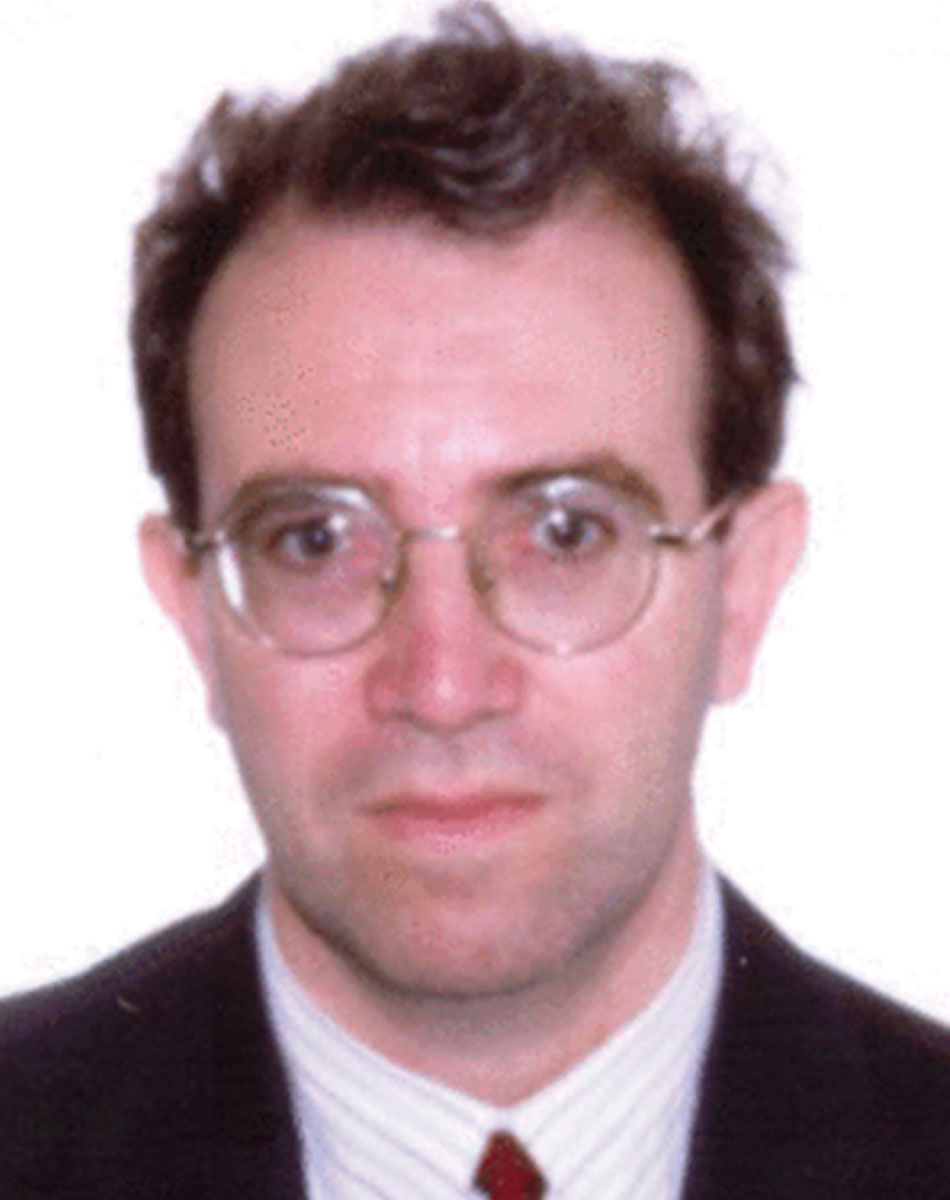}}]
{Gaetano Giunta} (Senior Member, IEEE) received
the M.S. degree electronic engineering from the University
of Pisa, Pisa, Italy and the Ph.D. degree in
information and communication engineering from the
University of Rome La Sapienza, Rome, Italy, in
1985 and 1990, respectively. During 1989-1990, he
was also a Research Fellow with Signal Processing
Laboratory, the \'Ecole Polytechnique F\'ed\'erale de Lausanne,
Lausanne, Switzerland. In 1992, he became
an Assistant Professor with the INFO-COM Department,
University of Rome La Sapienza. In 1998, he
taught signal processing with the University of Roma Tre, Rome, Italy. From
2001 to 2005, he was with the University of Roma Tre as an Associate Professor
of telecommunications. Since 2005, he has been a Full Professor of telecommunications
with the same University. His main research interests include signal
and image processing for mobile communications and wireless networks. He
is a Member of the IEEE Societies of Communications, Signal Processing, and
Vehicular Technology. He was also a Reviewer for various IEEE Transactions,
IET (formerly IEE) proceedings, and EURASIP journals, and a TPC Member
for various international conferences and symposia in the same fields.
\end{IEEEbiography}

\begin{IEEEbiography}
[{\includegraphics[width=1in,height=1.25in]{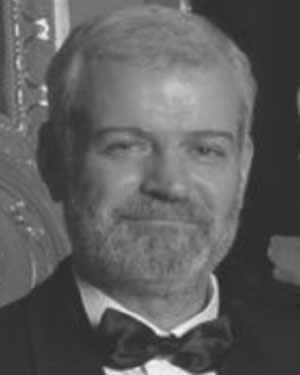}}]
{Alfonso Farina} (Life Fellow, IEEE) received the
Doctoral Laurea - degree in electronic engineering
from the University of Rome, Rome, Italy, in 1973.
In 1974, he joined SELENIA S.P.A., then Selex ES,
where he became the Director of the Analysis of
Integrated Systems Unit, and the Director of Engineering
of the Large Business Systems Division. In
2012, he was the Senior VP and the Chief Technology
Officer (CTO) of the Company, reporting directly to
the President. From 2013 to 2014, he was a Senior
Advisor to the CTO. He retired in October 2014.
From 1979 to 1985, he was also a Professor of radar techniques with the
University of Naples, Naples, Italy. He is currently a Visiting Professor with
the Department of Electronic and Electrical Engineering, University College
London, London, U.K., and with the Centre of Electronic Warfare, Information
and Cyber, Cranfield University, Cranfield, U.K., a Distinguished Lecturer of the
IEEE Aerospace and Electronic Systems Society and a Distinguished Industry
Lecturer for the IEEE Signal Processing Society during January 2018-December
2019. He is a Consultant to Leonardo S.p.A. Land and Naval Defence Electronics
Division, Rome. He has authored or coauthored more than 800 peer-reviewed
technical papers and books and monographs (published worldwide), some of
them also translated in to Russian and Chinese. Dr. Farina has been an IEEE
Signal Processing Magazine Senior Editorial Board Member (three-year term)
since 2017. He was the Conference General Chairman of the IEEE Radar Con
2008, Rome, Italy, May 2008, and the Executive Conference Chair at the
International Conference on Information Fusion, Florence, Italy, July 2006. He
is the Honorary Chair of IEEE Radar Conf September 2020, Florence. He was the
recipient of many awards, including Leader of the team that won the First Prize
of the first edition of the Finmeccanica Award for Innovation Technology, out
of more than 330 submitted projects by the Companies of Finmeccanica Group
in 2004, an International Fellow of the Royal Academy of Engineering, U.K.
in 2005, the fellowship was presented to him by HRH Prince Philip, the Duke
of Edinburgh, IEEE Dennis J. Picard Medal for Radar Technologies and Applications
for Continuous, Innovative, Theoretical, and Practical Contributions to
Radar Systems and Adaptive Signal Processing Techniques in 2010, IEEE Honor
Ceremony, Montreal, Canada, Oscar Masi Award for the AULOS green radar by
the Italian Industrial Research Association (AIRI) in 2012, IET Achievement
Medal for Outstanding contributions to radar system design, signal, data and
image processing, and data fusion in 2014, IEEE SPS Industrial Leader Award
for contributions to radar array processing and industrial leadership in 2017, 2019
Christian Hulsmeyer Award from the German Institute of Navigation (DGON),
with the motivation: In appreciation of his outstanding contribution to radar
research and education. He was the recipient of Best Paper Awards, such as the
B. Carlton of IEEE TRANSACTIONS ON AEROSPACE AND ELECTRONIC SYSTEMS
in 2001, 2003, and 2013, IET-Proceeding on Radar Sonar and Navigation during
2009-2010, and International Conference on Information Fusion in 2004. In the
past few decades, he has been collaborating with several professional journals
and conferences (mainly on radar and data fusion) as an Associate Editor,
Reviewer, organizer of special issues, Session Chairman, and Plenary Speaker.
He is a Fellow of the IET, a Fellow of the Royal Academy of Engineering, and a
Fellow of EURASIP. Since 2017, he has been the Chair of Italy Section Chapter,
IEEE AESS-10, and a Member of the IEEE AESS BoG, Jan 2019-Dec 2021. In
26 October 2018, he was interviewed, among few top managers of the Company,
at RAI Storia for the 70th anniversario di Leonardo Company (70th anniversary
of Leonardo Company). Since 2010, he has also been a Fellow of EURASIP.
Starting from 2020, he is a Member of European Academy of Science.
\end{IEEEbiography}

\begin{IEEEbiography}
[{\includegraphics[width=1in,height=1.25in]{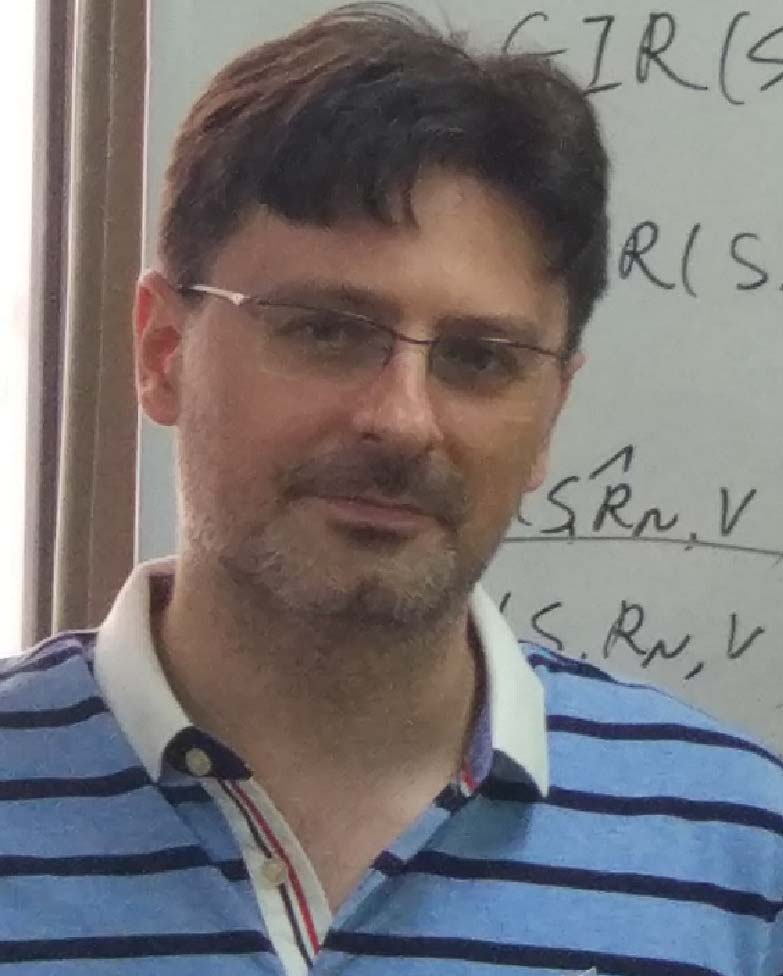}}]
{Danilo Orlando}
(Senior Member, IEEE) was born
in Gagliano del Capo, Italy, in August 1978.
He received the Dr. Eng. degree (Hons.) in computer
engineering and the Ph.D. degree (Hons.)
in information engineering from the University of
Salento (formerly University of Lecce), Lecce, Italy,
in 2004 and 2008, respectively. From July 2007 to
July 2010, he was with the University of Cassino,
Cassino, Italy, engaged in a research project on algorithms
for track-before-detect of multiple targets in
uncertain scenarios. From September to November
2009, he was a Visiting Scientist with the NATO Undersea Research Centre,
La Spezia, Italy. From September 2011 to April 2015, he was with Elettronica
S.p.A. engaged as a System Analyst in the field of electronic warfare. In May
2015, he joined the Universit\`a degli Studi Niccol\`o Cusano, Rome, Italy,
where he is currently an Associate Professor. In 2007, he has held visiting
positions with the Department of Avionics and Systems, ENSICA (now
Institut Sup\'erieur de l'A\'eronautique et de l'Espace, ISAE), Toulouse, France,
and from 2017 to 2019, he was with the Chinese Academy of Science, Beijing,
China. He is the author or coauthor of more than 150 scientific publications
in international journals, conferences, and books. His main research interests
include statistical signal processing with more emphasis on adaptive detection
and tracking of multiple targets in multisensor scenarios. He was a Senior
Area Editor of the IEEE TRANSACTIONS ON SIGNAL PROCESSING. He is
currently an Associate Editor for the IEEE OPEN JOURNAL ON SIGNAL
PROCESSING, {\em EURASIP Journal on Advances in Signal Processing}, and
{\em Remote Sensing} (MDPI).
\end{IEEEbiography}
\end{document}